\documentclass{article}

\usepackage{arxiv}
\usepackage[T1]{fontenc}
\usepackage[utf8]{inputenc}
\usepackage{textcomp}
\usepackage{amsmath}
\usepackage{amssymb}
\usepackage{graphicx}
\usepackage{booktabs}
\usepackage{siunitx}
\usepackage{xcolor}
\usepackage{listings}
\usepackage{float}
\usepackage[section]{placeins}
\usepackage{caption}
\usepackage{array}
\newcolumntype{P}[1]{>{\raggedright\arraybackslash}p{#1}}

\usepackage[numbers,sort&compress]{natbib}
\makeatletter
\newcommand{\arxivfrontmatter}[2]{%
  \begingroup
    \renewcommand{\thefootnote}{\fnsymbol{footnote}}%
    \long\def\@makefntext##1{%
      \parindent 1em\noindent
      \hbox to 1.8em{\hss $\m@th ^{\@thefnmark}$}##1
    }%
    \thispagestyle{empty}%
    \if@twocolumn
      \twocolumn[{%
        \@maketitle
        \centerline{\large\bfseries\scshape Abstract}%
        \vskip 0.5em
        {\par\leftskip 2em \rightskip 2em #1\par}%
        {\leftskip 2em \rightskip 2em \keywords{#2}\par}%
        \vskip 0.25in
      }]%
    \else
      \@maketitle
      \begin{abstract}#1\end{abstract}%
      \keywords{#2}%
    \fi
    \@thanks
  \endgroup
  \let\maketitle\relax
  \let\thanks\relax
}
\makeatother

\usepackage[hidelinks]{hyperref}
\usepackage{orcidlink}

\graphicspath{{figures/}}

\definecolor{efobsmod}{HTML}{2c7fb8}
\title{EuroFlood: A queryable index for satellite-derived flood depths in Europe}

\author{%
  J\"urgen Hackl~\orcidlink{0000-0002-8849-5751}\thanks{Corresponding author:
    \texttt{hackl@princeton.edu}} \\
  Complex Infrastructure Systems Group \\
  Princeton University \\
  Princeton, NJ 08544, USA
}
\date{\today}

\renewcommand{\shorttitle}{EuroFlood: indexing satellite flood-depth maps for Europe}
\renewcommand{\headeright}{}
\renewcommand{\undertitle}{}

\hypersetup{
  pdftitle={EuroFlood: A queryable index for satellite-derived flood depths in Europe},
  pdfauthor={Juergen Hackl},
  pdfkeywords={flood depth, Sentinel-1, open-source software, spatial indexing, flood recurrence, Copernicus EMS},
}

\begin{document}
\arxivfrontmatter{%
Satellite-derived observations of flood water depth support flood model validation and impact assessment, yet the only open continental-scale archive of such observations, the flood-depth maps of the Copernicus Emergency Management Service, is distributed as several thousand raster files whose sole spatial metadata is a coordinate encoded in each filename.
This paper presents EuroFlood, an open-source Python library and a published spatial index that make the archive queryable.
The index records which archived events inundated each grid cell, enabling event discovery, footprint reconstruction, and recurrence analysis directly from the index, while high-resolution depth rasters are retrieved only when needed.
It reproduces the archived event footprints losslessly at its grid.
A region-level corroboration audit against an independent flood-impact database shows stronger agreement for river floods, while results for flash floods are less conclusive.
Applications demonstrate continental recurrence mapping, comparison of observed with modelled flood extents, and event-based exposure assessment.
}{flood depth \and Sentinel-1 \and open-source software \and spatial indexing \and flood recurrence \and Copernicus EMS}

% ===========================================================================
% ===========================================================================

% ===========================================================================
\section{Introduction}
\label{sec:intro}

Floods are Europe's costliest weather-related hazard.
Between 1980 and 2024, weather- and climate-related extremes caused an estimated EUR 822 billion in economic losses across the European Union, with floods accounting for the largest share \citep{eea_losses}.
Individual events can have severe consequences; the July 2021 floods in western Germany, Belgium, and the Netherlands caused more than 220 fatalities and approximately EUR 46 billion in damage \citep{mohr2023part1,tradowsky2023}.
Reported flood frequency and losses have increased in recent decades, partly reflecting growing exposure and improved reporting \citep{jongman2012,delforge2025emdat}, while climate projections indicate further increases in flood risk \citep{dottori2020peseta}.

The EU Floods Directive requires member states to map flood hazard and risk at multiple probability levels and to update these maps regularly \citep{floodsdirective}.
Because these products are predominantly model-based, their reliability requires evaluation against observations \citep{trigg2016,bernhofen2018}.
Modelled hazard data are widely available \citep{dottori2016,sampson2015}, and satellite services increasingly provide observations of flood extent \citep{tellman2021}, but observations of water depth remain scarce.
This gap is important because inundation depth is a primary input to flood-damage estimation, with losses generally increasing with water depth \citep{huizinga2017,wing2020,jongman2012}.

\begin{figure}[htbp]\centering
\includegraphics[width=\linewidth]{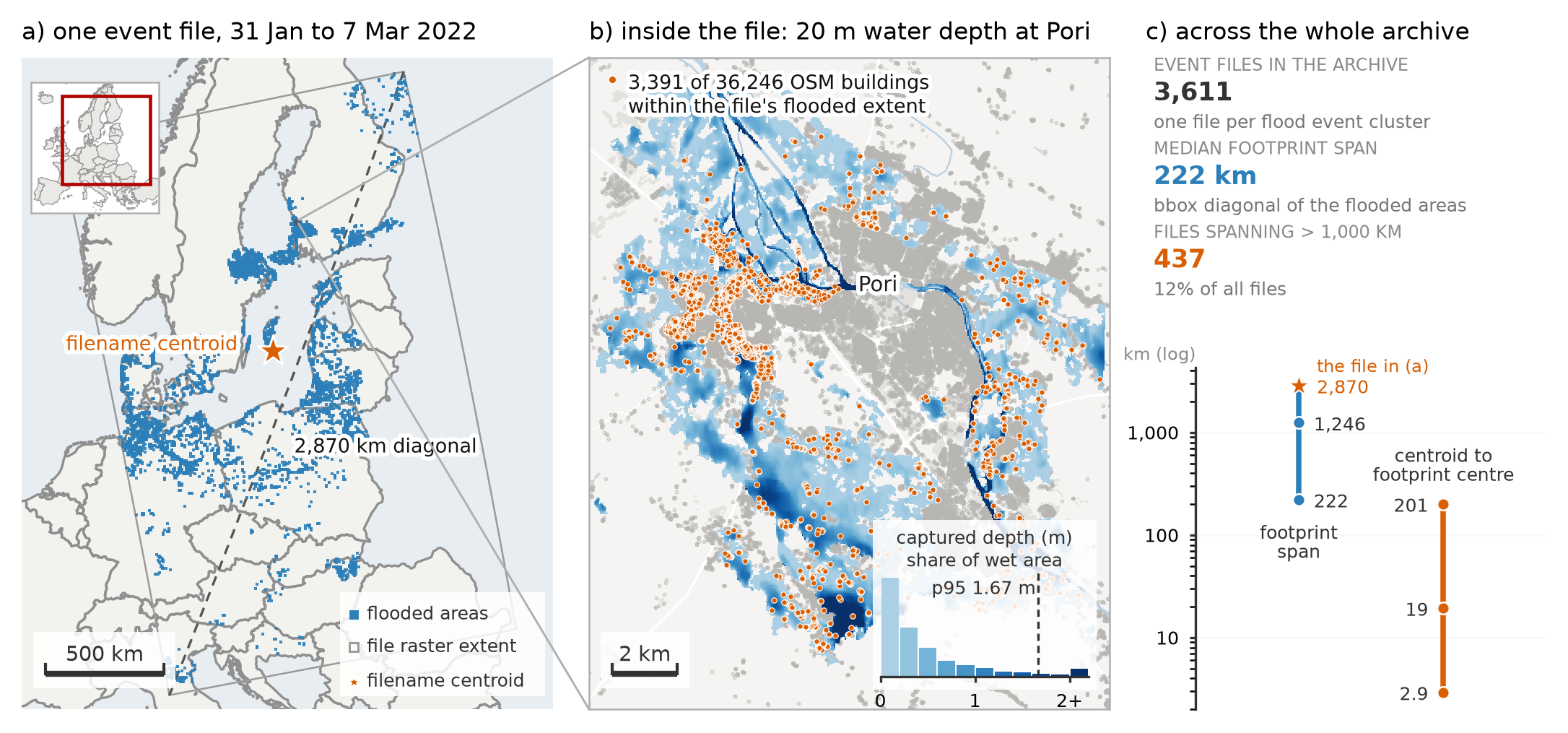}
\caption{A single event file can contain flooded areas scattered across Europe, making the centroid encoded in its filename an unreliable indicator of where flooding occurred. (a)~All flooded areas recorded in the archive's widest event file, a winter cluster observed between 31~January and 7~March 2022. The grey rectangle marks the extent of the file's raster tile, the dashed diagonal crosses the tile corner to corner, the flooded areas the file bundles span \qty{2870}{\km} along the diagonal, and the star marks the filename
centroid. This centroid, the only location metadata available without opening the file, lies in the open Baltic Sea. (b)~The archive's underlying product within the same file: per-pixel water depth at \qty{20}{\m} resolution along the Kokem\"aenjoki at Pori, with buildings inside the flooded extent marked in orange and counted in the panel. (c)~Median, 90th percentile, and maximum footprint span and centroid-to-footprint-centre distance across all 3,611 event files on a logarithmic scale. In total, 437 files span more than \qty{1000}{\km}, and the star marks the file shown in panel~(a).}
\label{fig:centroid}
\end{figure}

The CEMS-EFAS (Copernicus Emergency Management Service, European Flood Awareness System) satellite-derived flood-depth maps partly address this gap.
They provide per-pixel water depths reconstructed from Sentinel-1 flood extents for dated events across Europe from 2015--2025 at \qty{20}{\m} resolution \citep{betterle2025dataset,betterle2024flexth}.
Although openly available, the archive is difficult to query systematically.
Its 3,611 GeoTIFF files are distributed through a flat HTTP directory without an API, catalogue, or spatial index, and the only spatial metadata available without opening a file is a centroid encoded in its filename.
A single event file may contain flooded areas separated by up to \qty{2870}{\km}, making this centroid a poor representation of the event footprint (Fig.~\ref{fig:centroid}).
Filename-based filtering consequently misses many events relevant to a given location.
Collectively, the event footprints also encode information about observed flood recurrence, but extracting it requires identifying and processing the relevant files.

Existing tools address related but different problems.
Satellite services such as the Copernicus Global Flood Monitoring service provide observed flood extent \citep{krullikowski2023gfm,tellman2021}, while hazard-modelling frameworks and geospatial platforms provide access to modelled return-period scenarios \citep{aznar2019climada,gorelick2017gee}.
Domain-specific libraries such as \texttt{argopy}, \texttt{climate4R}, and \texttt{HyRiver} demonstrate how large environmental archives can be exposed through unified programming interfaces \citep{maze2020argopy,iturbide2019climate4r,chegini2021hyriver}.
However, no comparable programmatic access layer exists for the CEMS-EFAS archive of observed flood depth (Section~\ref{sec:related}).

To address this gap, this paper presents EuroFlood, an open-source Python library and spatial index that make the CEMS-EFAS archive discoverable, queryable, and analysis-ready.
At its core is an inverted raster index in which each $\sim$\qty{90}{\m} grid cell stores the identifiers of all archived events that inundated it.
This representation reconstructs event footprints and recurrence counts without accessing the corresponding depth rasters.
A two-stage Discover$\rightarrow$Extract interface first identifies events affecting an area of interest and then retrieves native-resolution raster data only for those events.
The same workflow also provides access to modelled CEMS-GLOFAS return-period hazard maps for comparison with observed inundation.

The design objectives are event discovery without scanning source rasters (O1), preservation of event membership at the index resolution with quantified resampling effects (O2), efficient construction and querying (O3), version-pinned reproducibility across delivery modes (O4), and characterisation of the archive's fitness for use (O5).
The primary contribution is the index and its software interface; characterisation of the underlying archive is secondary.
Specifically, the paper contributes:

\begin{enumerate}

\item[(i)] A versioned and citable inverted raster index that reconstructs archived event footprints exactly at its $\sim$\qty{90}{\m} grid and supports event discovery, footprint reconstruction, and recurrence analysis without accessing the source depth rasters (Section~\ref{sec:index}).

\item[(ii)] A Discover$\rightarrow$Extract API that binds analyses to specific library, index, and dataset versions, enabling reproducible online and offline workflows (Section~\ref{sec:generality}).

\item[(iii)] A fitness-for-use characterisation based on corroboration against the independent HANZE (Historical Analysis of Natural Hazards in Europe) flood-impact database, stratified by flood type, year, and country, together with case studies illustrating applications and detectability limits (Sections~\ref{sec:validation} and~\ref{sec:application}).

\end{enumerate}

The remainder of the paper is organised as follows.
Section~\ref{sec:data} introduces the observed and modelled data collections, and Section~\ref{sec:related} reviews related flood products and software tools.
Sections~\ref{sec:index} and~\ref{sec:software} present the index and the EuroFlood library, respectively.
Section~\ref{sec:validation} verifies the index, benchmarks its query performance, and evaluates corroboration against documented floods.
Sections~\ref{sec:application}--\ref{sec:conclusions} present the applications, discuss the archive's fitness for use and limitations, and summarise the main conclusions.
The flood-science interpretation of the archive is developed in the companion study \citep{betterle_sciadv}; the present paper provides the access and indexing infrastructure that makes such analyses practical and reproducible.

% ===========================================================================
\begin{table}[htbp]\centering
\caption{Flood data products related to the collections used in this work,
organised by information type (observed or modelled) and variable (extent or
per-pixel depth). RP denotes return-period scenarios.}
\label{tab:products}
\footnotesize
\setlength{\tabcolsep}{3pt}
\resizebox{\textwidth}{!}{%
\begin{tabular}{@{}llllll@{}}
\toprule
Product & Type & Variable & Access & Coverage & Reference \\
\midrule
CEMS-EFAS flood-depth maps & Observed & Depth  & CC-BY-4.0 & Europe, 2015--2025, \qty{20}{\m} & \citet{betterle2025dataset} \\
CSIRO Murray--Darling depth maps & Observed & Depth & CC-BY-4.0 & Murray--Darling, 1988--2022, \qty{30}{\m} & \citet{penton2023mdb} \\
ICEYE Flood Insights & Observed & Extent, depth & Commercial & Global, per event & \citet{ardila2022iceye} \\
\midrule
CEMS-GLOFAS river flood hazard (fluvial) & Modelled, RP & Depth & CC-BY-4.0 & Global, $\sim$\qty{90}{\m} & \citet{baugh2024floodhazard} \\
WRI Aqueduct (fluvial, coastal) & Modelled, RP & Depth & Open & Global, $\sim$\qty{1}{\km} & \citet{aqueduct} \\
Deltares coastal flood maps & Modelled, RP & Depth & Open & Global coastal, \qty{90}{\m}--\qty{5}{\km} & \citet{deltares2021coastal} \\
Fathom Global Flood Map (multi-peril) & Modelled, RP & Depth & Commercial & Global, $\sim$\qty{30}{\m} & \citet{wing2024fathom} \\
\midrule
GFM (Sentinel-1) & Observed & Extent & Open (registration) & Near-global, 2015--, \qty{20}{\m} & \citet{krullikowski2023gfm} \\
OPERA DSWx & Observed & Extent & Open & Near-global, 2023--, \qty{30}{\m} & \citet{opera_dswx} \\
Global Flood Database & Observed & Extent & CC-BY-NC-4.0 & Global, 2000--2018, \qty{250}{\m} & \citet{tellman2021} \\
DFO / NASA MODIS--VIIRS & Observed & Extent & Open & Global, $\sim$\qty{250}{\m} & \citet{dfo} \\
\bottomrule
\end{tabular}}
\end{table}
% ===========================================================================

% ===========================================================================
\section{Data}
\label{sec:data}

EuroFlood provides access to two complementary CEMS data collections: an archive of observed flood events and modelled flood-hazard scenarios.
This section summarises the provenance, generation, coverage, and principal limitations of both datasets; the index and software built on them are described in Sections~\ref{sec:index} and~\ref{sec:software}.

\subsection{Observed events: CEMS-EFAS satellite-derived flood-depth maps}

The CEMS-EFAS \emph{Satellite-Derived Flood Depth Maps for Europe} are, to our knowledge, the only open continental-scale archive providing per-pixel water depth for observed flood events \citep{betterle2025dataset}.
The Joint Research Centre (JRC) derives the archive from Sentinel-1 observations in three stages.
First, the Copernicus Global Flood Monitoring (GFM) service delineates inundation at \qty{20}{\m} resolution using an ensemble of three independent detection algorithms \citep{krullikowski2023gfm,wagner2026}.
These detections are then consolidated and clustered in space and time into flood events, with terrain information used to reduce under-detection and false positives \citep{betterle2025dataset}.
Finally, FLEXTH reconstructs water depth by interpolating terrain elevations sampled along the wet--dry boundary and subtracting the underlying terrain elevation \citep{betterle2024flexth}.
The method also extends inundation into adjacent masked areas lying below the estimated water surface.

Each event is distributed as a \qty{20}{\m} GeoTIFF of reconstructed water depth on the Equi7 Europe grid (EPSG:27704).
The archive event unit is a spatio-temporal cluster of observations and does not necessarily correspond to a single hydrological flood.
Filenames record the observation period, event duration, and one centroid coordinate.
The archive state analysed here contains 3,611 events; all but seven begin between 2015 and 2025, with the remaining seven beginning on 29~December 2014.
The 331 files added for the 2025 observation year were incorporated into the release update described in Section~\ref{sec:construction}.

Three properties are important for interpretation.
First, Sentinel-1 may miss inundation in urban areas, beneath vegetation, and during short-lived events that develop and recede between satellite acquisitions.
Absence from the archive therefore does not imply absence of flooding, as examined in Sections~\ref{sec:validation} and~\ref{sec:application}.
Second, water depth is reconstructed rather than measured directly.
FLEXTH reported root-mean-square errors of \qtyrange{0.28}{0.62}{\m} and mean absolute errors of \qtyrange{0.20}{0.40}{\m} against hydrodynamic reference simulations at two river sites, with larger errors expected in steep terrain \citep{betterle2024flexth}.
Depth estimates should therefore be interpreted at decimetre rather than centimetre precision.
Third, observation opportunities vary over time.
The Sentinel-1 constellation provided two-satellite coverage only from 2017 to 2021 before the loss of Sentinel-1B, while Sentinel-1C restored two-satellite operations during 2025 \citep{torres2012,esa2022s1b,esa2024s1c,copernicus2025s1c}.
Detection opportunities consequently vary among years and regions.

The archive structure also constrains access.
The 3,611 event rasters are hosted in a flat HTTP directory without an API, catalogue, or spatial index.
Files have a median size of \qty{4.4}{\mega\byte} and a mean of \qty{10.9}{\mega\byte}, and together occupy \qty{39.3}{\giga\byte}.
The only spatial metadata available without opening a raster is the centroid encoded in its filename.
Because a single event file may contain flooded areas separated by up to \qty{2870}{\km}, this centroid can be a poor representation of the event footprint (Fig.~\ref{fig:centroid}).

\subsection{Modelled hazard: CEMS-GLOFAS river flood hazard maps}

The second collection served by EuroFlood is the CEMS-GLOFAS global river flood hazard dataset.
It provides modelled riverine inundation depths for return periods of 10, 20, 50, 75, 100, 200, and 500 years \citep{baugh2024floodhazard}.
The maps are generated from LISFLOOD hydrological reanalysis, extreme-value analysis of river discharge, and two-dimensional LISFLOOD-FP inundation modelling using MERIT terrain data \citep{dottori2016,baugh2024floodhazard}.
They are distributed as GeoTIFF tiles at 3~arc-seconds, approximately \qty{90}{\m} resolution, with depth in metres; this study uses version 2.1.2.
The available return periods span the probability ranges considered in the EU Floods Directive \citep{floodsdirective}.

Several assumptions constrain interpretation.
The maps represent fluvial flooding only, omit pluvial and coastal processes, and assume no flood protection \citep{dottori2016}.
Basins smaller than \qty{500}{\km\squared} are excluded, limiting representation of small headwater systems.
A companion layer further identifies channels with upstream areas below \qty{3000}{\km\squared}, where simulated depths may be unreliable and should be masked in analyses of absolute water depth \citep{baugh2024floodhazard}.

The return-period layers represent statistical hazard scenarios rather than reconstructions of individual floods.
For example, RP100 denotes inundation associated with a 1\% annual exceedance probability, not an event expected to occur regularly once every 100 years.
Observed flood footprints may therefore extend beyond a given return-period envelope without contradicting the modelled probability.
EuroFlood consequently uses these maps as hazard scenarios for comparison with observed inundation rather than as reference data for formal validation.

The two collections are therefore complementary: CEMS-EFAS provides dated satellite-derived flood observations, whereas CEMS-GLOFAS provides modelled river-flood hazard scenarios defined by annual exceedance probability.
Section~\ref{sec:related} places both within the broader landscape of flood data products and access tools.

% ===========================================================================
% ===========================================================================

% ===========================================================================
\section{Related data products and tools}
\label{sec:related}

This section situates the two collections served by EuroFlood within the broader landscape of flood data products and environmental data-access software.
It first distinguishes products by the type of information they provide and then reviews the software through which comparable environmental archives are accessed.

\subsection{Related flood data products}

Flood data products can be classified along two dimensions: whether they represent observed events or modelled scenarios, and whether they provide flood extent or per-pixel water depth (Table~\ref{tab:products}).
The distinction between extent and depth is important for impact analysis because an extent mask identifies where flooding occurred, whereas per-pixel depth supports damage estimation through depth--damage relationships \citep{huizinga2017}.
Products exist in all four categories, but open datasets of observed water depth remain substantially less common than observed-extent or modelled-hazard products.

Observed flood extent is the most widely available category.
The Copernicus Global Flood Monitoring service derives flood extent from Sentinel-1 acquisitions at \qty{20}{\m} resolution and provides records from 2015 onwards \citep{krullikowski2023gfm,wagner2026}.
NASA's OPERA Dynamic Surface Water eXtent products map surface water from optical and radar observations at \qty{30}{\m} resolution \citep{opera_dswx}.
The Global Flood Database provides MODIS-derived extents for 913 events between 2000 and 2018 \citep{tellman2021} and has recently been extended into the Sentinel-1 era \citep{misra2025}.
Near-real-time services such as the Dartmouth Flood Observatory complement these retrospective archives by supporting operational flood monitoring \citep{dfo}.

Modelled flood depth is more commonly distributed as return-period scenarios.
Examples include the CEMS-GLOFAS river flood hazard maps \citep{dottori2016,baugh2024floodhazard}, the WRI Aqueduct flood maps \citep{aqueduct}, the Deltares global coastal flood maps \citep{deltares2021coastal}, and commercial products such as the Fathom Global Flood Map \citep{wing2024fathom}.
These products describe probabilistic hazard scenarios rather than dated observations of individual events.

Observed per-pixel water depth is available from fewer sources.
The CEMS-EFAS satellite-derived flood-depth maps introduced in Section~\ref{sec:data} provide this information at continental scale.
The CSIRO Murray--Darling Basin dataset provides two-monthly maximum water-depth maps for 1988--2022 at \qty{30}{\m} resolution, reconstructed from Landsat flood extents using the FwDET algorithm \citep{penton2023mdb,cohen2018fwdet}.
Commercial services such as ICEYE provide event-specific flood extent and depth derived from synthetic-aperture radar observations \citep{ardila2022iceye}, while rapid-mapping services such as UNOSAT publish floodwater-depth products for selected major events \citep{unosat2022pakistan}.
To our knowledge, the CEMS-EFAS archive remains the only open observed-depth product with continental, multi-country coverage.

\subsection{Related software tools}

Domain-specific access libraries are an established form of environmental research software.
Examples include \texttt{argopy} for the Argo float archive \citep{maze2020argopy}, \texttt{climate4R} for climate data collections \citep{iturbide2019climate4r}, \texttt{HyRiver} for United States hydrological data \citep{chegini2021hyriver}, \texttt{pDEMtools} for polar elevation models \citep{chudley2024pdemtools}, and \texttt{OSMnx} for urban networks derived from OpenStreetMap \citep{boeing2025osmnx}.
These libraries provide unified, analysis-ready interfaces to archives that would otherwise require users to manage heterogeneous files, formats, and access protocols directly.

Several related software families support flood and Earth-observation workflows, but they address different parts of the access problem.
\texttt{CLIMADA-petals} integrates hazard datasets, including GLOFAS maps, into a broader climate-risk modelling framework \citep{aznar2019climada}.
Google Earth Engine and the Microsoft Planetary Computer host curated geospatial catalogues and provide cloud-based analysis interfaces \citep{gorelick2017gee}, while GFM flood extents can be processed through openEO workflows \citep{schramm2021openeo}.
Generic clients such as \texttt{pystac-client} discover and access collections published through the SpatioTemporal Asset Catalog specification, whereas foundational libraries such as \texttt{rasterio} and \texttt{xarray} provide the raster and multidimensional-array functionality on which higher-level tools are built \citep{hoyer2017xarray}.

A more detailed comparison of these tools and their intended scope is provided in the Supplementary Information (Table~S11).
As of September 2026, no published software package provides a dedicated programmatic access layer for the CEMS-EFAS flood-depth archive.

% ===========================================================================
\section{The EuroFlood index}
\label{sec:index}

\begin{figure}[htbp]\centering
\includegraphics[width=\linewidth]{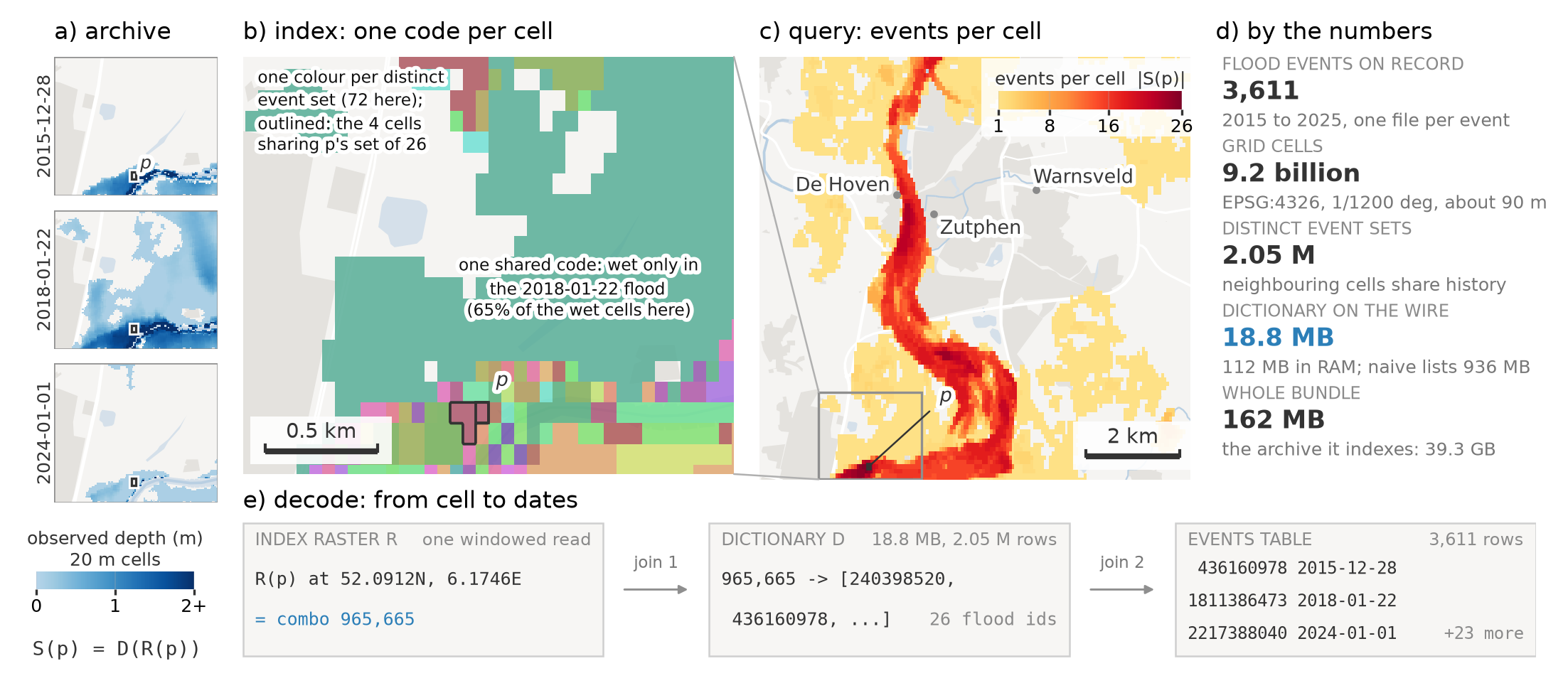}
\caption{Construction and decoding of the inverted raster index for an IJssel floodplain window south of Zutphen (see also Fig.~\ref{fig:zutphen}). (a)~Three dated \qty{20}{\m} source rasters containing inundation at cell $p$. (b)~The same observations encoded on the $\sim$\qty{90}{\m} index grid using the any-wet rule. Each cell stores a combination ID representing its complete event history; the outlined 11 cells share the 26-event history of $p$. (c)~Recurrence count across the wider query window. (d)~Archive and index dimensions: 3,611 events, 9.2 billion grid cells, 2.05 million distinct event combinations, and a \qty{162}{\mega\byte} index. (e)~Decoding the history of $p$: raster $R$ returns its combination ID, dictionary $D$ recovers the corresponding event IDs, and table $T$ provides their metadata. Thus, $S(p)=D(R(p))$ preserves event identity rather than only recurrence count.}
\label{fig:schematic}
\end{figure}

The source archive cannot directly support a basic spatial query: which archived events inundated a given location?
Answering this question from the source files alone would require identifying, downloading, and opening candidate rasters individually.
EuroFlood addresses this limitation by precomputing event membership on a continental grid and publishing the result as a compact spatial index.
Fig.~\ref{fig:schematic} illustrates the principle for one floodplain window: the archive stores separate dated depth rasters (Fig.~\ref{fig:schematic}a), the index converts them into a per-cell record of event membership (Fig.~\ref{fig:schematic}b), and recurrence counts and event footprints are derived directly from that record (Fig.~\ref{fig:schematic}c).

\subsection{Index structure}

The index is defined on a geographic grid $\mathcal{G}$ in EPSG:4326 at $1/1200^{\circ}$ resolution, equivalent to 3~arc-seconds or approximately \qty{90}{\m}.
It is a geographic raster rather than a discrete global grid system \citep{sahr2003dggs}.
Let $\mathcal{E}$ denote the 3,611 archived events.
The index represents the membership map $S:\mathcal{G}\rightarrow 2^{\mathcal{E}}$, assigning each grid cell the set of events in which it was inundated.

The map is materialised through three linked components.
(i) A sparse Cloud-Optimized GeoTIFF $R$ \citep{cogspec} stores one \texttt{uint32} combination identifier per cell, with each identifier representing a distinct event set and a reserved value for never-wet cells (Fig.~\ref{fig:schematic}b).
(ii) A dictionary $D$ maps each combination identifier to its sorted list of event identifiers, while (iii) an events table $T$ associates each event with metadata including observation dates, source file, and spatial cluster.
The representation is lossless at the index resolution.
For every cell $p\in\mathcal{G}$,

\[ S(p)=D\big(R(p)\big), \]

so both the recurrence count $|S(p)|$ and the footprint $\{p\in\mathcal{G}:e\in S(p)\}$ of any event $e\in\mathcal{E}$ can be reconstructed exactly.
Figure~\ref{fig:schematic} illustrates this decoding: $R$ returns the combination identifier of a cell, $D$ recovers its event identifiers, and $T$ supplies the corresponding metadata.
Unlike occurrence products that retain only the number of wet observations \citep{pekel2016}, EuroFlood preserves event identity, enabling spatial event discovery and recurrence analysis from the same representation.

The encoding is compact because neighbouring cells frequently share identical event histories.
Of the $9.2\times10^{9}$ cells in the continental grid, 89.6 million (1.0\%) were observed wet at least once, but these cells contain only approximately $2.05\times10^{6}$ distinct event combinations.
A dense 32-bit raster would require \qty{39.3}{\giga\byte}, while storing explicit event lists for all wet cells would require \qty{936}{\mega\byte}.
Combination encoding reduces $R$ to \qty{142}{\mega\byte}; together with $D$ (\qty{18.8}{\mega\byte}) and $T$ (\qty{0.4}{\mega\byte}), the published index occupies \qty{162}{\mega\byte}.
The representation combines established concepts from GIS overlay analysis, dictionary encoding and inverted indexes \citep{zobel2006,chambi2016roaring}, and recurrence mapping \citep{pekel2016} into a lossless and cloud-native materialisation of the event-membership map $S$.

\subsection{Hosting and query model}

The index components are published as static, versioned files: $R$ as a Cloud-Optimized GeoTIFF and $D$ and $T$ as GeoParquet tables \citep{euroflood_index}.
They are hosted on a public object store and accessed over HTTP without a server-side query engine\footnote{\url{https://source.coop/hackl/euroflood-index}}; the complete bundle can also be mirrored locally.
Each release is immutable, citable, and linked to the corresponding state of the source archive, allowing analyses to be reproduced against the same inputs.

A discovery query comprises a windowed read of $R$, expansion of the returned combination identifiers through $D$, and a metadata join with $T$ (Fig.~\ref{fig:schematic}e).
Because Cloud-Optimized GeoTIFFs support HTTP range requests, only tiles intersecting the query region are transferred.
Once $D$ and $T$ are available, event discovery therefore requires no access to the \qty{39.3}{\giga\byte} collection of source depth rasters.

A conventional per-event spatial catalogue can also identify events whose footprints intersect a query region, but does not encode event membership at the cell level.
Recurrence and related cell-level analyses therefore require additional processing of matched footprints or source rasters.
The inverted raster index supports both event discovery and cell-level analysis from the same representation.
Even continental queries remain bounded by the \qty{142}{\mega\byte} index raster, or its internal overviews, rather than the full source archive.
Section~\ref{sec:validation} quantifies these query regimes, and Section~\ref{sec:application} demonstrates their use in analysis.

\subsection{Ingest rules}

The index preserves the source archive's event definition, where each raster file is treated as one spatio-temporal event cluster and is not re-segmented into hydrologically distinct floods.
Each indexed event therefore remains directly traceable to a specific JRC source file.

Source rasters at \qty{20}{\m} resolution are mapped conservatively onto $\mathcal{G}$ using an any-wet rule.
An index cell is classified as wet if at least one wet source pixel falls within that cell.
This preserves all detected inundation at the index scale, but enlarges event footprints relative to the native-resolution rasters; the resulting area increase is quantified in Section~\ref{sec:validation}.

A source pixel is considered wet when its reconstructed depth is positive and differs from the declared no-data value.
No upper depth threshold is applied, so event membership depends only on inundation presence, not depth magnitude.
Source files without a declared coordinate reference system are excluded rather than assigned an assumed one, preventing spatial misregistration in the index.

\subsection{Construction and release}
\label{sec:construction}

Each release is constructed from the current state of the source archive.
New source rasters are retrieved and independently mapped onto the index grid to produce per-event cell-membership tables, which are then combined into the index raster $R$, combination dictionary $D$, and event table $T$.
The event-level ingest is resumable, so previously processed rasters can be reused when the archive is extended.
The final reduction is performed over the complete event collection because the combination identifiers represent globally defined event sets, requiring $R$, $D$, and $T$ to be regenerated together for each release.

Using a workstation with 12 CPUs, ingesting 3,611 source rasters required 21 minutes.
Constructing $R$, $D$, and $T$ required a further 60 seconds and approximately \qty{8}{\giga\byte} of memory.
The complete rebuild of the EuroFlood index took 24 minutes.
Additional benchmarks are reported in the Supplementary Information (Section~S2).

Thereby updates are additive.
Newly observed events can add event memberships to existing cells and populate previously empty cells without removing previously recorded memberships.
This preserves the information contained in earlier releases, while immutable versioned releases allow analyses based on previous archive states to remain reproducible.
Each release includes a manifest recording the source listing, software commit, and file checksums and is archived under a persistent identifier.

% ===========================================================================
% ===========================================================================

% ===========================================================================
\section{Software design and implementation}
\label{sec:software}

EuroFlood provides an open-source Python API and command-line interface for querying the index and retrieving associated raster data.
The software separates user-facing analysis from index maintenance: analysis functions discover events and extract data, while maintenance commands build, validate, publish, and audit index releases.

User access follows a two-stage Discover$\rightarrow$Extract workflow.
Discovery identifies which events affected a region, when they occurred, and how often cells were inundated using only the \qty{162}{\mega\byte} index.
Extraction then retrieves native-resolution depth rasters only for the selected events, allowing large-scale exploration before source rasters are downloaded.

The interface follows conventions established by domain-specific environmental software, where results are exposed through objects compatible with \texttt{geopandas} and \texttt{xarray} \citep{maze2020argopy,chegini2021hyriver,chudley2024pdemtools}, while common workflows are accessible through a small set of functions and parameters \citep{boeing2025osmnx}.

% ===========================================================================
\begin{table}[htbp]
\centering
\caption{EuroFlood public analysis interface for event discovery, raster retrieval, hazard access, and offline operation. Maintenance commands for constructing and publishing index releases are listed in the Supplementary Information.}
\label{tab:api}
\footnotesize
\setlength{\tabcolsep}{4pt}
\begin{tabular}{@{}p{0.32\linewidth}p{0.62\linewidth}@{}}
\toprule
Call & Functionality \\
\midrule
\multicolumn{2}{@{}l}{\emph{Python API (analysis)}} \\
\texttt{ef.floods(region, ...)} & discover archived events and return a \texttt{FloodFrame} with one row per matched event \\
\texttt{.plot()} & render a static recurrence map from the index without downloading source rasters \\
\texttt{.explore()} & render an interactive recurrence map from the index \\
\texttt{.footprints()} & reconstruct per-event footprint polygons and areas \\
\texttt{.stats()} & calculate per-event depth, inundated-area, and water-volume statistics \\
\texttt{.summary()} & calculate aggregate depth, inundated-area, and water-volume statistics \\
\texttt{.download(dir)} & download region-of-interest-cropped \qty{20}{\m} observed-depth rasters \\
\texttt{.depths()} & return downloaded rasters as depth-raster objects \\
\texttt{ef.hazard(region, return\_period)} & retrieve modelled scenarios for one, several, or all return periods \\
\texttt{ef.mirror(target, ...)} & stage the index, observed-depth rasters, or hazard layers for offline use \\
\texttt{ef.verify(target, ...)} & report the completeness and integrity of a staged data layer \\
\texttt{ef.offline()} & switch the current session to local-only operation \\
\midrule
\multicolumn{2}{@{}l}{\emph{Command-line interface (analysis)}} \\
\texttt{floods} & discover events for a region without downloading source rasters \\
\texttt{download} & retrieve observed-depth rasters for a query \\
\texttt{hazard} & retrieve modelled return-period scenarios \\
\texttt{mirror <index|floods|hazard|all>} & stage one or more data layers for offline use \\
\texttt{verify <index|floods|hazard|all|remote>} & check the completeness of a local mirror or audit a published release over HTTP \\
\bottomrule
\end{tabular}
\end{table}
% ===========================================================================

\subsection{The Discover$\rightarrow$Extract API}
\label{sec:api}

For routine analysis, EuroFlood exposes the Discover$\rightarrow$Extract workflow through the \texttt{floods()} and \texttt{hazard()} entry points and the \texttt{FloodFrame} class:

\begin{lstlisting}
import euroflood as ef

# Discover: one row per archived event
cat = ef.floods("Zutphen, Netherlands")
cat.explore()   # interactive recurrence map, no downloads

# Extract: ROI-cropped 20 m depth rasters
cat.download("data/")

# Retrieve a modelled return-period scenario
haz = ef.hazard("Zutphen, Netherlands",return_period=10)
\end{lstlisting}

The \texttt{floods()} function resolves the user input to a region of interest, reads the corresponding portion of the published index, and returns a \texttt{FloodFrame} containing one row per matched event without downloading any depth raster.
The region can be supplied as a place name, a coordinate pair with a search radius, a bounding box, a \texttt{shapely} geometry, a \texttt{GeoDataFrame}, or a vector file, and the result can be restricted by year or date range.
Coordinates default to EPSG:4326, with points ordered as (latitude, longitude) and bounding boxes as (west, south, east, north); an alternative coordinate reference system can be supplied explicitly:

\begin{lstlisting}
ef.floods(point=(52.14, 6.20), radius_m=5000)
ef.floods(bbox=(5.9, 51.9, 6.4, 52.3), year=2023)
ef.floods(shapefile="catchment.gpkg", crs="EPSG:27704")
\end{lstlisting}

The returned \texttt{FloodFrame} supports discovery and visualisation directly from the index.
The \texttt{download()} method fetches region-of-interest-cropped \qty{20}{\m} depth rasters for the selected events, while \texttt{hazard()} retrieves and mosaics modelled return-period scenarios over the same region.
Depth, inundated-area, and water-volume statistics are calculated from the downloaded rasters.
Table~\ref{tab:api} summarises the analysis interface;  maintenance commands for constructing and publishing index releases are listed in the Supplementary Information.

\texttt{FloodFrame} subclasses \texttt{geopandas.GeoDataFrame} \citep{jordahl2020geopandas}, so standard filtering, spatial joins, plotting, and file-format operations remain available.
Subsetting preserves the class and therefore retains the EuroFlood discovery and extraction methods.
Raster outputs are \texttt{xarray}-compatible \citep{hoyer2017xarray}, allowing direct integration with established multidimensional-array and geospatial workflows.

\subsection{Architecture and caching}

\begin{figure}[htbp]\centering
\includegraphics[width=\linewidth]{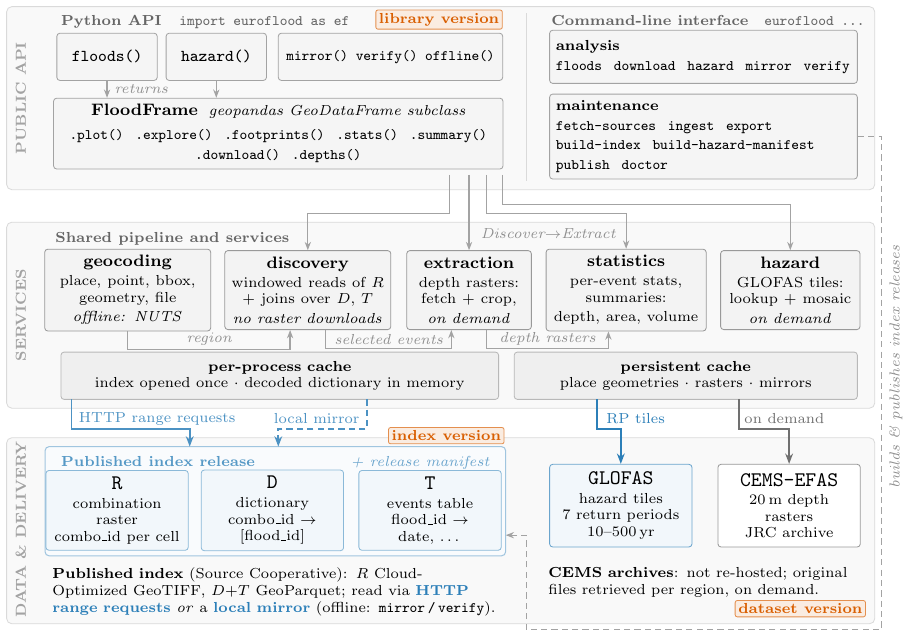}
%\vspace{-10mm}
\caption{EuroFlood's layered software architecture. The user-facing Python API, including the \texttt{FloodFrame} class, and the command-line interface delegate to pipeline and service layers responsible for region resolution, index access, raster retrieval, processing, and statistics. Caching at two timescales, per-process reuse and a persistent local cache, avoids repeated loading, decoding, and downloading, while discovery reads only the published index and extraction retrieves native-resolution depth rasters on demand.}
\label{fig:architecture}
\end{figure}

EuroFlood uses a layered architecture that separates the public interface from data access and processing logic (Fig.~\ref{fig:architecture}).
The Python API and command-line interface share the same pipeline and service layers for region resolution, index access, raster retrieval, processing, and statistical analysis.
This design allows online and offline operation to use identical analysis logic against different data locations, ensuring consistent results across interfaces and delivery modes.
Caching is implemented at both process and session scales.
Within a process, index artifacts are opened once and reused, with the decoded \qty{112}{\mega\byte} event-combination dictionary retained in memory.
Across sessions, resolved place geometries and downloaded rasters persist in a local cache.
The resulting reduction in query latency is quantified in Section~\ref{sec:validation}.
The public API is fully typed and tested under continuous integration, with a minimum statement coverage of 90\% and documentation examples included in the test suite.
Releases follow semantic versioning, are distributed through PyPI under the MIT licence, and are designed in accordance with FAIR principles for research software \citep{barker2022fair4rs}.

\subsection{Online and offline operation}
\label{sec:delivery}

EuroFlood supports both interactive online analysis and offline batch or cluster workflows.
In online mode, the index is accessed through HTTP range requests, allowing queries without authentication or bulk download.
Native-resolution depth rasters remain hosted by JRC and are retrieved from the source archive only when extraction is requested.
For offline operation, the \texttt{mirror} interface stages either the complete index or selected observed-depth and hazard rasters in a local cache.
The corresponding \texttt{verify} commands check the completeness and integrity of the staged data.
Offline mode then restricts all data access to these local copies and uses cached Eurostat NUTS boundaries for place-name resolution.
Online and offline modes use the same analysis functions and produce identical results, as verified in Section~\ref{sec:validation}.
Workflows developed interactively can therefore be transferred unchanged to batch, cluster, or disconnected computing environments.

\begin{lstlisting}
# Mirror once, then operate without network access
$ export EUROFLOOD_CACHE_DIR=/scratch/ef
$ euroflood mirror index
$ export EUROFLOOD_OFFLINE=1
$ euroflood floods "Zutphen, Netherlands"
\end{lstlisting}

% ===========================================================================
% ===========================================================================

\subsection{Transferability to other event-based raster archives}
\label{sec:generality}

Although EuroFlood was developed for the CEMS-EFAS archive, its indexing and query framework assumes only a collection of event rasters that can be mapped to a common grid.
The core representation is therefore independent of the specific flood-depth product.
Applying the same framework to archives such as GFM flood extents \citep{krullikowski2023gfm} or OPERA DSWx surface-water products \citep{opera_dswx} would require a source-specific ingestion layer defining event semantics, masks, units, and provenance, while retaining the same index structure and query interface.
Future work will test the framework on additional event-based raster archives to assess its transferability beyond CEMS-EFAS.

% ===========================================================================
% ===========================================================================

% ===========================================================================
\section{Verification and benchmarks}
\label{sec:validation}

A combined software-and-data contribution is useful only if its published artifacts reproduce the source data correctly, its access model is efficient, and the underlying archive is suitable for the intended analyses.
First, the published index is verified against two reconstructions of the source rasters, including one independent implementation (O1 and O2).
Second, query cost is benchmarked against alternative discovery strategies and across delivery modes (O3 and O4).
Third, correspondence with documented floods is assessed against chance-matching null models (O5).
All analyses can be reproduced using the archived materials described in the Software and data availability section.

\subsection{Index verification}

The published index was verified against the source archive using 110 events, sampled as ten events per archive year with a fixed random seed.
Event footprints decoded from the index were compared with two reconstructions of the corresponding \qty{20}{\m} source rasters on the index grid.
The first used the production ingest routine, while the second independently implemented the ingest rule in Section~\ref{sec:index} without importing EuroFlood code and evaluated the full continental grid.

Agreement was quantified using intersection over union (IoU).
All 110 events achieved $\mathrm{IoU}=1.000$ against both reconstructions, corresponding to 5.74 million cells with no omitted or additional memberships.
Eight additional events representing archive extremes, including the largest footprint and the geographic limits of the grid, were also reproduced exactly.
Full sampling and per-event results are reported in the Supplementary Information.

The index nevertheless differs systematically from the native \qty{20}{\m} rasters because of the conservative any-wet resampling rule.
Across the 110 verification events, indexed footprints were larger than the native wet area by a median factor of 1.50, with a 10th--90th percentile range of 1.28--1.94.
Index footprints are therefore appropriate for discovery, recurrence, and spatial screening, whereas analyses requiring native-resolution inundated area should use the \qty{20}{\m} source rasters.
Equivalent queries produced byte-identical results when the index was accessed from the public host, a local offline mirror, or a private HTTP replica, confirming that query results are independent of delivery mode.

\subsection{Query cost}

Query performance was evaluated systematically across 90 spatial windows sampled from 1,765 European administrative regions.
Up to 15 bounding boxes were drawn from each of six logarithmic area classes, spanning query areas from \qty{15}{\km\squared} to \qty{1.1e6}{\km\squared}.
For illustration at specific spatial scales, three additional queries were examined: Zutphen (33 matched events), Cologne (12 events), and the Netherlands (126 events) (Fig.~\ref{fig:query-sweep}a,b).
Detailed sampling and benchmarking procedures are provided in the Supplementary Information.

\begin{figure}[htbp]
\centering
\includegraphics[width=\textwidth]{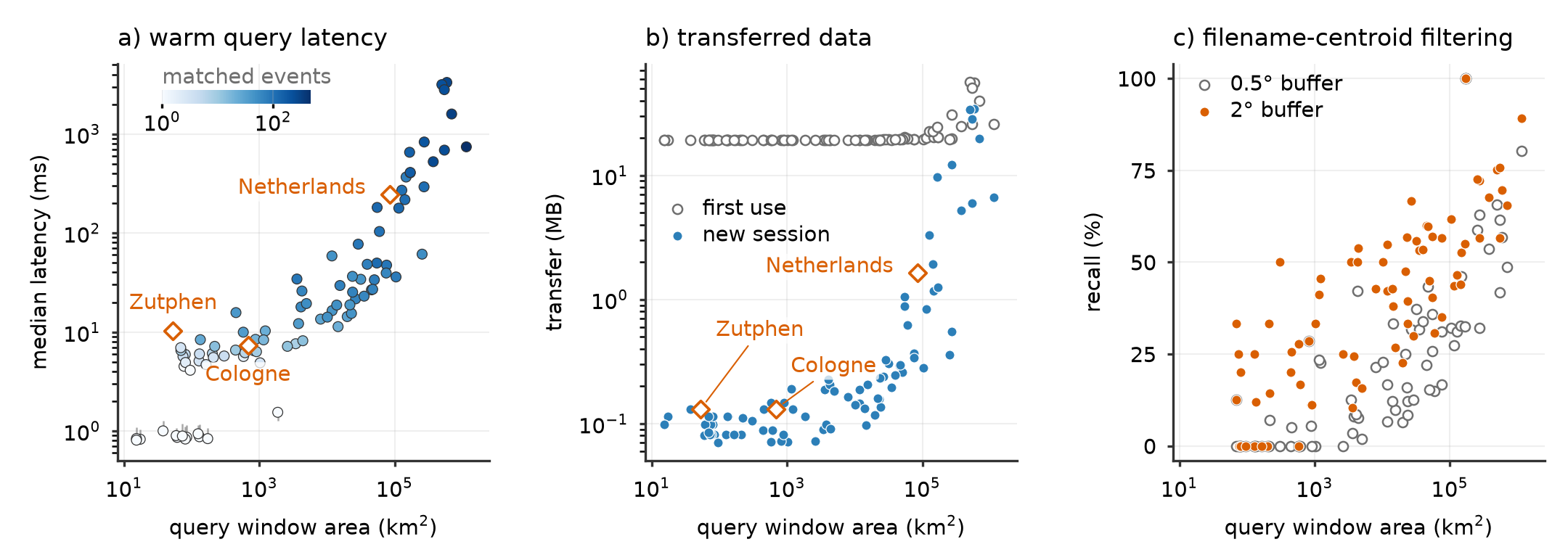}
\caption{Systematic query-performance evaluation across 90 spatial windows spanning \qty{15}{\km\squared} to \qty{1.1e6}{\km\squared}. (a)~Median latency of repeated queries; vertical bars show the interquartile range across repetitions and colour indicates the number of matched events. (b)~Data transferred by the first query with an empty cache and by a new process after the disk cache has been populated. (c)~Recall of filename-centroid filtering using \ang{0.5} and \ang{2} buffers; only windows containing at least one archived event are shown. Diamonds in (a) and (b) mark the illustrative Zutphen, Cologne, and Netherlands queries.}
\label{fig:query-sweep}
\end{figure}

Across the 90-window evaluation, the first query with an empty cache transferred a median of \qty{19.3}{\mega\byte}, including the one-time loading of the event-combination dictionary and events table.
Once cached, a new process transferred only \qty{0.14}{\mega\byte} at the median, while repeated queries completed in \qty{14}{\ms}.
Repeated-query latency and warmed-cache transfer increased sublinearly with query area, with fitted exponents of 0.59 and 0.38, respectively (Fig.~\ref{fig:query-sweep}a,b).
The three illustrative regions show the same pattern: warmed-cache queries transferred \qtyrange{0.13}{1.51}{\mega\byte}, compared with \qty{39.3}{\giga\byte} for the complete source archive, while repeated queries required \qtyrange{8}{278}{\ms}.
Cold queries against the live endpoint completed in \qtyrange{6}{8}{\s}.

To assess the spatial discovery capability of the source archive itself, we compared the index with filtering based on the centroid encoded in each filename, the only spatial metadata available without opening the rasters.
Across windows containing events, a \ang{2} buffer recovered a median of 42\% of the events whose footprints intersected the query window and achieved complete recall in only 2.6\% of cases.
With a \ang{0.5} buffer, median recall fell to 15\%.
The low recall reflects the inability of a single centroid to represent spatially dispersed event footprints (Fig.~\ref{fig:query-sweep}c).

To determine whether a simpler spatial catalogue would provide the same functionality, we also compared the inverted index with a per-event footprint catalogue.
The \qty{67}{\mega\byte} GeoParquet catalogue recovers all intersecting events by construction, although its coarsened polygons can introduce false positives, and does not retain cell-level event membership.
Consequently, computing flood recurrence over the Netherlands requires reading \qty{4.5}{\giga\byte} of matched source rasters, whereas the inverted index requires only a \qty{20.8}{\mega\byte} windowed read, a 218-fold reduction.
Thus, a footprint catalogue is sufficient for event discovery, while the inverted index additionally enables recurrence and other cell-level analyses without accessing the source rasters.

% ===========================================================================
% ===========================================================================

% ===========================================================================
% ===========================================================================

% ===========================================================================
% ===========================================================================

\subsection{Corroboration of the observed archive against documented floods}

The preceding tests verify the index but not the extent to which the underlying satellite archive captures known floods.
We therefore compared the archive with HANZE (Historical Analysis of Natural Hazards in Europe), an independent database of damaging floods compiled from documentary sources \citep{paprotny2018,paprotny2024essd,paprotny2024hanze}, which records 397 European floods during 2015--2025.
A HANZE flood was considered corroborated when the satellite archive contained an event within seven days of the documented flood that overlapped at least one reported affected region.
This indicates that flooding was observed in the same place and around the same time, but does not prove that the archived footprint represents the complete documented event.
For comparison, we also tested a broader match requiring only an archived event in the same country.
To estimate how many matches could occur by chance, we repeated the analysis 1,000 times after randomly shifting the documented flood dates while keeping their locations and durations unchanged.

\begin{figure}[htbp]\centering
\includegraphics[width=\linewidth]{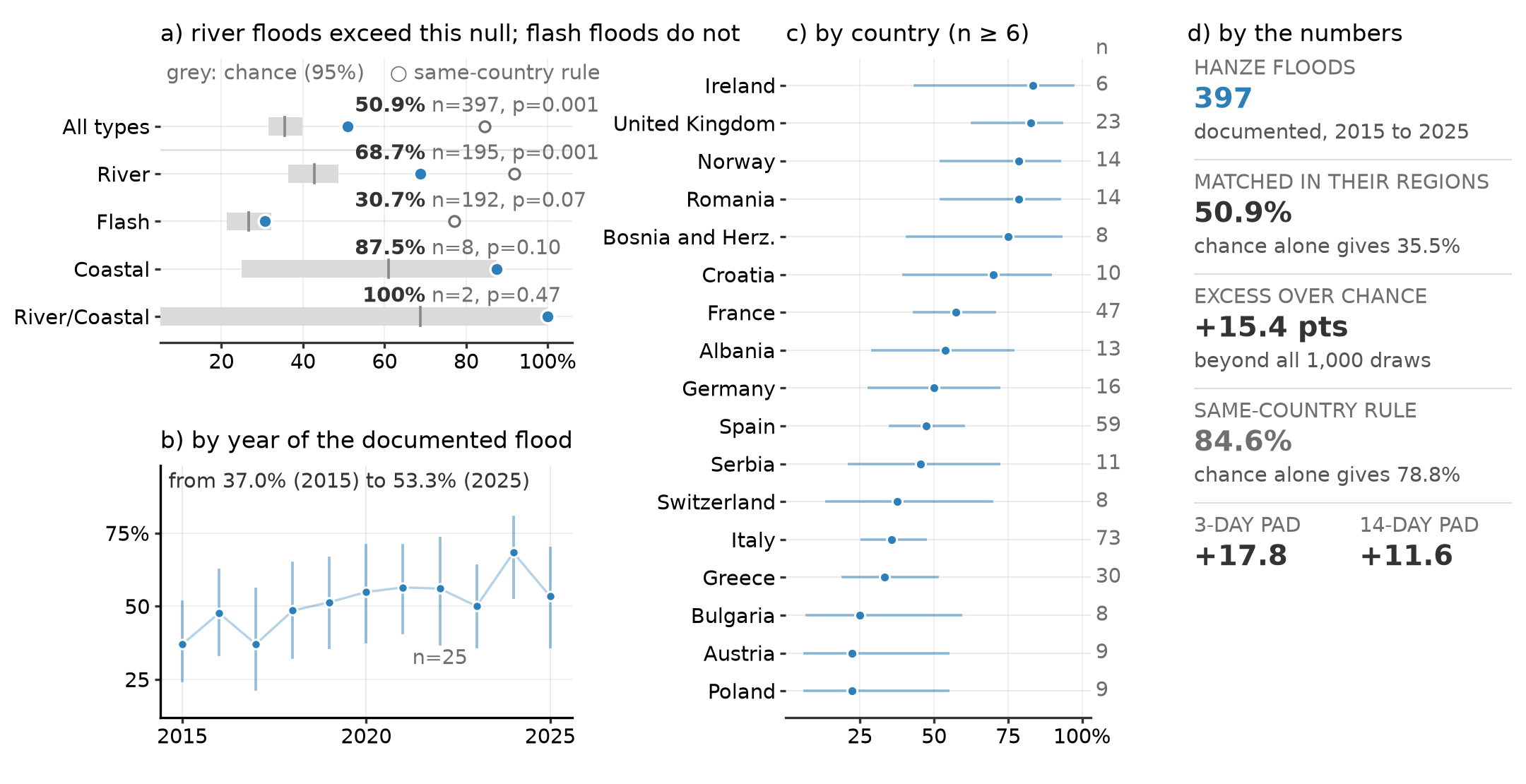}
\caption{Corroboration of 397 floods documented in HANZE during 2015--2025 by the CEMS-EFAS satellite archive. A flood is corroborated when an archived event occurs within seven days of the documented dates and intersects an affected region. (a)~Corroborated share by flood type compared with the central 95\% interval from 1,000 date-randomization permutations; open circles show the same-country criterion. (b)~Corroborated share by year with Wilson 95\% confidence intervals. (c)~Corroborated share by country for countries with at least six documented floods. (d)~Overall corroboration and excess above the null expectation, including sensitivity to the spatial criterion and temporal matching window.}
\label{fig:hanze-audit}
\end{figure}

Overall, 50.9\% of documented floods were corroborated, compared with 35.5\% under the null model, an excess of 15.4 percentage points that exceeded all 1,000 permutations ($p\approx0.001$).
Similar excesses of 15.0--15.4 percentage points were obtained under alternative null models preserving seasonality or temporal clustering.
The signal was dominated by river floods, of which 68.7\% were corroborated compared with 42.8\% expected by chance (Fig.~\ref{fig:hanze-audit}a).
Flash floods showed a smaller excess, 30.7\% versus 26.6\%, that was not significant under the primary null model ($p=0.07$) and was sensitive to the null specification.
The coastal categories contained too few events for reliable inference ($n=8$ and $n=2$).
These differences are consistent with the limited ability of Sentinel-1 to capture short-lived flooding between satellite acquisitions.
The coarser same-country criterion matched 84.6\% of documented floods, but only 60.1\% of these matches persisted at the region level.
Moreover, matched footprints covered a median of only 0.16\% of the documented regions, reinforcing that corroboration denotes co-occurrence rather than correspondence of complete flood extents.

Corroboration varied across both years and countries.
Annual rates increased from 37.0\% in 2015 to 68.3\% in 2024 and were 53.3\% in 2025 (Fig.~\ref{fig:hanze-audit}b), indicating that temporal changes in archive detections also reflect changing observation opportunities.
Across the 17 countries with at least six documented floods, corroboration declined with the proportion of flash floods ($r=-0.51$), consistent with lower detectability of rapid and topographically confined events.
The higher-than-chance corroboration was robust to the matching definition, remaining positive for temporal windows of 3--14 days and minimum detected-area thresholds up to \qty{5}{\km\squared} (Fig.~\ref{fig:hanze-audit}d).
Full sensitivity analyses and null-model specifications are provided in the Supplementary Information (Section~S1.8).

% ===========================================================================
\section{Applications and case studies}
\label{sec:application}

This section illustrates the analyses enabled by the EuroFlood index and library using only the public interface described in Section~\ref{sec:software}.
The four examples demonstrate continental recurrence analysis, comparison of observed and modelled flood extents, observation-based exposure assessment, and the detectability limit associated with a rapid-onset flooding.
These examples are intended as screening analyses rather than formal validation studies.

\subsection{Eleven years of observed flood recurrence in Europe}

\begin{figure}[htbp]\centering
\includegraphics[width=\linewidth]{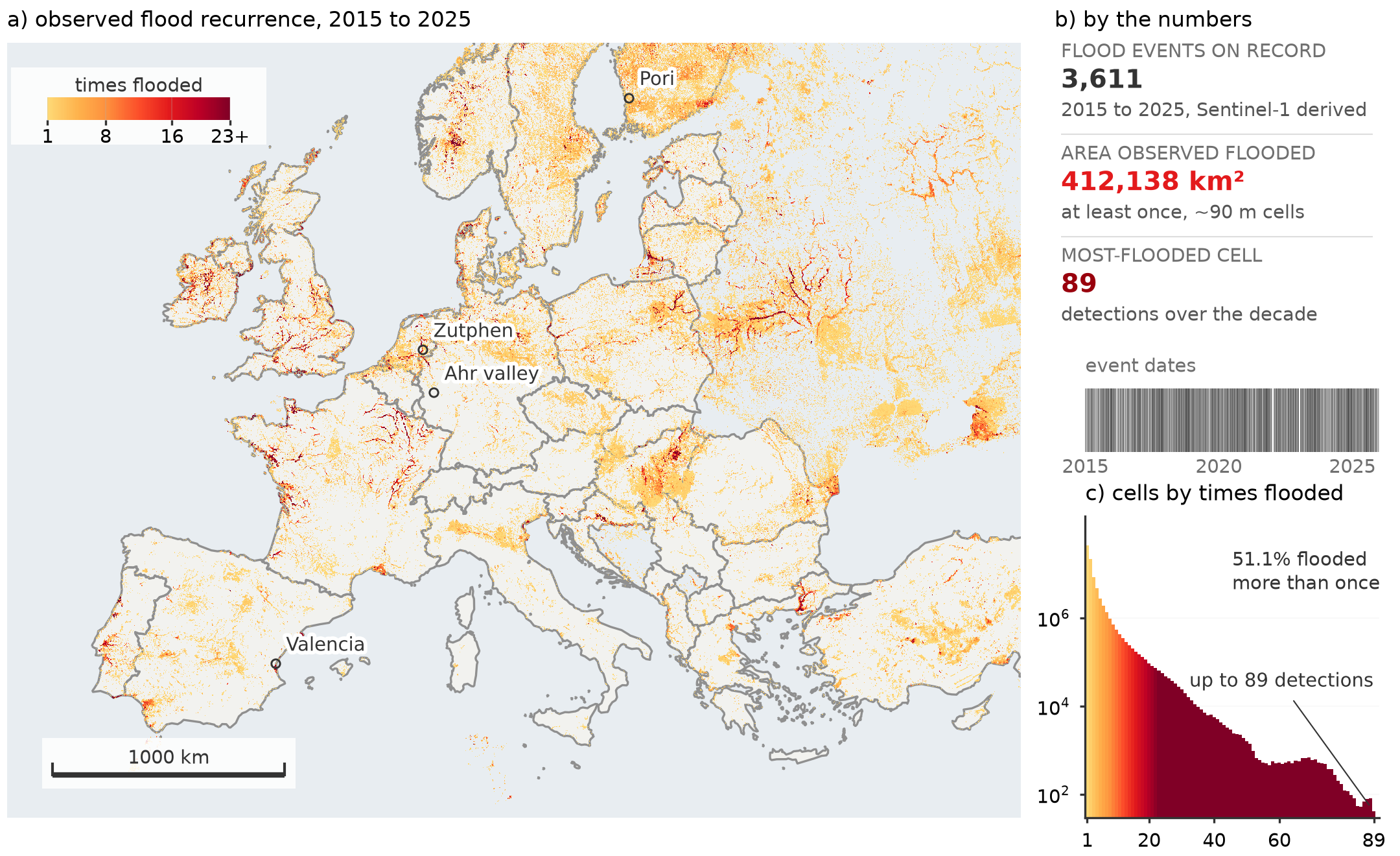}
\caption{Observed flood recurrence across Europe from 2015 to 2025, computed entirely from the inverted index. (a)~Number of archived events in which each cell was detected as wet, max-pooled for continental display and shown with the colour scale saturated at the 99th percentile of the pooled counts; markers identify the case-study locations examined in Figs.~\ref{fig:centroid} and~\ref{fig:zutphen}--\ref{fig:dana}. (b)~Summary of the archive and the temporal distribution of the event dates included in the record. (c)~Number of grid cells associated with each recurrence count on a logarithmic scale, coloured using the same scale as panel~(a). Recurrence denotes the number of archived event detections and should not be interpreted as a return period.}
\label{fig:recurrence}
\end{figure}

The first application examines how often individual locations across Europe were detected as inundated during 2015--2025.
Without the index, this analysis would require downloading, aligning, and stacking the complete archive of event rasters.
EuroFlood instead derives the continental recurrence surface directly from a single query:

\begin{lstlisting}
import euroflood as ef

# Continental recurrence from the index
frame = ef.floods(bbox=EUROPE)
frame.plot()
\end{lstlisting}

The query uses an internal overview for interactive visualisation, while the statistics and map in Fig.~\ref{fig:recurrence} were computed from the full-resolution index.
Across the archive period, \qty{412138}{\km\squared} of Europe was detected as inundated at least once; 51.1\% of wet cells occurred in more than one archived event, and the maximum recurrence was 89 events.
Area estimates account for the geodesic area of each geographic grid cell.

Observed recurrence is concentrated along major river corridors and lowland floodplains (Fig.~\ref{fig:recurrence}a).
The event-date distribution in Fig.~\ref{fig:recurrence}b shows the temporal coverage of the archive, while Fig.~\ref{fig:recurrence}c reveals a strongly right-skewed recurrence distribution: most inundated cells were detected in only one or two events, whereas a smaller subset accumulated repeated observations.

Recurrence should be interpreted as the number of archived satellite detections, not as an annual exceedance probability or hydrological return period \citep{pekel2016}.
Because the counting unit is the archive event file, a single physical flood represented by multiple files may contribute more than once.
The map therefore represents where satellite-observed inundation has repeatedly been recorded within the archive, rather than the underlying frequency of flooding itself.

\subsection{Observed footprints and modelled hazard extents}

EuroFlood enables direct comparison of dated flood observations with modelled return-period hazard scenarios corresponding to the probability ranges considered in the EU Floods Directive \citep{floodsdirective}.
A single \texttt{hazard()} query retrieves and mosaics all seven CEMS-GLOFAS return-period layers over the same region as the observed-event query.
We first illustrate this comparison at Zutphen on the river IJssel and then extend it to the same 110 events used for index verification, sampled across the 2015--2025 archive period.
A complementary analysis of Zutphen's multi-event water-depth record is provided in the Supplementary Information.

Figure~\ref{fig:zutphen} compares observed recurrence with modelled hazard within the Zutphen query window, where 33 archived events were identified between 2015 and 2025.
Of the 10,382 cells observed wet at least once, 82.5\% fall within the RP10 extent and 17.5\% lie outside it (Fig.~\ref{fig:zutphen}c).
Increasing the return period expands the modelled envelope and reduces the observed-only fraction to 5.1\% at RP500, although a distinct observed footprint northeast of the town remains outside the modelled extent (Fig.~\ref{fig:zutphen}d).

\begin{figure}[htbp]\centering
\includegraphics[width=\linewidth]{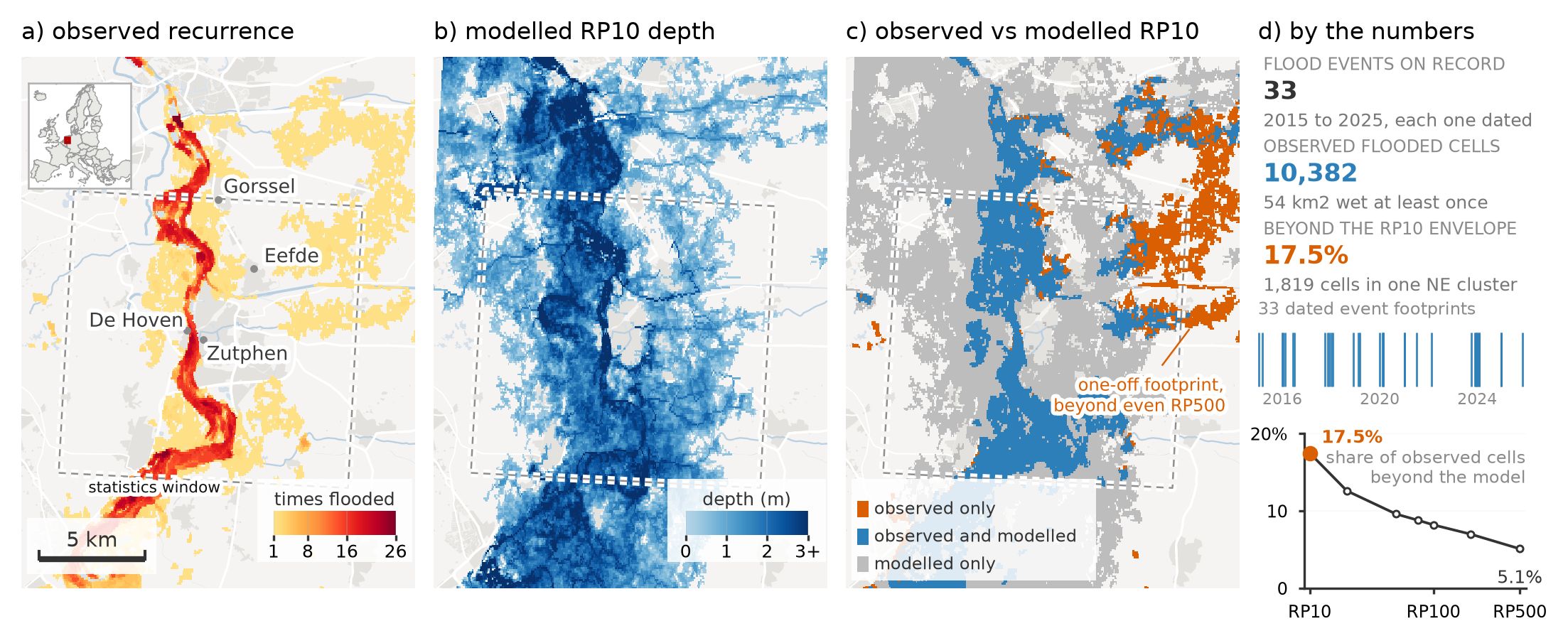}
\caption{Comparison of observed flood recurrence and modelled river-flood hazard at Zutphen on the IJssel. (a)~Observed recurrence from 33 archived events between 2015 and 2025. (b)~Modelled inundation depth for the 10-year return-period scenario. (c)~Classification of cells as observed only, both observed and modelled, or modelled only within the RP10 comparison; the analysis contains 10,382 cells observed wet at least once. (d)~Summary of the event record and the fraction of observed cells lying outside the modelled envelope across return periods from RP10 to RP500. Statistics are calculated within the dashed query window, while the maps show a slightly wider surrounding area.}
\label{fig:zutphen}
\end{figure}

Across the 110 sampled events, observed footprints show substantially lower and more variable overlap with the modelled hazard extents than at Zutphen.
At RP10, the median event has 41\% of its observed footprint within the modelled extent, and 56\% of events have less than half of their footprint represented.
Overlap varies widely among events, with a 10th--90th percentile range of 2--83\%.
Expanding the modelled envelope to RP500 increases the median overlap only to 46\%.
Repeating the RP10 comparison using the native \qty{20}{\m} observations increases the median to 45\%, indicating that the limited overlap is not primarily caused by resampling to the coarser index grid.

These differences should not be interpreted as model error alone.
The hazard maps represent fluvial flooding, omit pluvial and coastal processes, exclude small catchments, and assume no existing flood protection.
Zutphen lies on a river explicitly represented by the model, whereas the continental sample includes observations associated with processes and channels outside the modelled domain.
The comparison therefore characterises the correspondence between observed inundation and modelled hazard envelopes rather than formally validating either dataset.

\subsection{Observation-based exposure: the July 2021 western European floods}

\begin{figure}[htbp]\centering
\includegraphics[width=\linewidth]{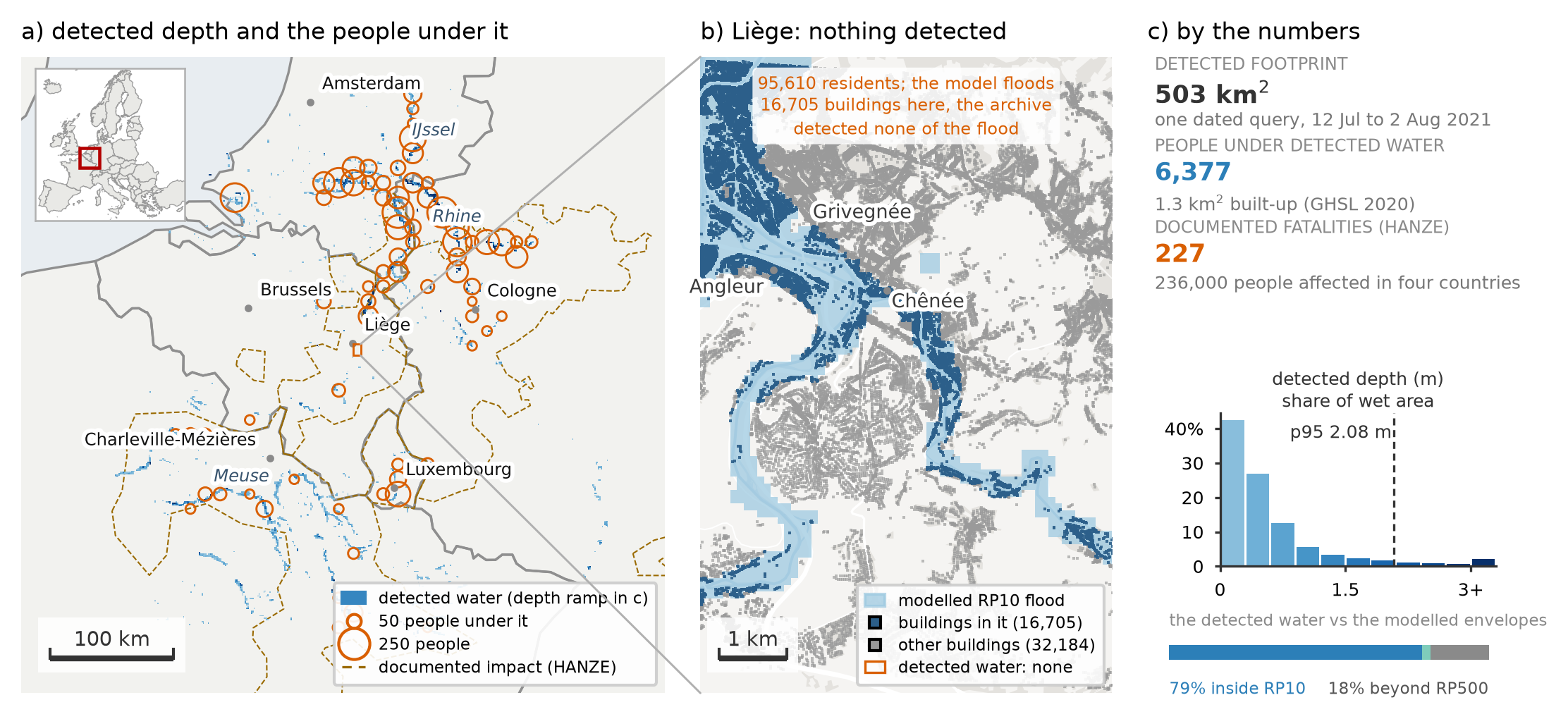}
\caption{Observation-based exposure assessment for the July 2021 western European floods. (a)~Detected flood-depth footprint with circles scaled to the population within the detected water, documented HANZE impact regions outlined, and major rivers labelled; the largest detected population exposures occur along the lower-gradient Meuse and Rhine corridors rather than in the steep valleys where the event caused its most severe impacts. (b)~Building-scale comparison at Li\`ege, where the modelled RP10 extent intersects 16,705 buildings but the satellite archive contains no detected flood footprint. (c)~Summary of the detected footprint and exposed population, distribution of detected water depth, and overlap between the detected footprint and the modelled return-period envelopes.}
\label{fig:impact}
\end{figure}

Flood-exposure estimates commonly rely on modelled scenarios because spatially explicit observations are rarely available for individual historical events.
For the July 2021 western European floods, a dated \texttt{floods()} query provides an observed inundation footprint that can be intersected directly with population and built-up density grids \citep{schiavina2023ghspop,pesaresi2023ghsbuilts} (Fig.~\ref{fig:impact}).

The detected footprint covers \qty{503}{\km\squared} and contains an estimated 6,377 residents and \qty{1.3}{\km\squared} of built-up surface (Fig.~\ref{fig:impact}a,c).
The largest detected population exposures occur along the lower-gradient Meuse and Rhine corridors rather than in the steep tributary valleys where impacts were most severe.
These estimates therefore represent lower bounds on event exposure rather than complete estimates of the affected population.

The Li\`ege example illustrates this limitation (Fig.~\ref{fig:impact}b).
The modelled RP10 extent intersects a densely developed area containing 16,705 buildings, yet no inundation is detected there in the satellite archive.
The flood developed and receded rapidly in the Vesdre valley before a suitable satellite acquisition, demonstrating that even spatially precise observations can omit major components of an event.

Across the detected footprint, 79\% of observed inundation lies within the modelled RP10 extent, while 18\% remains outside the RP500 envelope (Fig.~\ref{fig:impact}c).
The two datasets are therefore complementary: satellite observations provide dated evidence of detected inundation, whereas hazard maps identify areas susceptible to flooding that may not have been observed during the event.
The same workflow can be applied to other archived floods to derive observation-based lower bounds on population and built-environment exposure.

% ===========================================================================
% ===========================================================================

\subsection{Rapid-onset flooding at the detection boundary: the Valencia DANA}
\label{sec:limits}

\begin{figure}[htbp]\centering
\includegraphics[width=\linewidth]{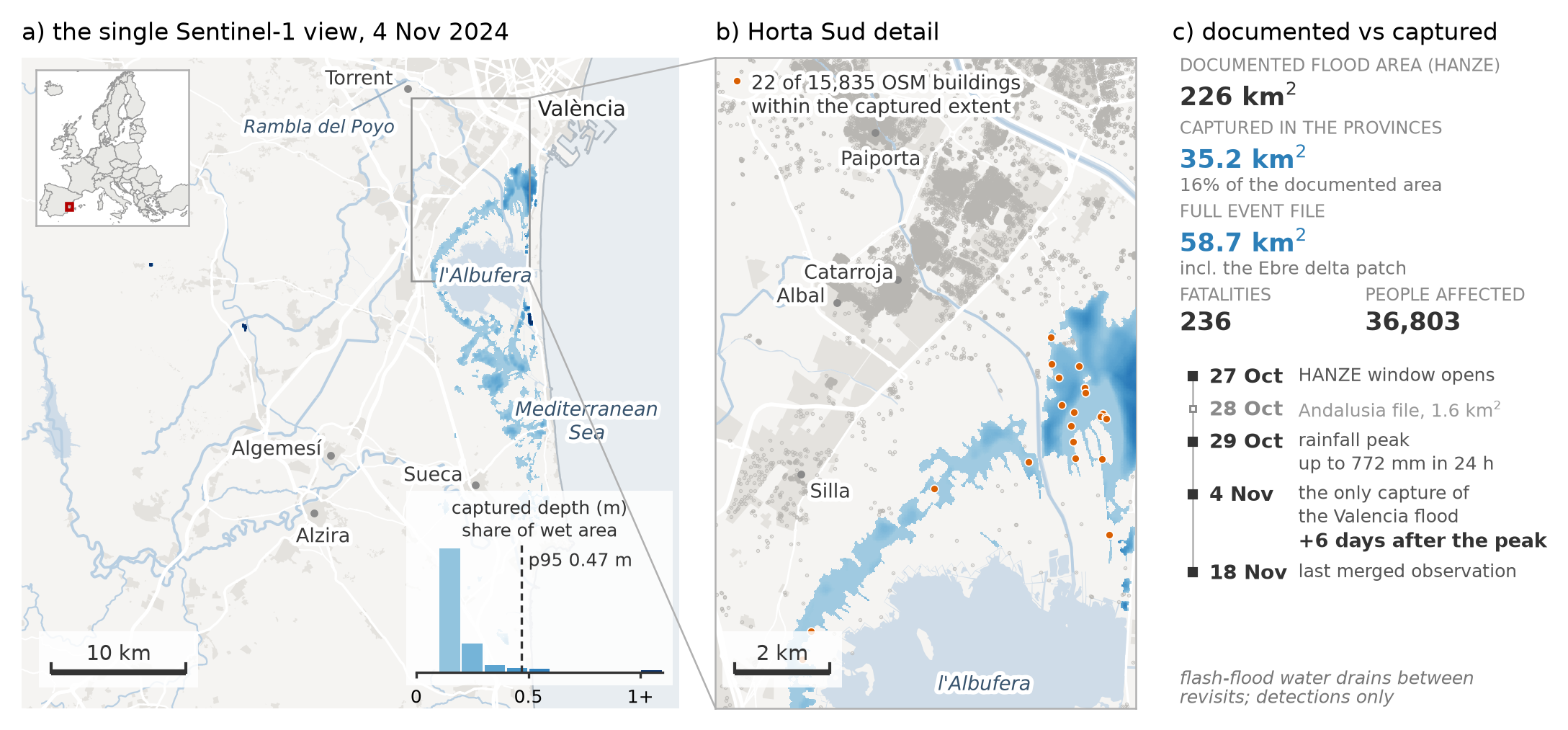}
\caption{The October 2024 Valencia DANA as an example of the detectability limit for rapid-onset flooding. (a)~The only Sentinel-1 observation of the Valencia flood contained in the archive, acquired on 4~November 2024, six days after the rainfall peak; the inset histogram shows that the captured water is predominantly shallow, and the rectangle marks the extent of panel~(b). (b)~Building-scale view of the Horta Sud area, where the detected water is concentrated in fields and rice paddies near l'Albufera rather than in the urban streets affected during the flood; 22 of 15,835 mapped buildings lie within the captured footprint. (c)~Comparison of the documented event with the satellite record and the event timeline. The archive captures \qty{35.2}{\km\squared} within the documented provinces, equivalent to 16\% of the documented \qty{226}{\km\squared}, while the complete event file contains \qty{58.7}{\km\squared} including separate patches along the Castell\'on coast and near the Ebro delta.}
\label{fig:dana}
\end{figure}

Rapid-onset floods may develop and recede between successive satellite acquisitions, imposing a fundamental detectability limit on Sentinel-1-based event archives.
The DANA that affected Valencia on 29~October 2024, causing 236 fatalities and an estimated \qty{226}{\km\squared} of documented inundation across four provinces, illustrates this limitation (Fig.~\ref{fig:dana}).

The archive contains one Sentinel-1 observation of the event, acquired on 4~November 2024, six days after the rainfall peak.
Detected inundation is predominantly shallow and concentrated south of Valencia around l'Albufera (Fig.~\ref{fig:dana}a).
At building scale, the footprint lies mainly across fields and rice paddies rather than the densely developed Horta Sud towns affected during the flood (Fig.~\ref{fig:dana}b); only 22 of 15,835 mapped buildings in the detailed frame intersect the detected water.
By the acquisition date, much of the short-lived urban inundation had receded, while residual water remained in lower-lying agricultural areas.

Within the documented provinces, the archive contains \qty{35.2}{\km\squared} of detected inundation, equivalent to 16\% of the documented \qty{226}{\km\squared} flood area (Fig.~\ref{fig:dana}c).
The complete archive file covers \qty{58.7}{\km\squared} because the spatio-temporal clustering procedure also includes geographically separate patches along the Castell\'on coast and near the Ebro delta.
These area comparisons illustrate the temporal and spatial selectivity of satellite detection rather than provide a direct validation of either estimate.

A small or absent archived footprint should therefore not be interpreted as evidence of a small event.
For rapidly evolving floods, the archive may primarily represent residual inundation remaining at the time of satellite acquisition.

\subsection{Further demonstrations}

Five additional case studies in the Supplementary Information extend these workflows to transboundary discovery during Storm Boris (Section~S1.1), continental detection climatology (S1.2), infrastructure screening in the Shannon callows (S1.3), exposure during the May 2023 Emilia-Romagna floods (S1.5), and the October 2023 Baltic storm surge (S1.7).

% ===========================================================================
\section{Discussion}
\label{sec:discussion}

\subsection{Fitness for use and interpretation}

The verification results establish the fidelity of the index, but the usefulness of any analysis ultimately depends on the observational limits of the underlying satellite archive.
The evidence consistently indicates that persistent, lowland river flooding is its strongest use case.
River floods show a robust excess of corroboration above chance, whereas the signal for flash floods depends on the null model.
The Valencia and Baltic case studies further show that rapid-onset and coastal events may be represented only by residual inundation.
Together, the corroboration analysis and case studies indicate systematic differences in detectability among flood processes.

Within these limits, the archive provides dated and georeferenced observations of inundation that support recurrence and infrastructure screening, observation-based lower bounds on exposure, and comparison with modelled hazard scenarios.
These analyses should nevertheless be interpreted as records of detected flooding rather than complete reconstructions of flood occurrence.
An absent or spatially limited footprint does not imply that flooding was absent or limited, particularly for short-lived, urban, vegetated, or topographically confined events.
Likewise, recurrence measures archived detections rather than hydrological frequency, and observed--modelled comparisons reflect differences between datasets with distinct scopes and assumptions rather than formal validation of either product.

Temporal comparisons require additional caution because archive detectability is not stationary.
Corroboration varies among years together with satellite acquisition opportunities and processing conditions, so trends in event counts, recurrence, or inundated area may reflect both changes in flooding and changes in observation.
Depth estimates are also reconstructed from flood extent and terrain rather than measured directly and should therefore be interpreted at decimetre rather than centimetre precision.
Where observations are incomplete or a more comprehensive event reconstruction is required, hydrodynamic modelling or depth-reconstruction methods such as FLEXTH and FwDET remain necessary \citep{betterle2024flexth,cohen2018fwdet}.

\subsection{Limitations}

The principal limitations originate in the source archive.
Sentinel-1 under-detects inundation beneath vegetation, within dense urban areas, and during events that develop and drain between satellite acquisitions, so portions of individual floods, and in some cases entire floods, may be absent.
The reconstructed depth fields introduce additional uncertainty, with reported root-mean-square errors of \qtyrange{0.28}{0.62}{\m} against hydrodynamic reference simulations \citep{betterle2024flexth}.
The archive's event unit is also a spatio-temporal cluster rather than necessarily one hydrological flood, so one physical event may span several files, while one file may contain observations separated in space or time.
Analyses that require hydrologically individuated events must therefore inspect and, where necessary, re-aggregate or subdivide the archived records.

The index introduces one deliberate spatial approximation.
Its conservative any-wet resampling retains every detected source pixel at the coarser $\sim$\qty{90}{\m} grid but enlarges the resulting footprints, producing a median area inflation factor of 1.50 relative to the native-resolution wet area.
Index-derived footprints are therefore appropriate for event discovery, recurrence screening, and spatial selection, while inundated area, depth, and water-volume statistics should be calculated from the downloaded \qty{20}{\m} rasters.

Finally, each index release is an immutable snapshot of a growing archive.
Newly published events appear only in subsequent releases, preserving reproducibility at the cost of automatic updating.
Reported quantities are therefore specific to the stated index release and observation period.
Retrieval of native-resolution rasters also depends initially on the availability of the JRC service; local mirrors remove this dependency only after the required files have been staged.

\subsection{Sustainability and outlook}

The published index is designed to remain usable independently of continued development of the EuroFlood Python package.
Its components are distributed in open, cloud-native formats, namely a Cloud-Optimized GeoTIFF and GeoParquet tables, that can be read directly by standard geospatial software \citep{geoparquet,abernathey2021arco}.
Each release is immutable, archived under a persistent identifier, and linked to the corresponding source-archive state.
Local mirrors further enable the same queries without network access.
Although software maintenance currently depends on a single developer, the published index releases remain accessible independently of the library itself.

New index releases are planned as the source archive is extended, typically following major upstream updates.
Each release is versioned, immutable, and accompanied by a manifest recording the source listing and software state, allowing analyses to identify and retrieve the exact inputs on which they were based.

Future releases will extend the observational record and may also reflect changes in satellite coverage.
The restoration of two-satellite operations with Sentinel-1C in 2025 \citep{esa2024s1c,copernicus2025s1c} may increase observation opportunities relative to earlier years.
Detectability and corroboration characteristics should therefore be reassessed as the archive grows.

Further development could improve interoperability and generality.
Publishing STAC metadata would facilitate discovery through standard geospatial catalogues \citep{stacspec}, while source-specific ingestion adapters could extend the Discover$\rightarrow$Extract workflow to event-based archives such as GFM and OPERA DSWx.
More broadly, the inverted raster representation provides a candidate approach for transforming spatially dispersed event rasters into compact, versioned, and queryable environmental data resources.

% ===========================================================================
\section{Conclusions}
\label{sec:conclusions}

EuroFlood transforms the CEMS-EFAS satellite-derived flood-depth archive from a collection of spatially opaque raster files into a queryable and reproducible data resource by materialising event membership at the grid-cell level.
The resulting index enables efficient event discovery, footprint reconstruction, and recurrence analysis without requiring bulk access to the source depth rasters.
The accompanying Discover$\rightarrow$Extract interface separates low-cost exploration from native-resolution data retrieval and provides observed flood events and modelled CEMS-GLOFAS hazard scenarios within the same reproducible workflow.
Verification and benchmarking show that the index reproduces the source archive faithfully at its grid resolution and supports efficient spatial queries across a wide range of scales.
Compared with the spatial information available directly from the source archive, it substantially improves event discovery while also enabling cell-level analyses that simpler event catalogues cannot provide directly.
The usefulness of the underlying archive nevertheless depends on the flood process and observation conditions.
Persistent flooding that remains observable across satellite acquisitions is captured more consistently than rapidly evolving or spatially constrained events.
Absence or limited extent in the archive should therefore not be interpreted as evidence that flooding was absent or limited.
Within these observational constraints, EuroFlood supports continental recurrence screening, event-based exposure and infrastructure analysis, and comparison of observed inundation with modelled hazard scenarios.
By publishing the index in open, versioned, and cloud-native formats, EuroFlood provides a reproducible framework for incorporating future observations and a candidate access model for other event-based environmental raster archives.

% ===========================================================================
\section*{Software and data availability}
\label{sec:availability}

EuroFlood is open-source software released under the MIT licence and available from PyPI\footnote{\url{https://pypi.org/project/euroflood/}}; source code, documentation, tests, and release history are hosted at GitHub\footnote{\url{https://github.com/cisgroup/euroflood}}.
The software release used in this study is archived under a persistent identifier \citep{euroflood_software}.
EuroFlood requires Python 3.13 or later and accesses the published index without authentication; the index can also be mirrored for fully offline use.
The versioned spatial index, comprising the Cloud-Optimized GeoTIFF, event-combination dictionary, events table, and release manifest, is available from Source Cooperative\footnote{\url{https://source.coop/hackl/euroflood-index}} and archived under a persistent identifier \citep{euroflood_index}.
The original CEMS-EFAS flood-depth rasters \citep{betterle2025dataset} and CEMS-GLOFAS hazard maps \citep{baugh2024floodhazard} remain hosted by the Joint Research Centre and are accessed on demand rather than redistributed.
All scripts, configuration files, derived data, and figure-generation workflows required to reproduce the verification, benchmarks, corroboration analysis, and case studies are archived online \citep{euroflood_scripts}.
Results were produced with EuroFlood 0.3.1 and index release v1.1.0.
The Supplementary Information provides extended analyses, maintenance procedures, construction benchmarks, and verification details.
All figures were generated through the archived Python workflows.
Unless stated otherwise, map backgrounds and building and road geometries are derived from OpenStreetMap under the Open Database Licence, and administrative boundaries from Eurostat GISCO.
Modelled flood depths use CEMS-GLOFAS version 2.1.2 \citep{baugh2024floodhazard}; exposure analyses use the 2020 GHS-POP and GHS-BUILT-S grids \citep{schiavina2023ghspop,pesaresi2023ghsbuilts}; and documented flood dates, regions, and impacts are from HANZE \citep{paprotny2018,paprotny2024essd,paprotny2024hanze}.
Dataset-specific processing choices and analysis regions are defined in the corresponding methods and figure captions.

% ===========================================================================
\section*{Declaration of generative AI and AI-assisted technologies in the manuscript preparation process}
During the preparation of this manuscript, the author used Claude (Anthropic) to check spelling and grammar and to correct typographical errors.
The tool was not used to generate scientific content, data, results, or interpretations.
After using this tool, the author reviewed and edited the text as needed and takes full responsibility for the content of the publication.

\section*{CRediT authorship contribution statement}
\textbf{J\"urgen Hackl:} Conceptualization, Methodology, Software, Validation, Formal analysis, Data curation, Writing -- original draft, Visualization.

\section*{Declaration of competing interest}
The author declares that he has no known competing financial interests or personal relationships that could have appeared to influence the work reported in this paper.

\section*{Funding}
This research did not receive any specific grant from funding agencies in the public, commercial, or not-for-profit sectors.

\section*{Acknowledgements}

The author acknowledges ICEYE's publicly available flood briefing reports as inspiration for the visual presentation and layout of several figures.
The author also thanks Source Cooperative for providing the public platform used to host and distribute the EuroFlood index.

% ===========================================================================

\bibliographystyle{unsrtnat}
\bibliography{references}

\begin{thebibliography}{61}
\providecommand{\natexlab}[1]{#1}
\providecommand{\url}[1]{\texttt{#1}}
\expandafter\ifx\csname urlstyle\endcsname\relax
  \providecommand{\doi}[1]{doi: #1}\else
  \providecommand{\doi}{doi: \begingroup \urlstyle{rm}\Url}\fi

\bibitem[{European Environment Agency}(2025)]{eea_losses}
{European Environment Agency}.
\newblock Economic losses from weather- and climate-related extremes in
  {Europe}.
\newblock EEA indicator, European Environment Agency, Copenhagen, 2025.
\newblock URL
  \url{https://www.eea.europa.eu/en/analysis/indicators/economic-losses-from-climate-related}.
\newblock Accessed 13 July 2026.

\bibitem[Mohr et~al.(2023)Mohr, Ehret, Kunz, Ludwig, Caldas-Alvarez, Daniell,
  et~al.]{mohr2023part1}
Susanna Mohr, Uwe Ehret, Michael Kunz, Patrick Ludwig, Alberto Caldas-Alvarez,
  James~E. Daniell, et~al.
\newblock A multi-disciplinary analysis of the exceptional flood event of
  {July} 2021 in central {Europe} -- {Part} 1: Event description and analysis.
\newblock \emph{Natural Hazards and Earth System Sciences}, 23\penalty0
  (2):\penalty0 525--551, 2023.
\newblock \doi{10.5194/nhess-23-525-2023}.

\bibitem[Tradowsky et~al.(2023)Tradowsky, Philip, Kreienkamp, Kew, Lorenz,
  Arrighi, et~al.]{tradowsky2023}
Jordis~S. Tradowsky, Sjoukje~Y. Philip, Frank Kreienkamp, Sarah~F. Kew, Philip
  Lorenz, Julie Arrighi, et~al.
\newblock Attribution of the heavy rainfall events leading to severe flooding
  in {Western} {Europe} during {July} 2021.
\newblock \emph{Climatic Change}, 176\penalty0 (7):\penalty0 90, 2023.
\newblock \doi{10.1007/s10584-023-03502-7}.

\bibitem[Jongman et~al.(2012)Jongman, Kreibich, Apel, Barredo, Bates, Feyen,
  Gericke, Neal, Aerts, and Ward]{jongman2012}
B.~Jongman, H.~Kreibich, H.~Apel, J.~I. Barredo, P.~D. Bates, L.~Feyen,
  A.~Gericke, J.~Neal, J.~C. J.~H. Aerts, and P.~J. Ward.
\newblock Comparative flood damage model assessment: towards a {European}
  approach.
\newblock \emph{Natural Hazards and Earth System Sciences}, 12\penalty0
  (12):\penalty0 3733--3752, 2012.
\newblock \doi{10.5194/nhess-12-3733-2012}.

\bibitem[Delforge et~al.(2025)Delforge, Wathelet, Below, Lanfredi~Sofia,
  Tonnelier, van Loenhout, and Speybroeck]{delforge2025emdat}
Damien Delforge, Valentin Wathelet, Regina Below, Cinzia Lanfredi~Sofia, Margo
  Tonnelier, Joris A.~F. van Loenhout, and Niko Speybroeck.
\newblock {EM-DAT}: the emergency events database.
\newblock \emph{International Journal of Disaster Risk Reduction},
  124:\penalty0 105509, 2025.
\newblock \doi{10.1016/j.ijdrr.2025.105509}.

\bibitem[Dottori et~al.(2020)Dottori, Mentaschi, Bianchi, Alfieri, and
  Feyen]{dottori2020peseta}
Francesco Dottori, Lorenzo Mentaschi, Alessandra Bianchi, Lorenzo Alfieri, and
  Luc Feyen.
\newblock Adapting to rising river flood risk in the {EU} under climate change.
\newblock Technical Report EUR 29955 EN, JRC118425, European Commission, Joint
  Research Centre (JRC), Luxembourg, 2020.
\newblock JRC PESETA IV project, Task 5.

\bibitem[{European Union}(2007)]{floodsdirective}
{European Union}.
\newblock Directive 2007/60/{EC} of the {European} parliament and of the
  council of 23 {October} 2007 on the assessment and management of flood risks
  (floods directive).
\newblock Official Journal of the European Union, L 288, 27--34, 2007.
\newblock URL \url{https://eur-lex.europa.eu/eli/dir/2007/60/oj}.

\bibitem[Trigg et~al.(2016)Trigg, Birch, Neal, Bates, Smith, Sampson,
  et~al.]{trigg2016}
M.~A. Trigg, C.~E. Birch, J.~C. Neal, P.~D. Bates, A.~Smith, C.~C. Sampson,
  et~al.
\newblock The credibility challenge for global fluvial flood risk analysis.
\newblock \emph{Environmental Research Letters}, 11\penalty0 (9):\penalty0
  094014, 2016.
\newblock \doi{10.1088/1748-9326/11/9/094014}.

\bibitem[Bernhofen et~al.(2018)Bernhofen, Whyman, Trigg, Sleigh, Smith,
  Sampson, et~al.]{bernhofen2018}
Mark~V. Bernhofen, Charlie Whyman, Mark~A. Trigg, P.~Andrew Sleigh, Andrew~M.
  Smith, Christopher~C. Sampson, et~al.
\newblock A first collective validation of global fluvial flood models for
  major floods in {Nigeria} and {Mozambique}.
\newblock \emph{Environmental Research Letters}, 13\penalty0 (10):\penalty0
  104007, 2018.
\newblock \doi{10.1088/1748-9326/aae014}.

\bibitem[Dottori et~al.(2016)Dottori, Salamon, Bianchi, Alfieri, Hirpa, and
  Feyen]{dottori2016}
Francesco Dottori, Peter Salamon, Alessandra Bianchi, Lorenzo Alfieri,
  Feyera~Aga Hirpa, and Luc Feyen.
\newblock Development and evaluation of a framework for global flood hazard
  mapping.
\newblock \emph{Advances in Water Resources}, 94:\penalty0 87--102, 2016.
\newblock \doi{10.1016/j.advwatres.2016.05.002}.

\bibitem[Sampson et~al.(2015)Sampson, Smith, Bates, Neal, Alfieri, and
  Freer]{sampson2015}
Christopher~C. Sampson, Andrew~M. Smith, Paul~D. Bates, Jeffrey~C. Neal,
  Lorenzo Alfieri, and Jim~E. Freer.
\newblock A high-resolution global flood hazard model.
\newblock \emph{Water Resources Research}, 51\penalty0 (9):\penalty0
  7358--7381, 2015.
\newblock \doi{10.1002/2015WR016954}.

\bibitem[Tellman et~al.(2021)Tellman, Sullivan, Kuhn, Kettner, Doyle,
  Brakenridge, Erickson, and Slayback]{tellman2021}
B.~Tellman, J.~A. Sullivan, C.~Kuhn, A.~J. Kettner, C.~S. Doyle, G.~R.
  Brakenridge, T.~A. Erickson, and D.~A. Slayback.
\newblock Satellite imaging reveals increased proportion of population exposed
  to floods.
\newblock \emph{Nature}, 596\penalty0 (7870):\penalty0 80--86, 2021.
\newblock \doi{10.1038/s41586-021-03695-w}.

\bibitem[Huizinga et~al.(2017)Huizinga, de~Moel, and Szewczyk]{huizinga2017}
Jan Huizinga, Hans de~Moel, and Wojciech Szewczyk.
\newblock Global flood depth-damage functions: Methodology and the database
  with guidelines.
\newblock Technical Report EUR 28552 EN, JRC105688, European Commission, Joint
  Research Centre (JRC), Luxembourg, 2017.

\bibitem[Wing et~al.(2020)Wing, Pinter, Bates, and Kousky]{wing2020}
Oliver E.~J. Wing, Nicholas Pinter, Paul~D. Bates, and Carolyn Kousky.
\newblock New insights into {US} flood vulnerability revealed from flood
  insurance big data.
\newblock \emph{Nature Communications}, 11\penalty0 (1):\penalty0 1444, 2020.
\newblock \doi{10.1038/s41467-020-15264-2}.

\bibitem[Betterle and Salamon(2025)]{betterle2025dataset}
Andrea Betterle and Peter Salamon.
\newblock Satellite-derived flood depth maps for {Europe}.
\newblock European Commission, Joint Research Centre (JRC) / Copernicus EMS,
  2025.

\bibitem[Betterle and Salamon(2024)]{betterle2024flexth}
Andrea Betterle and Peter Salamon.
\newblock Water depth estimate and flood extent enhancement for satellite-based
  inundation maps.
\newblock \emph{Natural Hazards and Earth System Sciences}, 24\penalty0
  (8):\penalty0 2817--2836, 2024.
\newblock \doi{10.5194/nhess-24-2817-2024}.

\bibitem[Krullikowski et~al.(2023)Krullikowski, Chow, Wieland, Martinis,
  Bauer-Marschallinger, Roth, Matgen, Chini, et~al.]{krullikowski2023gfm}
Christian Krullikowski, Candace Chow, Marc Wieland, Sandro Martinis, Bernhard
  Bauer-Marschallinger, Florian Roth, Patrick Matgen, Marco Chini, et~al.
\newblock Estimating ensemble likelihoods for the {Sentinel-1}-based global
  flood monitoring product of the {Copernicus} emergency management service.
\newblock \emph{IEEE Journal of Selected Topics in Applied Earth Observations
  and Remote Sensing}, 16:\penalty0 6917--6930, 2023.
\newblock \doi{10.1109/JSTARS.2023.3292350}.

\bibitem[Aznar-Siguan and Bresch(2019)]{aznar2019climada}
Gabriela Aznar-Siguan and David~N. Bresch.
\newblock {CLIMADA} v1: a global weather and climate risk assessment platform.
\newblock \emph{Geoscientific Model Development}, 12\penalty0 (7):\penalty0
  3085--3097, 2019.
\newblock \doi{10.5194/gmd-12-3085-2019}.

\bibitem[Gorelick et~al.(2017)Gorelick, Hancher, Dixon, Ilyushchenko, Thau, and
  Moore]{gorelick2017gee}
Noel Gorelick, Matt Hancher, Mike Dixon, Simon Ilyushchenko, David Thau, and
  Rebecca Moore.
\newblock {Google} {Earth} {Engine}: Planetary-scale geospatial analysis for
  everyone.
\newblock \emph{Remote Sensing of Environment}, 202:\penalty0 18--27, 2017.
\newblock \doi{10.1016/j.rse.2017.06.031}.

\bibitem[Maze and Balem(2020)]{maze2020argopy}
Guillaume Maze and Kevin Balem.
\newblock argopy: A {Python} library for {Argo} ocean data analysis.
\newblock \emph{Journal of Open Source Software}, 5\penalty0 (53):\penalty0
  2425, 2020.
\newblock \doi{10.21105/joss.02425}.

\bibitem[Iturbide et~al.(2019)Iturbide, Bedia, Herrera, Ba{\~n}o-Medina,
  Fern{\'a}ndez, Fr{\'i}as, Manzanas, San-Mart{\'i}n, Cimadevilla, Cofi{\~n}o,
  and Guti{\'e}rrez]{iturbide2019climate4r}
M.~Iturbide, J.~Bedia, S.~Herrera, J.~Ba{\~n}o-Medina, J.~Fern{\'a}ndez, M.~D.
  Fr{\'i}as, R.~Manzanas, D.~San-Mart{\'i}n, E.~Cimadevilla, A.~S. Cofi{\~n}o,
  and J.~M. Guti{\'e}rrez.
\newblock The {R}-based climate4r open framework for reproducible climate data
  access and post-processing.
\newblock \emph{Environmental Modelling \& Software}, 111:\penalty0 42--54,
  2019.
\newblock \doi{10.1016/j.envsoft.2018.09.009}.

\bibitem[Chegini et~al.(2021)Chegini, Li, and Leung]{chegini2021hyriver}
Taher Chegini, Hong-Yi Li, and L.~Ruby Leung.
\newblock {HyRiver}: Hydroclimate data retriever.
\newblock \emph{Journal of Open Source Software}, 6\penalty0 (66):\penalty0
  3175, 2021.
\newblock \doi{10.21105/joss.03175}.

\bibitem[Betterle and Salamon(2026)]{betterle_sciadv}
Andrea Betterle and Peter Salamon.
\newblock Ten years of floods across {Europe} mapped from space with
  reconstructed water depths.
\newblock EGU General Assembly 2026, abstract EGU26-7998, 2026.

\bibitem[Penton et~al.(2023)Penton, Teng, Ticehurst, Marvanek, Freebairn,
  Mateo, Vaze, Yang, Khanam, Sengupta, and Pollino]{penton2023mdb}
David~J. Penton, Jin Teng, Catherine Ticehurst, Steve Marvanek, Andrew
  Freebairn, Cherry Mateo, Jai Vaze, Ang Yang, Fathaha Khanam, Ashmita
  Sengupta, and Carmel Pollino.
\newblock The floodplain inundation history of the {Murray--Darling Basin}
  through two-monthly maximum water depth maps.
\newblock \emph{Scientific Data}, 10\penalty0 (1):\penalty0 652, 2023.
\newblock \doi{10.1038/s41597-023-02559-4}.

\bibitem[Ardila et~al.(2022)Ardila, Laurila, Kourkouli, and
  Strong]{ardila2022iceye}
Juan Ardila, Pekka Laurila, Penelope Kourkouli, and Shay Strong.
\newblock Persistent monitoring and mapping of floods globally based on the
  {ICEYE} {SAR} imaging constellation.
\newblock In \emph{IGARSS 2022 -- IEEE International Geoscience and Remote
  Sensing Symposium}, pages 6296--6299, 2022.
\newblock \doi{10.1109/IGARSS46834.2022.9883587}.

\bibitem[Baugh et~al.(2024)Baugh, Colonese, D'Angelo, Dottori, Neal, Prudhomme,
  and Salamon]{baugh2024floodhazard}
Calum Baugh, Juan Colonese, Claudia D'Angelo, Francesco Dottori, Jeffrey Neal,
  Christel Prudhomme, and Peter Salamon.
\newblock Global river flood hazard maps.
\newblock European Commission, Joint Research Centre (JRC) / Copernicus EMS,
  2024.
\newblock URL
  \url{https://data.jrc.ec.europa.eu/dataset/jrc-floods-floodmapgl_rp50y-tif}.
\newblock Version 2.1.2. Accessed 17 July 2026.

\bibitem[Ward et~al.(2020)Ward, Winsemius, Kuzma, Bierkens, Bouwman, de~Moel,
  et~al.]{aqueduct}
Philip~J. Ward, Hessel~C. Winsemius, Samantha Kuzma, Marc F.~P. Bierkens, Arno
  Bouwman, Hans de~Moel, et~al.
\newblock Aqueduct floods methodology.
\newblock Technical note, World Resources Institute (WRI), Washington, DC,
  2020.
\newblock URL \url{https://www.wri.org/research/aqueduct-floods-methodology}.

\bibitem[{Deltares}(2021)]{deltares2021coastal}
{Deltares}.
\newblock Planetary computer and {Deltares} global data: Flood hazard maps.
\newblock Technical Report 11206409-003-ZWS-0003, Deltares, 2021.
\newblock URL
  \url{https://planetarycomputer.microsoft.com/dataset/deltares-floods}.
\newblock Accessed 17 July 2026.

\bibitem[Wing et~al.(2024)Wing, Bates, Quinn, Savage, Uhe, Cooper, Collings,
  Addor, Lord, Hatchard, Hoch, Bates, Probyn, Himsworth,
  Rodr{\'i}guez~Gonz{\'a}lez, Brine, Wilkinson, Sampson, Smith, Neal, and
  Haigh]{wing2024fathom}
Oliver E.~J. Wing, Paul~D. Bates, Niall~D. Quinn, James T.~S. Savage, Peter~F.
  Uhe, Anthony Cooper, Thomas~P. Collings, Nans Addor, Natalie~S. Lord, Simbi
  Hatchard, Jannis~M. Hoch, Joe Bates, Izzy Probyn, Sam Himsworth, Josu{\'e}
  Rodr{\'i}guez~Gonz{\'a}lez, Malcolm~P. Brine, Hamish Wilkinson,
  Christopher~C. Sampson, Andrew~M. Smith, Jeffrey~C. Neal, and Ivan~D. Haigh.
\newblock A 30\,m global flood inundation model for any climate scenario.
\newblock \emph{Water Resources Research}, 60\penalty0 (8):\penalty0
  e2023WR036460, 2024.
\newblock \doi{10.1029/2023WR036460}.

\bibitem[{NASA Jet Propulsion Laboratory}(2023)]{opera_dswx}
{NASA Jet Propulsion Laboratory}.
\newblock {OPERA} dynamic surface water extent ({DSWx}) product suite.
\newblock Observational Products for End-Users from Remote Sensing Analysis
  (OPERA), NASA, 2023.
\newblock URL \url{https://www.jpl.nasa.gov/go/opera}.
\newblock Accessed 13 July 2026.

\bibitem[Brakenridge(2024)]{dfo}
G.~R. Brakenridge.
\newblock Global active archive of large flood events.
\newblock Dartmouth Flood Observatory, University of Colorado, Boulder, 2024.
\newblock URL \url{https://floodobservatory.colorado.edu/}.
\newblock Accessed 13 July 2026.

\bibitem[Wagner et~al.(2026)Wagner, Bauer-Marschallinger, Roth, Raiger-Stachl,
  Reimer, McCormick, Matgen, Chini, et~al.]{wagner2026}
Wolfgang Wagner, Bernhard Bauer-Marschallinger, Florian Roth, Tobias
  Raiger-Stachl, Christoph Reimer, Niall McCormick, Patrick Matgen, Marco
  Chini, et~al.
\newblock The fully-automatic {Sentinel-1} global flood monitoring service:
  Scientific challenges and future directions.
\newblock \emph{Remote Sensing of Environment}, 333:\penalty0 115108, 2026.
\newblock \doi{10.1016/j.rse.2025.115108}.

\bibitem[Torres et~al.(2012)Torres, Snoeij, Geudtner, Bibby, Davidson, Attema,
  Potin, Rommen, Floury, Brown, Traver, Deghaye, Duesmann, Rosich, Miranda,
  Bruno, L'Abbate, Croci, Pietropaolo, Huchler, and Rostan]{torres2012}
Ramon Torres, Paul Snoeij, Dirk Geudtner, David Bibby, Malcolm Davidson, Evert
  Attema, Pierre Potin, Bj{\"o}rn Rommen, Nicolas Floury, Mike Brown,
  Ignacio~Navas Traver, Patrick Deghaye, Berthyl Duesmann, Betlem Rosich, Nuno
  Miranda, Claudio Bruno, Michelangelo L'Abbate, Renato Croci, Andrea
  Pietropaolo, Markus Huchler, and Friedhelm Rostan.
\newblock {GMES} {Sentinel-1} mission.
\newblock \emph{Remote Sensing of Environment}, 120:\penalty0 9--24, 2012.
\newblock \doi{10.1016/j.rse.2011.05.028}.

\bibitem[{European Space Agency}(2022)]{esa2022s1b}
{European Space Agency}.
\newblock Mission ends for {Copernicus Sentinel-1B} satellite.
\newblock ESA news release, 3 August 2022, 2022.
\newblock URL
  \url{https://www.esa.int/Applications/Observing_the_Earth/Copernicus/Sentinel-1/Mission_ends_for_Copernicus_Sentinel-1B_satellite}.
\newblock Accessed 17 July 2026.

\bibitem[{European Space Agency}(2024)]{esa2024s1c}
{European Space Agency}.
\newblock Double win for {Europe}: {Sentinel-1C} and {Vega-C} take to the
  skies.
\newblock ESA press release 70--2024, 6 December 2024, 2024.
\newblock URL
  \url{https://www.esa.int/Newsroom/Press_Releases/Double_win_for_Europe_Sentinel-1C_and_Vega-C_take_to_the_skies}.
\newblock Accessed 20 July 2026.

\bibitem[{Copernicus Data Space Ecosystem}(2025)]{copernicus2025s1c}
{Copernicus Data Space Ecosystem}.
\newblock {Sentinel-1C} products are now calibrated.
\newblock Copernicus Data Space Ecosystem news, 21 May 2025, 2025.
\newblock URL
  \url{https://dataspace.copernicus.eu/news/2025-5-21-sentinel-1c-products-are-now-calibrated}.
\newblock Accessed 18 September 2026.

\bibitem[Misra et~al.(2025)Misra, White, Nsutezo, Straka, and
  Lavista]{misra2025}
Amit Misra, Kevin White, Simone~Fobi Nsutezo, William Straka, and Juan Lavista.
\newblock Mapping global floods with 10 years of satellite radar data.
\newblock \emph{Nature Communications}, 16\penalty0 (1):\penalty0 5762, 2025.
\newblock \doi{10.1038/s41467-025-60973-1}.

\bibitem[Cohen et~al.(2018)Cohen, Brakenridge, Kettner, Bates, Nelson,
  McDonald, Huang, Munasinghe, and Zhang]{cohen2018fwdet}
Sagy Cohen, G.~Robert Brakenridge, Albert Kettner, Bradford Bates, Jonathan
  Nelson, Richard McDonald, Yu-Fen Huang, Dinuke Munasinghe, and Jiaqi Zhang.
\newblock Estimating floodwater depths from flood inundation maps and
  topography.
\newblock \emph{JAWRA Journal of the American Water Resources Association},
  54\penalty0 (4):\penalty0 847--858, 2018.
\newblock \doi{10.1111/1752-1688.12609}.

\bibitem[{UNOSAT}(2022)]{unosat2022pakistan}
{UNOSAT}.
\newblock Floodwater depth in {Sindh} and {Balochistan} provinces, {Pakistan}
  as of 29 {November} 2022.
\newblock United Nations Satellite Centre (UNOSAT), UNITAR, 2022.
\newblock URL \url{https://unosat.org/products/3457}.
\newblock Accessed 17 July 2026.

\bibitem[Chudley and Howat(2024)]{chudley2024pdemtools}
Thomas~R. Chudley and Ian~M. Howat.
\newblock pdemtools: conveniently search, download, and process {ArcticDEM} and
  {REMA} products.
\newblock \emph{Journal of Open Source Software}, 9\penalty0 (102):\penalty0
  7149, 2024.
\newblock \doi{10.21105/joss.07149}.

\bibitem[Boeing(2025)]{boeing2025osmnx}
Geoff Boeing.
\newblock Modeling and analyzing urban networks and amenities with {OSMnx}.
\newblock \emph{Geographical Analysis}, 57\penalty0 (4):\penalty0 567--577,
  2025.
\newblock \doi{10.1111/gean.70009}.

\bibitem[Schramm et~al.(2021)Schramm, Pebesma, Milenkovi{\'c}, Foresta, Dries,
  Jacob, Wagner, Mohr, Neteler, Kadunc, Miksa, Kempeneers, Verbesselt,
  G{\"o}{\ss}wein, Navacchi, Lippens, and Reiche]{schramm2021openeo}
Matthias Schramm, Edzer Pebesma, Milutin Milenkovi{\'c}, Luca Foresta, Jeroen
  Dries, Alexander Jacob, Wolfgang Wagner, Matthias Mohr, Markus Neteler, Miha
  Kadunc, Tomasz Miksa, Pieter Kempeneers, Jan Verbesselt, Bernhard
  G{\"o}{\ss}wein, Claudio Navacchi, Stefaan Lippens, and Johannes Reiche.
\newblock The {openEO} {API}--harmonising the use of {Earth} observation cloud
  services using virtual data cube functionalities.
\newblock \emph{Remote Sensing}, 13\penalty0 (6):\penalty0 1125, 2021.
\newblock \doi{10.3390/rs13061125}.

\bibitem[Hoyer and Hamman(2017)]{hoyer2017xarray}
Stephan Hoyer and Joe Hamman.
\newblock xarray: {N}-{D} labeled arrays and datasets in {Python}.
\newblock \emph{Journal of Open Research Software}, 5\penalty0 (1):\penalty0
  10, 2017.
\newblock \doi{10.5334/jors.148}.

\bibitem[Sahr et~al.(2003)Sahr, White, and Kimerling]{sahr2003dggs}
Kevin Sahr, Denis White, and A.~Jon Kimerling.
\newblock Geodesic discrete global grid systems.
\newblock \emph{Cartography and Geographic Information Science}, 30\penalty0
  (2):\penalty0 121--134, 2003.
\newblock \doi{10.1559/152304003100011090}.

\bibitem[{Open Geospatial Consortium}(2023)]{cogspec}
{Open Geospatial Consortium}.
\newblock {OGC} cloud optimized {GeoTIFF} standard.
\newblock OGC 21-026, version 1.0, 2023.
\newblock URL \url{https://docs.ogc.org/is/21-026/21-026.html}.

\bibitem[Pekel et~al.(2016)Pekel, Cottam, Gorelick, and Belward]{pekel2016}
Jean-Fran{\c c}ois Pekel, Andrew Cottam, Noel Gorelick, and Alan~S. Belward.
\newblock High-resolution mapping of global surface water and its long-term
  changes.
\newblock \emph{Nature}, 540\penalty0 (7633):\penalty0 418--422, 2016.
\newblock \doi{10.1038/nature20584}.

\bibitem[Zobel and Moffat(2006)]{zobel2006}
Justin Zobel and Alistair Moffat.
\newblock Inverted files for text search engines.
\newblock \emph{ACM Computing Surveys}, 38\penalty0 (2):\penalty0 6, 2006.
\newblock \doi{10.1145/1132956.1132959}.

\bibitem[Chambi et~al.(2016)Chambi, Lemire, Kaser, and
  Godin]{chambi2016roaring}
Samy Chambi, Daniel Lemire, Owen Kaser, and Robert Godin.
\newblock Better bitmap performance with {Roaring} bitmaps.
\newblock \emph{Software: Practice and Experience}, 46\penalty0 (5):\penalty0
  709--719, 2016.
\newblock \doi{10.1002/spe.2325}.

\bibitem[Hackl(2026{\natexlab{a}})]{euroflood_index}
J{\"u}rgen Hackl.
\newblock {EuroFlood}: a queryable cloud-native index for the {CEMS-EFAS}
  satellite-derived flood depth maps (v1.1.0).
\newblock Zenodo, 2026{\natexlab{a}}.
\newblock Concept DOI 10.5281/zenodo.21284459 resolves to the current release.

\bibitem[Jordahl et~al.(2020)Jordahl, Van~den Bossche, Fleischmann, Wasserman,
  McBride, Gerard, et~al.]{jordahl2020geopandas}
Kelsey Jordahl, Joris Van~den Bossche, Martin Fleischmann, Jacob Wasserman,
  James McBride, Jeffrey Gerard, et~al.
\newblock geopandas/geopandas: v0.8.1.
\newblock Zenodo, 2020.

\bibitem[Barker et~al.(2022)Barker, Chue~Hong, Katz, Lamprecht, Martinez-Ortiz,
  Psomopoulos, Harrow, Castro, Gruenpeter, Martinez, and
  Honeyman]{barker2022fair4rs}
Michelle Barker, Neil~P. Chue~Hong, Daniel~S. Katz, Anna-Lena Lamprecht, Carlos
  Martinez-Ortiz, Fotis Psomopoulos, Jennifer Harrow, Leyla~Jael Castro, Morane
  Gruenpeter, Paula~Andrea Martinez, and Tom Honeyman.
\newblock Introducing the {FAIR} principles for research software.
\newblock \emph{Scientific Data}, 9\penalty0 (1):\penalty0 622, 2022.
\newblock \doi{10.1038/s41597-022-01710-x}.

\bibitem[Paprotny et~al.(2018)Paprotny, Morales-N{\'a}poles, and
  Jonkman]{paprotny2018}
Dominik Paprotny, Oswaldo Morales-N{\'a}poles, and Sebastiaan~N. Jonkman.
\newblock {HANZE}: a pan-{European} database of exposure to natural hazards and
  damaging historical floods since 1870.
\newblock \emph{Earth System Science Data}, 10\penalty0 (1):\penalty0 565--581,
  2018.
\newblock \doi{10.5194/essd-10-565-2018}.

\bibitem[Paprotny et~al.(2024)Paprotny, Terefenko, and
  {\'S}ledziowski]{paprotny2024essd}
Dominik Paprotny, Pawe{\l} Terefenko, and Jakub {\'S}ledziowski.
\newblock {HANZE} v2.1: an improved database of flood impacts in {Europe} from
  1870 to 2020.
\newblock \emph{Earth System Science Data}, 16\penalty0 (11):\penalty0
  5145--5170, 2024.
\newblock \doi{10.5194/essd-16-5145-2024}.

\bibitem[Paprotny(2026)]{paprotny2024hanze}
Dominik Paprotny.
\newblock {HANZE} database of historical flood impacts in {Europe}, 1870--2025
  (v3.0.1 beta).
\newblock Zenodo, 2026.

\bibitem[Schiavina et~al.(2023)Schiavina, Freire, Carioli, and
  MacManus]{schiavina2023ghspop}
Marcello Schiavina, Sergio Freire, Alessandra Carioli, and Kytt MacManus.
\newblock {GHS-POP} r2023a -- {GHS} population grid multitemporal (1975--2030).
\newblock European Commission, Joint Research Centre (JRC), 2023.

\bibitem[Pesaresi and Politis(2023)]{pesaresi2023ghsbuilts}
Martino Pesaresi and Panagiotis Politis.
\newblock {GHS-BUILT-S} r2023a -- {GHS} built-up surface grid, derived from
  {Sentinel-2} composite and {Landsat}, multitemporal (1975--2030).
\newblock European Commission, Joint Research Centre (JRC), 2023.

\bibitem[{GeoParquet contributors}(2024)]{geoparquet}
{GeoParquet contributors}.
\newblock {GeoParquet} specification.
\newblock Community specification, version 1.1.0, 2024.
\newblock URL \url{https://geoparquet.org}.

\bibitem[Abernathey et~al.(2021)Abernathey, Augspurger, Banihirwe,
  Blackmon-Luca, Crone, Gentemann, Hamman, Henderson,
  et~al.]{abernathey2021arco}
Ryan~P. Abernathey, Tom Augspurger, Anderson Banihirwe, Charles~C.
  Blackmon-Luca, Timothy~J. Crone, Chelle~L. Gentemann, Joseph~J. Hamman, Naomi
  Henderson, et~al.
\newblock Cloud-native repositories for big scientific data.
\newblock \emph{Computing in Science \& Engineering}, 23\penalty0 (2):\penalty0
  26--35, 2021.
\newblock \doi{10.1109/MCSE.2021.3059437}.

\bibitem[{STAC Contributors}(2024)]{stacspec}
{STAC Contributors}.
\newblock {SpatioTemporal} asset catalog ({STAC}) specification.
\newblock Community specification, version 1.1.0, 2024.
\newblock URL \url{https://stacspec.org}.

\bibitem[Hackl(2026{\natexlab{b}})]{euroflood_software}
J{\"u}rgen Hackl.
\newblock {EuroFlood}: query {Europe}'s satellite flood-depth maps by place and
  time.
\newblock Zenodo, 2026{\natexlab{b}}.
\newblock URL \url{https://github.com/cisgroup/euroflood}.
\newblock Concept DOI, which always resolves to the current release; version
  0.3.1 is archived as 10.5281/zenodo.22863140.

\bibitem[Hackl(2026{\natexlab{c}})]{euroflood_scripts}
J{\"u}rgen Hackl.
\newblock {EuroFlood} paper reproduction package: figures, derived data, and
  analysis code.
\newblock Zenodo, 2026{\natexlab{c}}.
\newblock Concept DOI, which always resolves to the current release; this paper
  is reproduced by version 2.

\end{thebibliography}

% ===== SUPPLEMENTARY INFORMATION (from manuscript/supplementary.tex) =====
\clearpage
\onecolumn
\setcounter{section}{0}\renewcommand{\thesection}{S\arabic{section}}
\setcounter{figure}{0}\renewcommand{\thefigure}{S\arabic{figure}}
\setcounter{table}{0}\renewcommand{\thetable}{S\arabic{table}}
\setcounter{equation}{0}\renewcommand{\theequation}{S\arabic{equation}}
\begin{center}{\LARGE\scshape Supplementary Information}\end{center}
\vskip 0.15in

% ==========================================================================
\section{Extended illustrative case studies}
\label{si:sec:cases}

\subsection*{Overview and common methodology}
\label{si:sec:cases-overview}

This section presents a set of exploratory case studies that demonstrate the range of analyses enabled by the EuroFlood access model.
The studies span different regions, flood processes, spatial scales, downstream applications, and user communities, and collectively illustrate both the archive's principal strengths and the conditions under which its outputs require cautious interpretation.
Several cases are summarised in the main text and documented here in greater detail: the observed--modelled comparison at Zutphen (Section~7.2 of the main text), the July 2021 exposure analysis (Section~7.3), the Valencia DANA detectability example (Section~7.4), and the HANZE corroboration audit (Section~6).
The remaining cases are mentioned only briefly in the main text and are included here as additional demonstrations of analytical scope and fitness for use.

All case studies follow a common workflow.
Event discovery begins with the index-only query \texttt{floods(region, \dots)}, which reads a small portion of the versioned index and returns a \texttt{FloodFrame}, a subclass of \texttt{geopandas.GeoDataFrame}.
Per-event footprints are reconstructed with \texttt{.footprints()}, native-resolution water-depth rasters are retrieved only when required with \texttt{download()}, and modelled hazard layers are obtained with \texttt{hazard(return\_period=\dots)}.
Because \texttt{FloodFrame} inherits the standard \texttt{GeoDataFrame} interface, the archive can be combined directly with external open datasets, including administrative boundaries, road networks, population and built-up grids, and documentary flood records, using conventional spatial operations.
Each analysis first confirms archive coverage through a low-cost discovery query before downloading any source rasters, and each case study is reproduced by a dedicated script, using the public index and openly available external data.

Unless stated otherwise, observed extents are derived from the conservative any-wet index footprint at approximately \qty{90}{\m} resolution.
These footprints are affected by two opposing sources of bias: synthetic-aperture radar under-detection omits inundation that the sensor cannot observe, whereas spatial aggregation enlarges the detected footprint relative to the native \qty{20}{\m} source, with a median area inflation factor of 1.50 (Section~6 of the main text).
All reported depth statistics are therefore recalculated from the native-resolution \qty{20}{\m} rasters.
The source archive covers 2015--2025 and inherits the acquisition frequency and detectability characteristics of Sentinel-1 and the Copernicus Global Flood Monitoring product.

Table~\ref{si:tab:case-overview} summarises the purpose and principal result of each case study.
The following subsections present each study using a common structure comprising motivation, data and methods, results, and a critical evaluation of what the findings support and how they should be interpreted.
The case studies are ordered from core discovery and analysis capabilities through process-specific detectability limits to a systematic cross-dataset audit and final fitness-for-use synthesis.

\begin{table}[htbp]
\centering
\caption{Overview of the extended illustrative case studies. The analytical dimension identifies the principal capability or fitness-for-use question examined by each study.}
\label{si:tab:case-overview}
\small
\begin{tabular}{@{}P{2.5cm}P{2.7cm}P{2.2cm}P{2.6cm}P{3.6cm}@{}}
\toprule
Study & Region / event & Flood process & Analytical dimension & Principal result \\
\midrule
Storm Boris (\ref{si:sec:boris}) & Central and Eastern Europe, Sept.--Nov. 2024 & Fluvial & Transboundary discovery & One query resolves a \qty{1253}{\km\squared} footprint across nine countries \\
\addlinespace[3pt]
Detection climatology (\ref{si:sec:clim}) & Continental Europe & Multiple processes & Temporal coverage & Feb.--Apr. accounts for 54\% of mapped water but 26\% of 3,246 events \\
\addlinespace[3pt]
Shannon callows (\ref{si:sec:shannon}) & Ireland & Fluvial / pluvial & Infrastructure screening & 61 events intersect \qty{26.0}{\km} of roads and repeatedly affect selected settlements \\
\addlinespace[3pt]
Zutphen record (\ref{si:sec:zutphen}) & IJssel at Zutphen, 2015--2025 & Fluvial & Multi-event place analysis & 33 events produce recurrence of up to 26 detections per cell \\
\addlinespace[3pt]
Exposure (\ref{si:sec:exposure}) & July 2021 Meuse--Rhine; Emilia-Romagna 2023 & Fluvial & Population and built environment & Observed event footprints provide lower-bound exposure estimates \\
\addlinespace[3pt]
Valencia DANA (\ref{si:sec:dana}) & Spain, Oct. 2024 & Flash / pluvial & Detectability limit & \qty{35.2}{\km\squared} is captured six days after the peak, equivalent to 16\% of the documented area \\
\addlinespace[3pt]
Baltic surge (\ref{si:sec:baltic}) & Germany and Denmark, Oct. 2023 & Coastal surge & Detectability limit & \qty{54.5}{\km\squared} of residual inundation is captured nine days after the surge \\
\addlinespace[3pt]
HANZE audit (\ref{si:sec:hanze}) & Continental Europe & Multiple processes & Cross-dataset corroboration & Corroboration varies systematically by flood type, year, and country \\
\bottomrule
\end{tabular}
\end{table}

\newpage

% ==========================================================================

\subsection{Transboundary discovery: Storm Boris, September 2024}
\label{si:sec:boris}

\paragraph{Motivation.}
Transboundary flood analysis is complicated by the fact that a single meteorological episode may affect several countries, while flood records, impact data, and mapping products are typically compiled within national systems using different formats and spatial reference systems.
Storm Boris, which affected Central and Eastern Europe between 9 and 16~September 2024, provides a representative example: persistent frontal rainfall over the Odra, Morava, and upper Danube basins caused severe flooding across Poland, Czechia, Austria, Slovakia, and downstream Danube states.
This case study examines whether a single EuroFlood query can recover the geographically distributed archive footprint and attribute detected flood area by country without first harmonising national flood maps.

\begin{figure}[htbp]
\centering
\includegraphics[width=\linewidth]{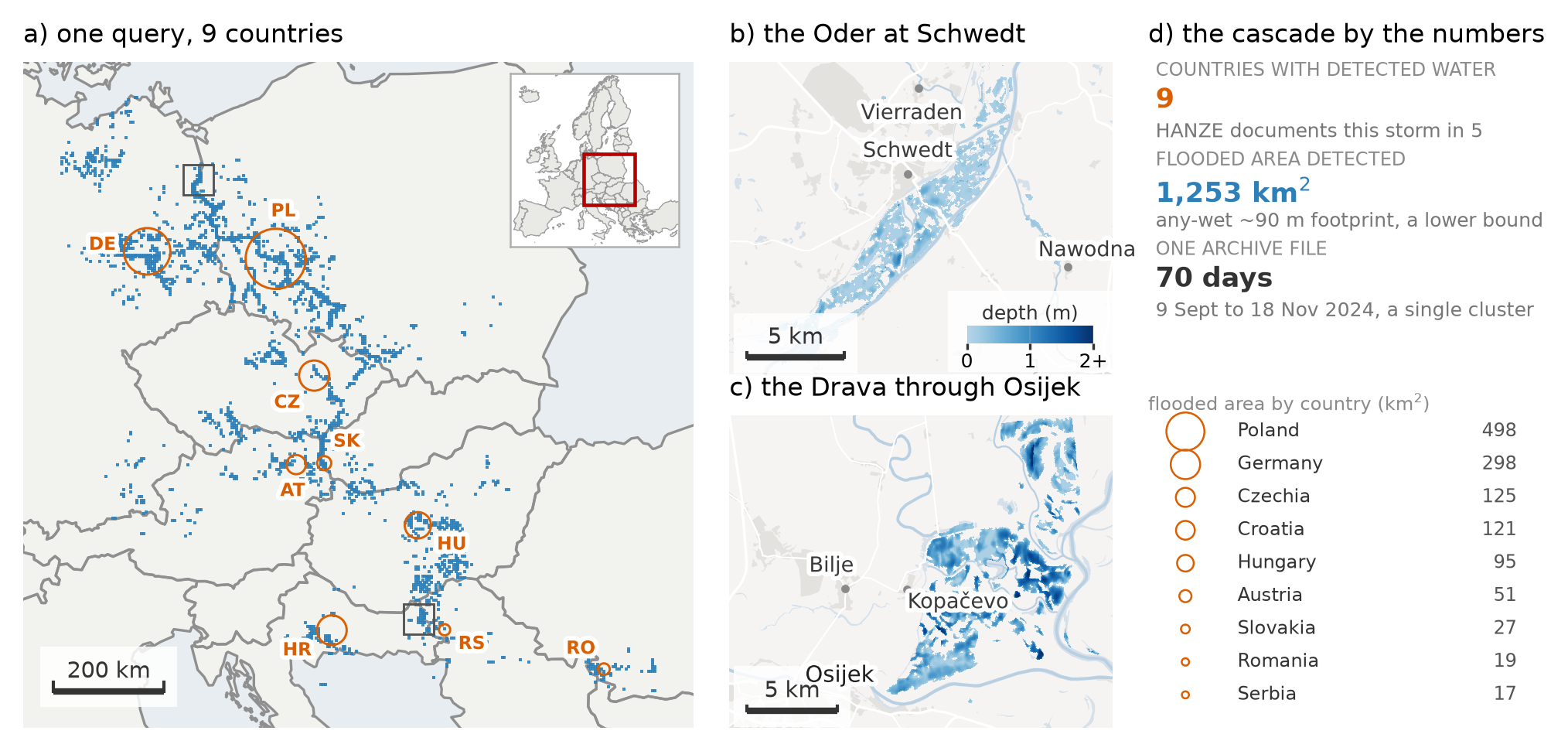}
\caption{Transboundary archive cluster beginning with Storm Boris in September 2024. (a)~The approximately \qty{90}{\m} index footprint across nine countries, with orange circles scaled to the detected area within each country and centred on the corresponding portion of the footprint; the rectangles mark the extents of panels~(b) and~(c). (b)~Native-resolution detected flood depth along the lower Oder at Schwedt, where the river forms part of the German--Polish border and inundation is recorded on both banks. (c)~Native-resolution detected depth in the Drava floodplain at Kopa\v{c}ki Rit near Osijek, representing part of the later downstream flooding included in the same archive cluster. (d)~Summary of the detected area, countries, 70-day cluster duration, and country-level attribution. Detected extents remain lower bounds because of satellite under-detection, while areas derived from the approximately \qty{90}{\m} index footprint are enlarged by conservative any-wet resampling.}
\label{si:fig:boris}
\end{figure}

\paragraph{Data and method.}
A single query, \texttt{floods(bbox=(8.0,\,43.5,\,24.0,\,55.0), start=\,\allowbreak 2024\text{-}09\text{-}01, end=\,2024\text{-}09\text{-}30)}, returns three archive clusters whose start dates fall within September 2024 and whose footprints intersect the Central and Eastern European study window.
The largest cluster, index event 772644797, is converted to a footprint using \texttt{.footprints()}, reprojected to the equal-area ETRS89-LAEA coordinate system (EPSG:3035), and intersected with GISCO NUTS level~0 country boundaries from the 2021 release at 1:10\,million scale.
Detected area is then summed by country from the approximately \qty{90}{\m} any-wet index footprint, while the native-resolution \qty{20}{\m} depth rasters are used for the local detail panels in Fig.~\ref{si:fig:boris}.

\paragraph{Results.}
The selected archive cluster contains \qty{1253}{\km\squared} of detected flooding distributed across nine countries (Table~\ref{si:tab:boris}; Fig.~\ref{si:fig:boris}).
The largest shares occur in Poland, with \qty{497.7}{\km\squared} concentrated principally along the Odra and Nysa systems, and Germany, with \qty{297.5}{\km\squared} primarily along the Elbe and lower Oder.
Additional detected areas extend through Czechia, Austria, Slovakia, Hungary, Croatia, Romania, and Serbia, forming a geographically coherent sequence across Central and Eastern Europe.
The local views illustrate both the transboundary character of the footprint at the German--Polish border near Schwedt and the later downstream signal in the Drava floodplain near Osijek.
Once the footprint has been retrieved, the country attribution requires only a spatial intersection with harmonised administrative boundaries.

\begin{table}[htbp]
\centering
\caption{Detected flood area by country for the archive cluster associated with Storm Boris, obtained by intersecting the approximately \qty{90}{\m} any-wet index footprint of event 772644797 with GISCO NUTS level~0 boundaries. The national areas sum to \qty{1251}{\km\squared} of the event's \qty{1253}{\km\squared} footprint; the remainder falls outside every national boundary, mostly along coasts and rivers that form borders.}
\label{si:tab:boris}
\small
\begin{tabular}{@{}lr@{\hspace{2em}}lr@{}}
\toprule
Country & Area (\unit{\km\squared}) & Country & Area (\unit{\km\squared}) \\
\midrule
Poland   & 497.7 & Austria  & 50.8 \\
Germany  & 297.5 & Slovakia & 27.1 \\
Czechia  & 125.3 & Romania  & 18.7 \\
Croatia  & 121.1 & Serbia   & 17.3 \\
Hungary  & 95.1 & & \\
\midrule
\multicolumn{3}{@{}l}{Sum across the nine countries} & 1,251 \\
\bottomrule
\end{tabular}
\end{table}

\paragraph{Evaluation.}
The case study demonstrates that one index query can retrieve a geographically distributed archive cluster and produce an auditable country-level attribution without requiring the prior collection and harmonisation of national flood datasets.
Three qualifications are important for interpreting the result.
First, the returned record is an archive-defined spatio-temporal cluster spanning 9~September to 18~November 2024, a duration of 70 days, and therefore combines the initial Storm Boris flooding with later autumn inundation along the wider Danube system.
The detected areas in downstream countries should consequently not be interpreted as footprints of the September peak alone.
This illustrates a central property of the archive: an indexed event corresponds to a source-file cluster rather than necessarily to one hydrologically individuated flood, so analyses requiring physical event attribution must inspect the temporal metadata and re-segment or aggregate records according to an appropriate event definition.
Second, query date filters operate on the cluster start date, so a narrowly specified interval such as 10--30~September would omit this record because its start date is 9~September.
Robust event discovery should therefore use a suitably padded date window and inspect the returned \texttt{date} and \texttt{end\_date} fields.
Third, the country areas inherit opposing spatial biases: synthetic-aperture radar under-detection omits inundation that was not observed, while conservative resampling to the approximately \qty{90}{\m} index grid enlarges the detected footprint, with a median inflation factor of 1.50 relative to the native \qty{20}{\m} wet area.

\newpage
% ==========================================================================
\subsection{Detection climatology of the observed archive}
\label{si:sec:clim}

\paragraph{Motivation.}
The continental recurrence map in Section~7.1 of the main text summarises where flooding was repeatedly detected but does not describe when those detections occurred.
This case study adds a temporal dimension by examining the seasonal and interannual distribution of archived events and the geographical variation in their timing.
It is framed throughout as a \emph{detected-flood climatology}: a description of the archive's observation record rather than a climatology of precipitation, discharge, or physical flood occurrence.

\paragraph{Data and method.}
A single index query over the continental window (\ang{-25}, \ang{34}) to (\ang{45}, \ang{72}) returns 3,253 archive events, of which 3,246 are dated between 2015 and 2025 and enter the analysis.
The seven remaining events returned by the query are dated 2014, while the other archived events fall outside the continental analysis window.
All archived events carry a parseable date.
Monthly event counts, mapped inundated area from the catalogue field \texttt{area\_km2}, and event durations are derived from this query.

Spatial seasonality is evaluated on a \ang{1.5} grid.
Each event contributes to every grid cell intersected by its reconstructed footprint, avoiding the use of a single centroid to represent events that may extend over hundreds of kilometres.
The footprints are read from the overview pyramid of the index Cloud-Optimized GeoTIFF at approximately \qty{720}{\m} resolution, which is substantially finer than the \ang{1.5} analysis grid.
Event location is therefore derived from the reconstructed footprint rather than from the common query geometry attached to the returned rows.

The regional summaries in Table~\ref{si:tab:clim} are calculated separately using one index query for each macro-region bounding box: Iberia, Italy and the Adriatic, France and the Benelux, the United Kingdom and Ireland, Central and Eastern Europe, and the Nordic and Baltic countries.
These bounding boxes are intended only as broad regional summaries and neither partition nor fully cover the continental domain.

\begin{figure}[htbp]
\centering
\includegraphics[width=\linewidth]{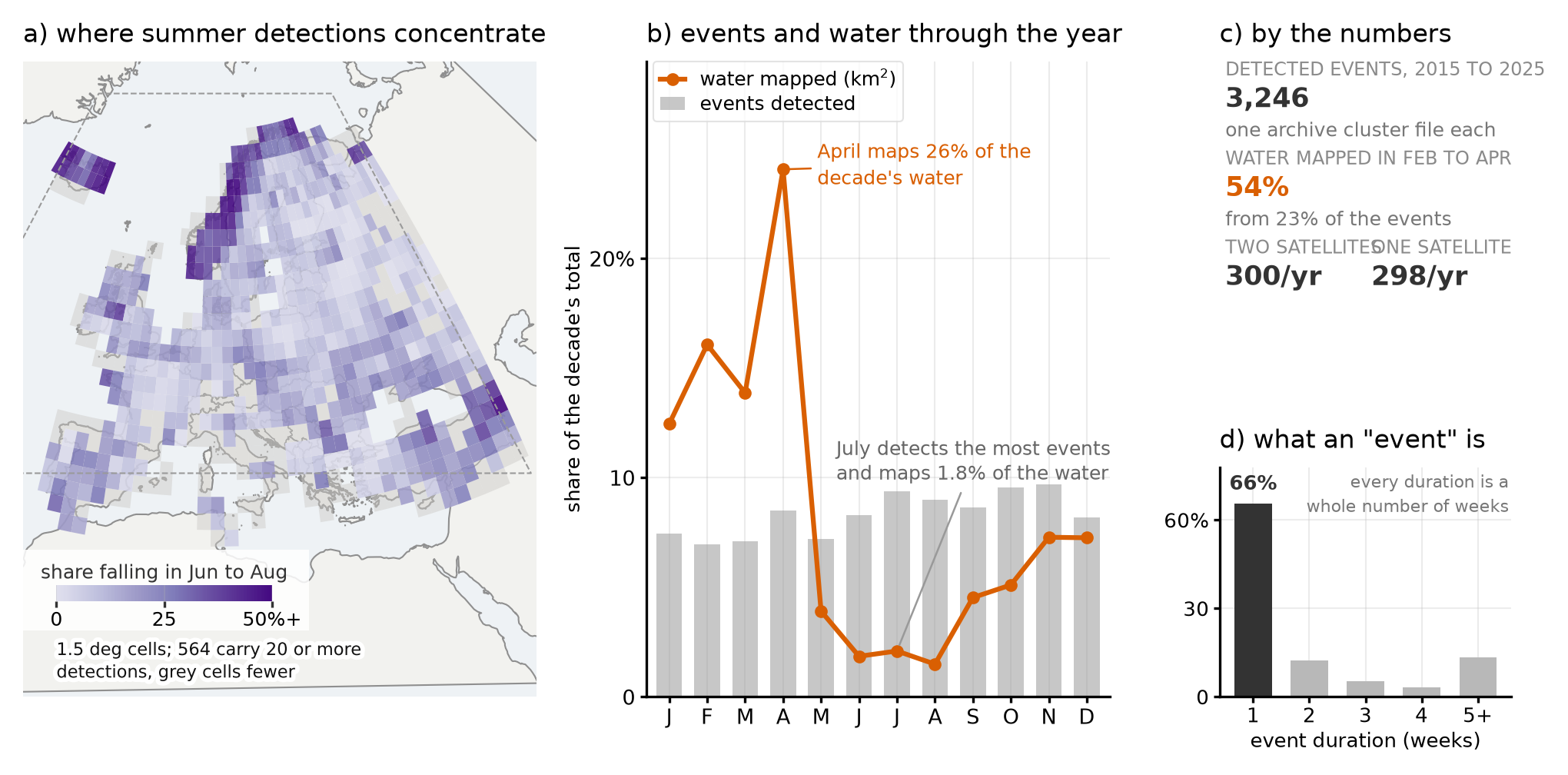}
\caption{Seasonality and temporal structure of the European flood detections archived from 2015 to 2025, derived from the index without downloading source rasters. (a)~Share of detections in each \ang{1.5} grid cell occurring from June through August. Each event contributes to every cell intersected by its footprint; cells with fewer than 20 detections are shown in grey, and the dashed boundary marks the continental query window. (b)~Monthly event counts and mapped inundated area, each expressed as a share of its respective total. Event counts are comparatively uniform, whereas mapped area is concentrated in late winter and spring: February through April contributes 54\% of the mapped area from 23\% of the events, and April alone contributes 24\%. (c)~Summary of the 3,246 events and the mean annual event count in the years with two operating satellites and in those with one. (d)~Distribution of archive-event duration; all durations are whole numbers of weeks and 66\% are one week, reflecting the upstream clustering procedure rather than the duration of hydrologically individuated floods. The seasonal patterns combine physical flood occurrence with acquisition timing, sensor detectability, vegetation conditions, and archive clustering and should not be interpreted as an uncorrected flood climatology.}
\label{si:fig:clim}
\end{figure}

\paragraph{Results.}
Event counts and mapped inundated area produce markedly different seasonal representations of the archive (Fig.~\ref{si:fig:clim}b).
Monthly event counts are comparatively uniform, ranging from 7.0\% of the archive total in February to 9.7\% in November.
Mapped inundated area is much more seasonal: February through April accounts for 54\% of the mapped area but only 23\% of the events, whereas June through August accounts for 5\% of the mapped area from 27\% of the events.
April alone contributes 24\% of the total mapped area.
Event counts should therefore not be treated as proportional to the amount of inundation represented in the archive.

The event definition introduces an additional source of temporal structure.
Every archived event has a duration equal to a whole number of weeks, and 66\% last exactly one week (Fig.~\ref{si:fig:clim}d).
An archive event is consequently a processing and clustering unit rather than necessarily one hydrological flood episode.
Months containing more events tend to contain shorter clusters ($r=-0.60$ across the twelve monthly aggregates), suggesting that part of the seasonal variation in event counts arises from how observations are divided among archive files.

Raw annual event counts are also similar under one-satellite and two-satellite operations, averaging 300 events per year in the five years when Sentinel-1A and Sentinel-1B were both operational (2017 to 2021) and 298 events per year in the four years covered by a single satellite (2015 and 2022 to 2024), with the two commissioning years 2016 and 2025 excluded from both averages (Fig.~\ref{si:fig:clim}c).
This similarity does not imply a constant correspondence with documented floods, because event counts and corroboration rates measure different properties.
The geographical pattern in Fig.~\ref{si:fig:clim}a shows relatively high summer shares along parts of the Norwegian and Fennoscandian coasts and across eastern Europe, and lower shares across Iberia, France, and the British Isles.
The coarse regional summaries in Table~\ref{si:tab:clim} provide a complementary but less spatially resolved view of this variation.

\begin{table}[htbp]
\centering
\caption{Detected events by macro-region from 2015 to 2025 and the calendar month containing the largest number of detections. The regions are coarse, overlapping bounding boxes that neither partition nor fully cover Europe: 1,431 of the 3,246 events fall outside all six regions, while 389 events intersect more than one region. The counts therefore should not be summed, and the monthly maxima should be interpreted as descriptive summaries of the archive rather than estimates of regional flood seasonality.}
\label{si:tab:clim}
\small
\begin{tabular}{@{}lrl@{}}
\toprule
Macro-region & Events & Month with most detections \\
\midrule
Nordic and Baltic & 604 & October \\
Central and Eastern Europe & 430 & December \\
United Kingdom and Ireland & 377 & August \\
France and the Benelux & 349 & November \\
Iberia & 308 & January \\
Italy and the Adriatic & 273 & May \\
\bottomrule
\end{tabular}
\end{table}

\paragraph{Evaluation.}
The analysis provides both a descriptive result and a methodological caution.
The concentration of mapped inundated area in late winter and spring, together with broad regional differences in summer share, indicates that the archive retains meaningful seasonal structure.
However, the monthly event count is influenced by acquisition opportunities and by the division of observations into weekly and multi-week clusters, and therefore cannot be interpreted directly as flood frequency.
The apparent regional peak months in Table~\ref{si:tab:clim}, particularly for smaller or more convective regimes, are also sensitive to the short eleven-year record, coarse overlapping regions, and process-dependent detectability.

The strong divergence between monthly event counts and mapped area illustrates why archive events should not be weighted equally without considering their spatial extent and observation context.
The same qualification applies to the recurrence map in Section~7.1 of the main text: its values count archived event detections, while the effective observation effort is determined by satellite acquisitions and the upstream clustering process rather than by a fixed number of independent opportunities per year.
The comparatively stable annual event counts do not contradict the non-stationary corroboration rates reported in Section~\ref{si:sec:hanze}, because the two analyses have different denominators and measure different aspects of the archive.
This case study should therefore be read as a characterisation of the archive's temporal sampling and biases, not as new evidence of trends or seasonality in European flood occurrence.
The practical implication is that raw archive counts form a detection record rather than a flood-frequency series, and temporal comparisons require explicit consideration of changes in acquisition, processing, and completeness.

% \newpage

% ==========================================================================
\subsection{Infrastructure exposure: the Shannon callows}
\label{si:sec:shannon}

\paragraph{Motivation.}
Infrastructure flood screening commonly relies on modelled hazard scenarios, even where repeated observations of historical inundation are available.
This case study demonstrates an observation-based alternative by using eleven years of dated flood footprints to identify road segments that intersect detected inundation and settlements repeatedly approached by floodwater.
The Shannon callows in the Irish midlands provide a suitable example because their broad, persistent flooding represents a process that Sentinel-1 observes comparatively well.
The analysis also demonstrates how EuroFlood outputs can be combined with a user-supplied infrastructure network using standard geospatial operations.

\paragraph{Data and method.}
A single query, \texttt{floods(bbox=(-8.6,\,52.7,\,-7.9,\,53.5))}, returns 61 archived events between 2015 and 2025.
Their index footprints are combined to form a \qty{238}{\km\squared} ever-flooded mask, while the individual footprints are retained to calculate how many events intersect each location.
Road centrelines from motorway through tertiary classes and settlement points classified as cities, towns, or villages are retrieved from OpenStreetMap through the Overpass API and reprojected to the Irish Transverse Mercator coordinate system (EPSG:2157).
A road segment is classified as exposed when it intersects at least one detected flood footprint, and its recurrence value is the number of distinct archived events whose footprints intersect it.
A settlement is classified as exposed when detected floodwater occurs within \qty{1}{\km} of its mapped point, and the number of events meeting this criterion is recorded.
OpenStreetMap retrieval uses a descriptive \texttt{User-Agent} and a mirror fallback for robustness.

\begin{figure}[htbp]
\centering
\includegraphics[width=\linewidth]{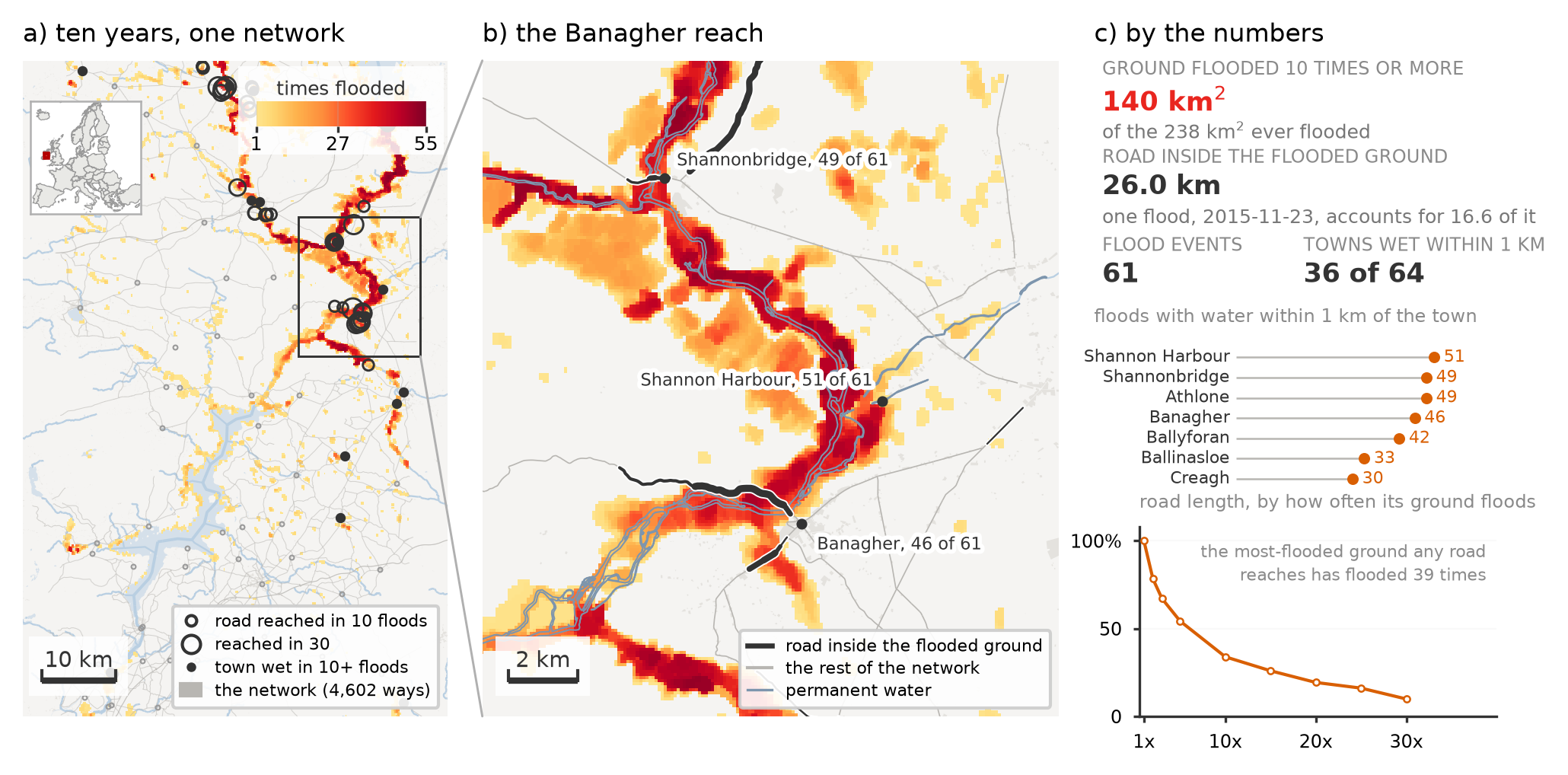}
\caption{Observation-based infrastructure screening in the Shannon callows using 61 archived events from 2015 to 2025. (a)~Per-cell flood recurrence together with the complete mapped road network; rings identify road locations intersected by at least ten event footprints and are scaled by recurrence, while filled points identify settlements with detected floodwater within \qty{1}{\km} in at least ten events. The rectangle marks the extent of panel~(b). (b)~The Banagher reach, showing roads within the ever-flooded footprint, the remaining road network, permanent water, and the recurrence of nearby settlements. (c)~Summary of repeatedly flooded ground, exposed road length, settlement proximity, and the decline in road length as the required recurrence threshold increases. The results identify potential infrastructure exposure but do not establish road closure, loss of accessibility, or settlement inundation.}
\label{si:fig:shannon}
\end{figure}

\paragraph{Results.}
Across the eleven-year record, detected flood footprints intersect \qty{26.0}{\km} of mapped roads (Table~\ref{si:tab:shannon}; Fig.~\ref{si:fig:shannon}).
Tertiary and secondary roads account for the largest exposed lengths, at 11.1 and \qty{9.9}{\km}, respectively, while trunk roads have the highest exposed proportion of the locally mapped network, at 6.0\%.
The union statistic is strongly influenced by one event: the footprint dated 23~November 2015 intersects \qty{16.6}{\km} of the total \qty{26.0}{\km}, whereas a typical event intersects approximately \qty{1.4}{\km}.

The dated archive reveals where exposure is recurrent rather than isolated.
Of the \qty{238}{\km\squared} observed flooded at least once, \qty{109}{\km\squared} is contained in ten or more event footprints.
The most repeatedly reached road location intersects footprints from 39 of the 61 events, and the total road length exposed decreases rapidly as the required recurrence threshold increases (Fig.~\ref{si:fig:shannon}c).
This pattern indicates that repeatedly exposed road locations form a relatively concentrated subset of the wider network.

Of the 64 mapped settlements, 36 have detected floodwater within \qty{1}{\km} in at least one event, and 16 meet this criterion in ten or more events.
The highest recurrence counts occur at Shannon Harbour, Shannonbridge, Athlone, Banagher, and Ballyforan, with water detected within \qty{1}{\km} in 42--51 of the 61 archived events.
These results identify locations for more detailed infrastructure and accessibility assessment rather than demonstrating that the settlements themselves were inundated or isolated.

\begin{table}[htbp]
\centering
\caption{Mapped road length intersecting the eleven-year union of detected flood footprints in the Shannon callows, grouped by OpenStreetMap highway class. Local length denotes the total mapped length of each class within the study window.}
\label{si:tab:shannon}
\small
\begin{tabular}{@{}lrrr@{}}
\toprule
Highway class & Intersected (km) & Local length (km) & Intersected (\%) \\
\midrule
Motorway  & 2.1  & 221.9   & 1.0 \\
Trunk     & 0.8  & 13.2    & 6.0 \\
Primary   & 2.1  & 129.8   & 1.6 \\
Secondary & 9.9  & 853.7   & 1.2 \\
Tertiary  & 11.1 & 1,123.3 & 1.0 \\
\midrule
Total     & 26.0 & & \\
\bottomrule
\end{tabular}
\end{table}

\paragraph{Evaluation.}
This analysis is an infrastructure-exposure screen rather than a road-accessibility or network-disruption model.
Intersection between a road centreline and a detected flood footprint does not establish that the road was impassable, damaged, or closed, and proximity of floodwater to a settlement does not demonstrate isolation or direct inundation.
Evaluating accessibility would require event-specific road conditions, network connectivity, destinations such as hospitals or emergency services, and a routing model that removes or penalises affected edges.
The \qty{1}{\km} settlement threshold is a screening heuristic, and the resulting settlement list should therefore be treated as a set of candidates for more detailed investigation.

The results also inherit limitations from both input datasets.
The observed flood footprints may omit inundation that Sentinel-1 did not detect, while conservative resampling to the approximately \qty{90}{\m} index grid enlarges the mapped footprint relative to the native \qty{20}{\m} observations.
OpenStreetMap coverage and road classification may also vary spatially.
Within these constraints, the case study demonstrates that repeated observed flood footprints can provide an empirical basis for prioritising road segments and settlements for subsequent engineering and accessibility analysis, complementing conventional screening based solely on modelled hazard scenarios.

% ==========================================================================
\subsection{Eleven years of observed flooding at one location: Zutphen on the IJssel}
\label{si:sec:zutphen}

\paragraph{Motivation.}
EuroFlood enables repeated observations of flood depth at one location to be compared with a modelled return-period hazard spectrum.
Zutphen on the river IJssel is selected because 33 archived events intersect the same river reach between 2015 and 2025, providing a comparatively dense multi-event record.
Section~7.2 of the main text compares the cumulative observed extent with the modelled hazard envelopes; this case study examines the corresponding event-level depth record in greater detail.

\paragraph{Data and method.}
A single index query, \texttt{floods("Zutphen, Netherlands", shape="bbox")}, returns 33 archived events between 2015 and 2025.
A rectangular region of interest is used because the administrative boundary of Zutphen bisects the IJssel and would exclude part of the river corridor.
Native-resolution depth rasters were available from the JRC service for all 33 events at the time of analysis, so the depth statistics and the recurrence record cover the same events.
Per-cell recurrence is calculated directly from the approximately \qty{90}{\m} index, while the 95th-percentile and maximum observed depths are calculated from the positive-depth cells of each downloaded \qty{20}{\m} raster within the common study window.
The modelled comparison is obtained with \texttt{hazard(return\_period=[10,\allowbreak 20,\allowbreak 50,\allowbreak 75,\allowbreak 100,\allowbreak 200,\allowbreak 500],\allowbreak shape="bbox")\allowbreak .download()\allowbreak .stats()}, using the same spatial window and calculating the corresponding 95th-percentile and maximum depths for each return-period scenario.
The 95th percentile provides a comparatively robust measure of high water depth across each footprint, while the maximum is retained to identify isolated local extremes.

\begin{figure}[htbp]
\centering
\includegraphics[width=\linewidth]{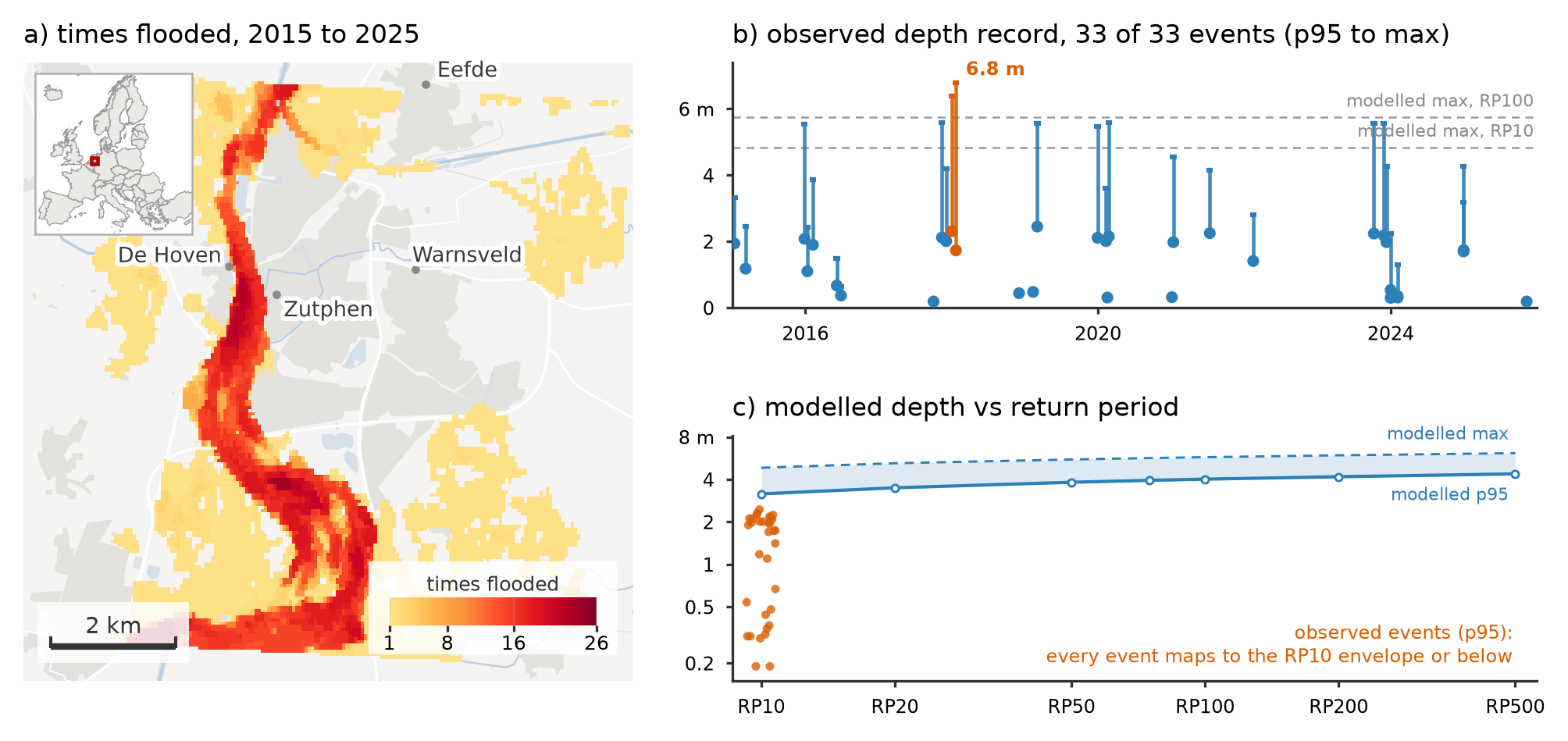}
\caption{Multi-event flood record at Zutphen on the IJssel from 2015 to 2025. (a)~Per-cell recurrence from the index, showing the number of archived events associated with each location. (b)~Observed depth statistics for all 33 events; each vertical segment spans the event-level 95th-percentile to maximum depth. Dashed lines show the modelled RP10 and RP100 maximum depths, and the two observed events whose maxima exceed the RP100 maximum are highlighted. (c)~Modelled 95th-percentile and maximum depth across return periods RP10--RP500, together with the observed event-level 95th-percentile depths. All observed 95th-percentile values lie at or below the lowest available modelled scenario, RP10; observed points are horizontally offset near RP10 for visibility, and depth is shown on a logarithmic scale.}
\label{si:fig:zutphen}
\end{figure}

\paragraph{Results.}
Between 2015 and 2025, individual cells along the IJssel corridor are associated with as many as 26 archived events (Fig.~\ref{si:fig:zutphen}a).
Across the 33 events, the median event-level 95th-percentile depth is \qty{1.73}{\m}, while the largest reconstructed depth is \qty{6.8}{\m} in the event dated 22~January 2018 (Fig.~\ref{si:fig:zutphen}b).
All observed event-level 95th-percentile depths are at or below the modelled RP10 95th-percentile depth of \qty{3.16}{\m}.
The modelled 95th-percentile depth increases from \qty{3.16}{\m} at RP10 to \qty{4.38}{\m} at RP500, whereas two observed events contain local maximum depths exceeding the modelled RP100 maximum.
These findings describe different properties of the spatial depth fields: the 95th percentile characterises broadly elevated depths within an event footprint, while the maximum may represent a highly localised peak.
Within the study window, the observed events generally contain shallower depths distributed across the detected floodplain, whereas the undefended fluvial scenarios concentrate greater depths within the channel and modelled inundation envelope.

\paragraph{Evaluation.}
This comparison is a diagnostic screening analysis rather than a validation of the hazard maps or an estimate of flood frequency.
Observed recurrence counts archived event detections rather than statistically independent floods and cannot be converted into annual exceedance probabilities or return periods.
An eleven-year observational record is also insufficient to estimate the return periods represented by the modelled RP50--RP500 scenarios.
The observed event points in Fig.~\ref{si:fig:zutphen}c therefore indicate only how their depth summaries compare with the available modelled envelope; their horizontal position does not assign a return period to the events.

The observed and modelled depths are produced by different methods and are calculated over spatial fields with different extents.
Observed depths are reconstructed from Sentinel-1 detections and terrain, may omit inundation that was not visible to the sensor, and carry decimetre-scale uncertainty.
The CEMS-GLOFAS scenarios are fluvial, assume no flood protection, and may therefore inundate protected land that is not wet in the observed record.
Maximum depths are additionally sensitive to isolated pixels, terrain errors, and reconstruction artefacts, so the two exceedances of the modelled RP100 maximum should be interpreted more cautiously than the 95th-percentile comparison.
Within these limitations, the study demonstrates a reproducible workflow for combining eleven years of dated observed depth fields with the modelled design-hazard spectrum at one location.

% ==========================================================================
\subsection{Observation-based exposure: July 2021 and Emilia-Romagna 2023}
\label{si:sec:exposure}

\paragraph{Motivation.}
Flood-exposure assessments commonly rely on modelled hazard scenarios because spatially explicit observations of water depth are rarely available for individual historical events.
EuroFlood enables an observation-based alternative in which dated flood-depth rasters are intersected with population and built-up layers and compared with modelled return-period envelopes over the same region.
The July 2021 western European floods provide the principal example and are summarised in Section~7.3 of the main text.
The May 2023 Emilia-Romagna floods provide a second case that tests whether the workflow transfers to a Mediterranean event represented by several archive rasters.
In both cases, the resulting exposure estimates describe only the inundation detected by Sentinel-1 and should therefore be interpreted as observation-based lower bounds on total event exposure.

\paragraph{Data and method.}
The July 2021 analysis uses the spatial window \qtyrange{3.0}{9.0}{\degree}E and \qtyrange{48.5}{52.5}{\degree}N and the date interval 1~July--15~August 2021.
The Emilia-Romagna analysis uses the spatial window \qtyrange{9.5}{13.5}{\degree}E and \qtyrange{43.5}{46.0}{\degree}N and the date interval 1~April--30~June 2023.
All area and exposure statistics are restricted to these respective analysis windows.

All matching native-resolution depth rasters are downloaded for each case.
Where several rasters represent the selected period, they are combined by taking the maximum reconstructed depth at each cell, producing a maximum-observed-depth composite.
This operation retains the greatest depth detected at each location but does not represent a simultaneous inundation surface or conserve event water volume when several rasters are combined.
The July 2021 case resolves to one source raster, whereas the Emilia-Romagna case combines three rasters.

Modelled river-flood hazard is retrieved for return periods of 10, 20, 50, 75, 100, 200, and 500 years.
Population and built-up exposure are derived from the 2020 JRC Global Human Settlement Layer products GHS-POP and GHS-BUILT-S, release R2023A, at \qty{100}{\m} resolution.
All observed, modelled, population, and built-up layers are reprojected to the equal-area ETRS89-LAEA coordinate system (EPSG:3035) and evaluated on a common \qty{100}{\m} grid.

Containment of the detected footprint within each modelled envelope is evaluated using three weightings.
The cell-based measure assigns equal weight to every detected wet cell, the population-based measure weights each cell by the population beneath the detected water, and the volume-based measure weights each cell by reconstructed water volume.
Modelled-envelope exposure is also calculated to describe the scale of each hazard scenario, but these values are not directly comparable with the exposure of one observed historical event.

\paragraph{Results: July 2021.}
The July 2021 detected footprint covers \qty{503}{\km\squared} and has a mean reconstructed depth of \qty{0.61}{\m}, a 95th-percentile depth of \qty{2.08}{\m}, and a 99th-percentile depth of \qty{3.89}{\m} (Table~\ref{si:tab:exposure-obs}).
The footprint comprises 83,707 wet analysis cells and approximately 306 million m$^3$ of reconstructed water.
It contains an estimated 6,377 residents and \qty{1.3}{\km\squared} of built-up surface.

The modelled hazard envelopes cover substantially larger areas because they represent regional statistical scenarios rather than the realised July 2021 event.
For example, the RP100 envelope covers \qty{18436.1}{\km\squared} and contains approximately 9.80 million residents and \qty{1133.2}{\km\squared} of built-up surface (Table~\ref{si:tab:exposure-rp}).
These values describe potential exposure within the entire modelled envelope and should not be interpreted as an estimate of the July 2021 event impact.

\begin{figure}[htbp]
\centering
\includegraphics[width=\linewidth]{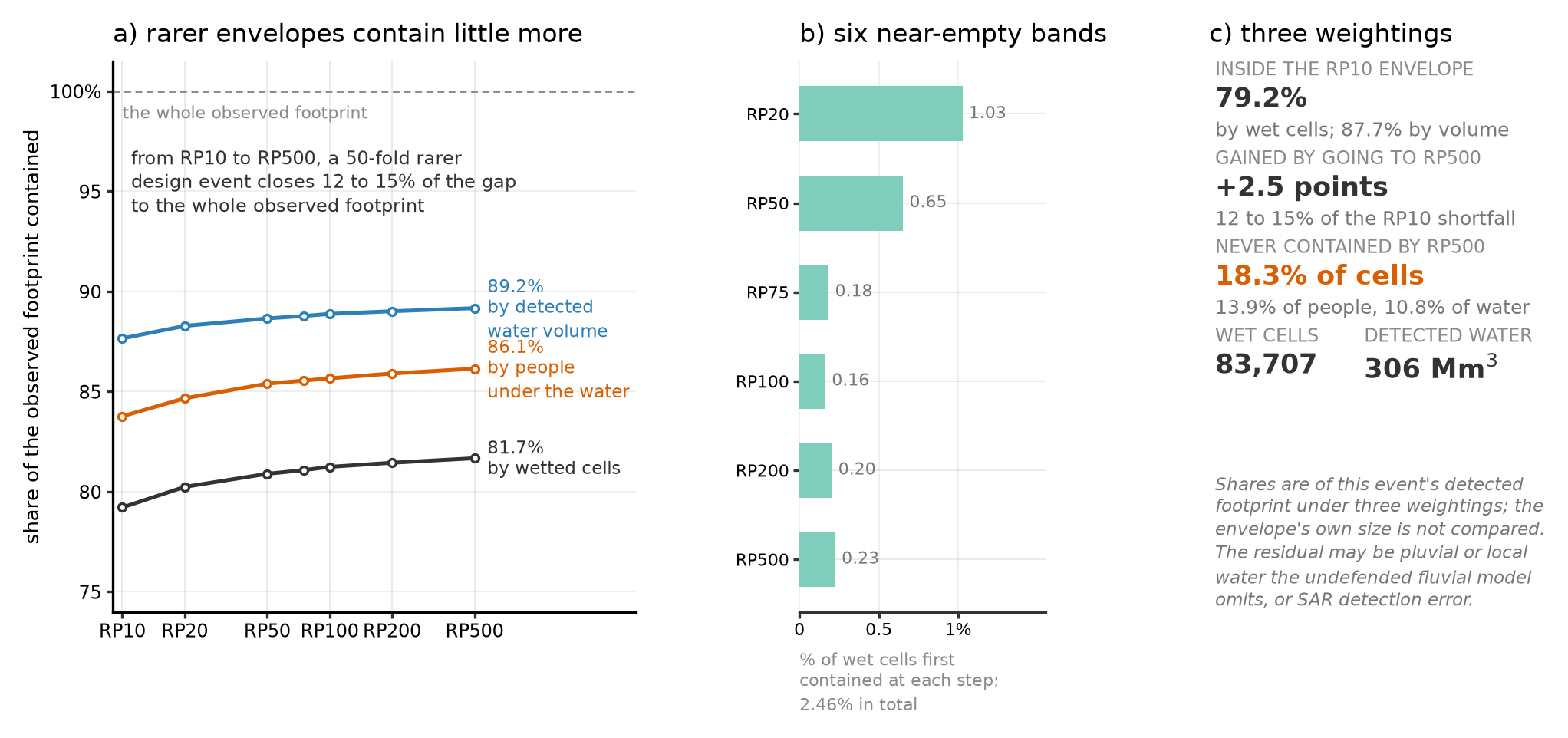}
\caption{Containment of the detected July 2021 footprint within the modelled river-flood hazard envelopes. (a)~Share of the detected footprint contained by each return-period envelope when weighted by wet-cell count, population beneath the detected water, and reconstructed water volume; the dashed line represents complete containment. (b)~Share of detected wet cells first contained by each successive envelope above RP10. No individual band adds more than 1.1\%, and the six bands from RP20 to RP500 together add 2.46\%. (c)~Summary of containment under the three weightings. The RP10 envelope contains 79.2\% of detected wet cells, while 18.3\% remain outside RP500; the residual decreases to 13.9\% under population weighting and 10.8\% under volume weighting. All percentages refer to the detected event footprint, and the size of the modelled envelope itself is not evaluated in this figure. The uncontained cells may reflect pluvial or local flooding, omitted small channels, differences in flood-protection assumptions, spatial resolution, or errors in satellite classification and depth reconstruction; the comparison cannot distinguish among these explanations.}
\label{si:fig:exposure-rp}
\end{figure}

The fraction of the detected footprint contained within the modelled envelopes changes little across the return-period spectrum (Fig.~\ref{si:fig:exposure-rp}).
The RP10 envelope contains 79.2\% of the detected wet cells and 87.7\% of the reconstructed water volume.
Increasing the return period from RP10 to RP500 adds only 2.5 percentage points to the cell-based containment, leaving 18.3\% of the wet cells outside the RP500 envelope.
The uncontained share is smaller when weighted by consequence, amounting to 13.9\% of the population beneath the detected water and 10.8\% of the reconstructed water volume.
No individual band added between RP20 and RP500 first contains more than 1.1\% of the detected wet cells, and all six bands together add only 2.46\%.
The modelled hierarchy therefore captures most of the detected water already at RP10, while the residual lies predominantly outside even the RP500 envelope rather than between successive return-period levels.

\begin{figure}[htbp]
\centering
\includegraphics[width=\linewidth]{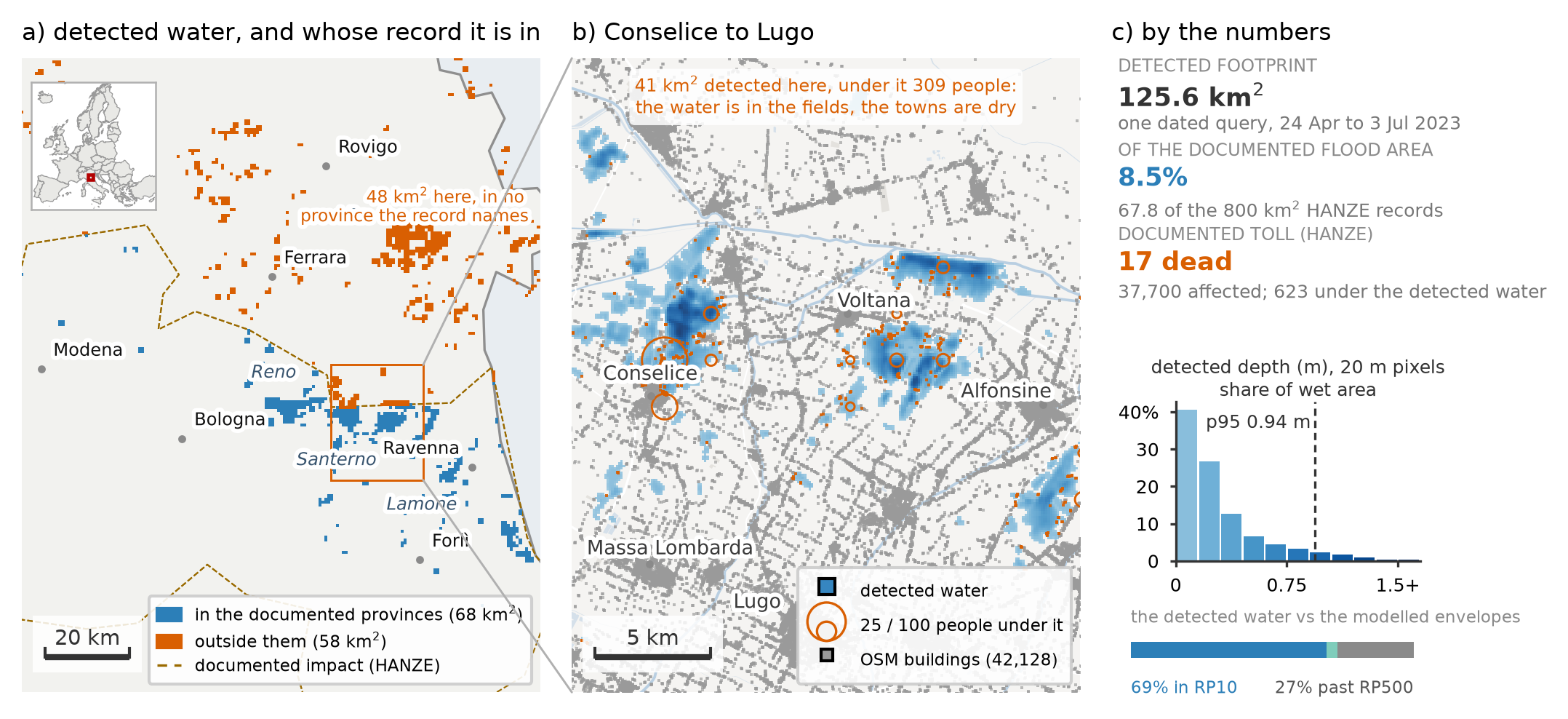}
\caption{Observation-based exposure analysis for the May 2023 Emilia-Romagna floods. (a)~Maximum-depth composite of three archive rasters, divided into detected water within the provinces named in the HANZE record and water outside them; the dashed boundaries show the documented impact regions, and the rectangle marks the extent of panel~(b). (b)~Detected depth and population exposure in the Conselice--Lugo lowland, with circles scaled to the population beneath detected water and OpenStreetMap buildings shown in grey. Detected inundation is concentrated largely on agricultural land, while the principal town footprints remain mostly dry in the archive record. (c)~Summary of the \qty{125.6}{\km\squared} detected footprint, its relation to the documented flood area, reconstructed-depth distribution, and containment within the modelled return-period envelopes. The detected extent may omit inundation that Sentinel-1 did not observe, reconstructed depths carry additional uncertainty, and the 42-day composite spans both documented May episodes rather than representing either episode at one time.}
\label{si:fig:emilia}
\end{figure}

\paragraph{Results: Emilia-Romagna 2023.}
The maximum-depth composite for Emilia-Romagna covers \qty{125.6}{\km\squared} and has a mean depth of \qty{0.30}{\m}, a 95th-percentile depth of \qty{0.94}{\m}, and a 99th-percentile depth of \qty{1.27}{\m} (Table~\ref{si:tab:exposure-obs}).
The detected footprint contains an estimated 623 residents and \qty{0.26}{\km\squared} of built-up surface.
Approximately 69\% of the detected wet cells fall within the RP10 envelope, 72.2\% fall within RP100, and 27\% remain outside RP500 (Fig.~\ref{si:fig:emilia}).

Only \qty{67.8}{\km\squared} of the detected footprint lies within the provinces named in the HANZE record, equivalent to 8.5\% of its documented \qty{800}{\km\squared} flood area.
A further approximately \qty{58}{\km\squared} lies outside those provinces, including a large patch in Ferrara and Rovigo where the archive records recurrent spring inundation.
The spatial correspondence between the composite and the documented May flood is therefore incomplete.

Within the Conselice--Lugo lowland, detected water is concentrated primarily on agricultural land rather than within the larger town footprints.
This spatial distribution explains why the estimated population and built-up exposure are small despite extensive mapped inundation.
The archive combines the observations into a 42-day record spanning both documented May flood episodes, so the composite represents the maximum water detected during the full period rather than the state of either episode at a single time.

\begin{table}[htbp]
\centering
\caption{Detected inundation and reconstructed depth for the two exposure case studies. Wet area is calculated from the approximately \qty{90}{\m} any-wet index footprint, while depth statistics are calculated from the native-resolution \qty{20}{\m} rasters. The Emilia-Romagna values describe a maximum-depth composite of three rasters rather than a simultaneous event surface.}
\label{si:tab:exposure-obs}
\small
\begin{tabular}{@{}lrrrr@{}}
\toprule
Event & Wet area (\unit{\km\squared}) & Mean depth (m) & p95 depth (m) & p99 depth (m) \\
\midrule
July 2021 (Meuse--Rhine) & 503.0 & 0.61 & 2.08 & 3.89 \\
Emilia-Romagna, May 2023 & 125.6 & 0.30 & 0.94 & 1.27 \\
\bottomrule
\end{tabular}
\end{table}

\begin{table}[htbp]
\centering
\caption{July 2021 modelled river-flood hazard envelopes and the fraction of the detected event footprint contained within each envelope. Modelled population and built-up exposure describe the complete regional hazard envelope and are not event-specific estimates. The detected July 2021 footprint contains 6,377 residents and \qty{1.3}{\km\squared} of built-up surface.}
\label{si:tab:exposure-rp}
\footnotesize
\begin{tabular}{@{}rrrrr@{}}
\toprule
RP (yr) & Modelled area (\unit{\km\squared}) & Footprint contained (fraction) & Modelled population & Modelled built-up (\unit{\km\squared}) \\
\midrule
10  & 13,357.9 & 0.792 & 6,385,862  & 769.6 \\
20  & 15,082.2 & 0.802 & 7,526,189  & 888.5 \\
50  & 17,061.9 & 0.809 & 8,866,122  & 1,028.7 \\
75  & 17,874.2 & 0.811 & 9,447,752  & 1,093.3 \\
100 & 18,436.1 & 0.812 & 9,804,202  & 1,133.2 \\
200 & 19,705.9 & 0.814 & 10,737,355 & 1,233.2 \\
500 & 21,121.6 & 0.817 & 12,003,200 & 1,352.8 \\
\bottomrule
\end{tabular}
\end{table}

\paragraph{Evaluation.}
The large difference between exposure beneath the detected footprints and exposure within the modelled hazard envelopes must be interpreted cautiously.
The former represents water observed during a particular historical period, while the latter represents all locations inundated under a regional statistical scenario.
Their direct comparison is therefore useful for contextualisation but does not demonstrate that the hazard model overpredicts event exposure.

Two findings are nevertheless informative.
First, containment of the July 2021 detected footprint changes little between RP10 and RP500.
The residual outside RP500 may arise from flood processes or channels omitted by the fluvial model, differences in flood-protection assumptions, spatial resolution, or errors in the observed reconstruction.
Because the comparison combines products with different scopes and uncertainties, the residual should not be interpreted as a direct measure of model error or skill.

Second, compositing all relevant source rasters is essential when an analysis period is represented by several archive records.
Using only the first Emilia-Romagna raster would underestimate the \qty{125.6}{\km\squared} maximum-observed footprint by more than an order of magnitude and would propagate the same omission into the population and built-up exposure estimates.
The resulting composite is appropriate for screening the maximum area detected during the analysis period, but it should not be interpreted as a simultaneous inundation map or used to estimate a physically conserved event water volume.

In both case studies, the reported exposure is an observation-based lower bound because the archive may omit floodwater that was not visible during a suitable satellite acquisition.
No monetary loss is estimated here because doing so would additionally require asset characteristics and vulnerability or depth--damage functions.
The workflow instead provides a reproducible basis for identifying detected population and built-environment exposure and for comparing that evidence with modelled hazard envelopes.
% ==========================================================================
\subsection{Detectability limit I: the Valencia DANA, October 2024}
\label{si:sec:dana}

\paragraph{Motivation.}
A data-access framework for observational products must characterise the conditions under which the underlying archive is incomplete, not only demonstrate successful applications.
The DANA that affected the Valencia region on 29~October 2024 was one of Europe's deadliest recent flood disasters, with extreme convective rainfall reaching locally reported totals of up to \qty{772}{\mm} in 24~hours.
The event belongs to the rapid pluvial and flash-flood regime that is particularly difficult for Sentinel-1 to observe because inundation can develop and drain between successive satellite acquisitions.
This case study examines the timing and spatial coverage of the archive record by comparing the detected inundation with an independent documentary account of the event.

\paragraph{Data and method.}
The archive is queried over a spatial window covering the documented impact regions and the surrounding event footprint, using a date range spanning late October to mid-November 2024.
All matched native-resolution depth rasters are downloaded and inspected.
Because the date filter applies to the start date of each archive cluster, two neighbouring records are excluded from the DANA analysis.
A pre-event cluster ending on 28~October records recurrent standing water in the l'Albufera wetlands and coastal rice-growing areas before the rainfall peak, while a cluster beginning on 25~November corresponds to a subsequent rainfall episode.
The retained archive cluster begins on 4~November and contains observations through 18~November.

The primary comparison is restricted to the union of the four NUTS level~3 regions associated with the event in HANZE (ES421, ES423, ES523, and ES617), ensuring that the detected and documented areas refer to the same geographical domain.
Detected wet area within these regions is reported as the principal archive capture, while the footprint of the complete source file is reported separately to show observations included outside the documented regions.
The independent reference is the HANZE v3.0.1b record for the 2024 Valencia flood, which provides the documented event dates, affected area, fatalities, and number of people affected.
The analysis is reproduced by the corresponding script in the \texttt{scripts/} directory.

\paragraph{Results.}
HANZE records an event window from 27~October to 4~November 2024, with 236 fatalities, 36,803 people affected, and \qty{226}{\km\squared} of documented flood area.
The archive contains one Sentinel-1 acquisition associated with the Valencia flood, dated 4~November 2024, six days after the rainfall peak.
This acquisition records \qty{35.2}{\km\squared} of detected inundation within the four documented regions, equivalent to approximately 16\% of the HANZE affected-area estimate.
The complete archive file contains \qty{58.7}{\km\squared} of detected water because the source cluster also includes geographically separate patches along the Castell\'on coast and near the Ebro delta.
Although the archive cluster spans observations from 4 to 18~November, the date assigned to the record is the first acquisition date.
The archive therefore represents a single, delayed observation of residual inundation rather than the maximum spatial extent reached during the event.
Table~\ref{si:tab:limits} compares these results with the corresponding coastal-surge case study.

\paragraph{Evaluation.}
This case study documents a substantive limitation of the source archive rather than a failure of the EuroFlood index or query interface.
It illustrates a mechanism consistent with the lower corroboration of flash floods reported in Section~6 of the main text and Section~\ref{si:sec:hanze}: rapid inundation can peak and recede between satellite acquisitions, leaving only persistent or residual water visible when the area is next observed.
The approximately 16\% value should not be interpreted as a formal detection rate or validation metric because the two areas represent different quantities.
The satellite footprint measures standing water detected during a delayed acquisition, whereas the HANZE estimate represents the wider area documented as affected during the event.
Differences in spatial definition, timing, sensor visibility, and index resolution all contribute to the comparison, so the percentage is best interpreted as an order-of-magnitude indication of the archive's limited representation of this rapid-onset flood.

The practical implication is that a small or absent archive footprint does not imply that the underlying flood was spatially limited or low in impact.
Rapid pluvial and flash-flood records may represent only the water remaining at the time of acquisition and should be supplemented with documentary evidence, rapid-mapping products, or event-specific modelling where a complete reconstruction is required.
Section~7.4 of the main text presents this case as a concise illustration of the archive's detectability boundary, while the full methodological and interpretive details are documented here.

% ==========================================================================
\subsection{Detectability limit II: the October 2023 Baltic storm surge}
\label{si:sec:baltic}

\paragraph{Motivation.}
The Valencia DANA illustrates the detectability limit for rapid pluvial and flash flooding, whereas this case study examines the same limitation for a physically distinct process: coastal storm surge.
Storm Babet generated an exceptional Baltic surge along the Schleswig-Holstein coast of Germany and the southern Danish coast on 20--21~October 2023.
Because coastal inundation can recede rapidly after the surge peak, the event provides a second test of whether the archive captures the maximum extent or only residual water observed during a later satellite acquisition.

\paragraph{Data and method.}
The archive is queried over the October 2023 surge period, and all matching native-resolution depth rasters are downloaded and inspected using the same workflow as in Section~\ref{si:sec:dana}.
The principal result is the detected wet area within the union of the nine NUTS level~3 regions in Germany and Denmark associated with the event in HANZE, ensuring that the satellite-derived area and documentary record refer to the same geographical domain.
The independent reference is formed by combining the HANZE records for the German and Danish components of the October 2023 coastal surge.
Unlike the Valencia case, these records report affected population but no documented flood area, so no area-based capture ratio can be calculated.

\begin{figure}[htbp]
\centering
\includegraphics[width=\linewidth]{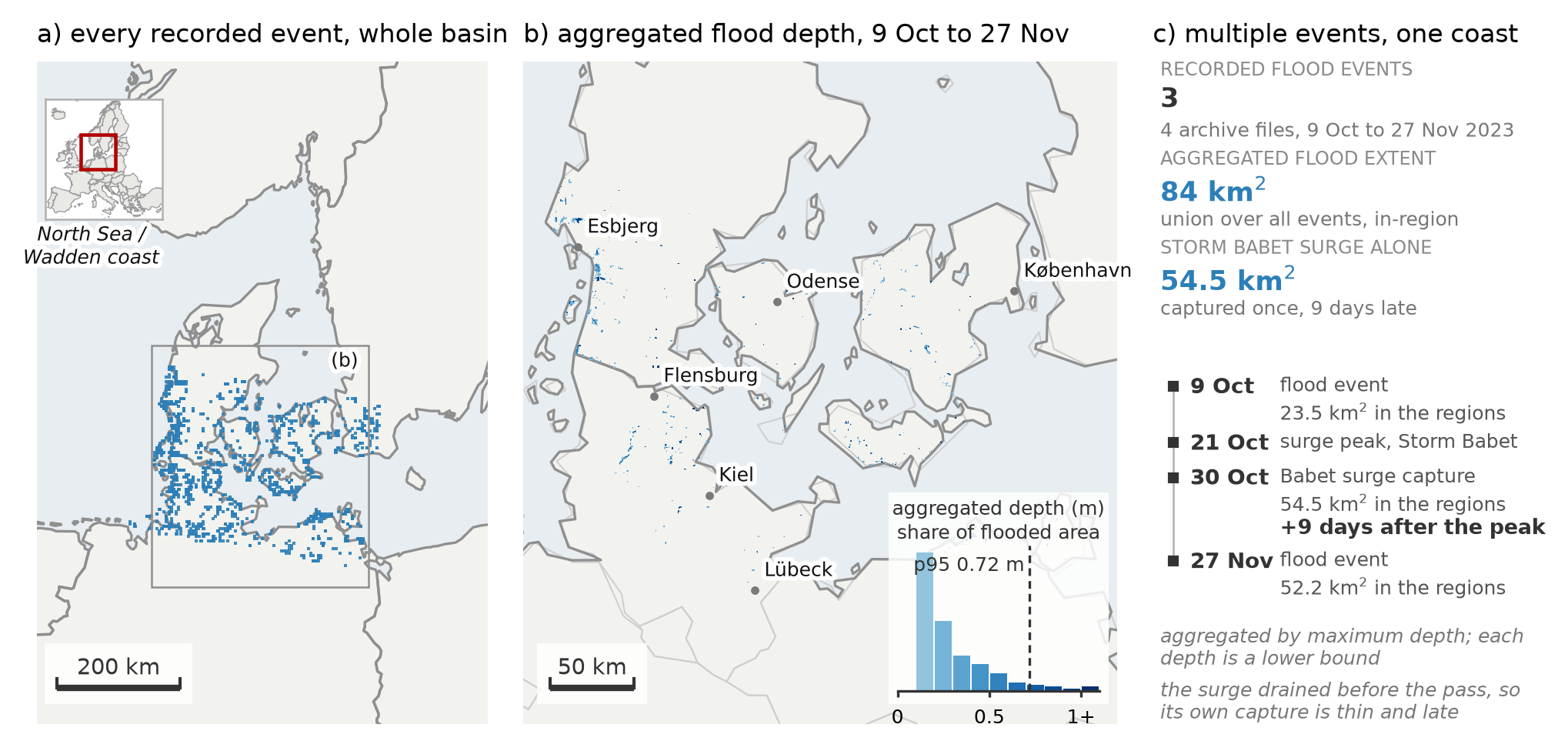}
\caption{Satellite record of the October 2023 Baltic storm surge and surrounding autumn flood detections. (a)~Dated archive events intersecting the German and Danish Baltic coast during autumn 2023, shown across the wider analysis region; the rectangle marks the extent of panel~(b). (b)~Maximum reconstructed depth within the documented regions across three events represented by four archive files between 9~October and 27~November 2023; the inset histogram shows that the detected water is predominantly shallow. (c)~Summary of the surge record: the first relevant acquisition occurs on 30~October, nine days after the 20--21~October peak, and records \qty{54.5}{\km\squared} of detected water within the documented regions, while the multi-event autumn composite covers \qty{84}{\km\squared}. The single acquisition represents delayed residual inundation rather than the maximum surge extent, and the multi-event composite is not a simultaneous flood surface.}
\label{si:fig:baltic}
\end{figure}

The matched archive record is part of a much larger Storm Babet cluster that also contains North Sea coastal and inland flooding outside the documented Baltic regions.
Within the wider analysis window of \qtyrange{6}{16}{\degree}E and \qtyrange{53}{59}{\degree}N, the cluster contains \qty{384}{\km\squared} of indexed flooding.
This spatial breadth illustrates the archive's source-file event semantics: one record may combine geographically and hydrologically distinct observations collected within the same upstream cluster.
For the broader seasonal comparison shown in Fig.~\ref{si:fig:baltic}, three matched events represented by four archive files between 9~October and 27~November are combined by taking the maximum reconstructed depth at each cell.
This composite describes the maximum water detected during the analysis period and should not be interpreted as a simultaneous inundation surface.

\paragraph{Results.}
HANZE represents the surge as separate German and Danish coastal events affecting a combined 2,140 people.
The archive contains one acquisition associated with the surge, dated 30~October 2023, nine days after the 20--21~October peak.
This acquisition records \qty{54.5}{\km\squared} of detected inundation within the documented regions (Table~\ref{si:tab:limits}; Fig.~\ref{si:fig:baltic}).
When all three autumn events are combined, the maximum-observed composite covers \qty{84}{\km\squared} within the same regions.
The detected water is predominantly shallow and spatially fragmented, consistent with an observation made after the principal surge had receded.

\begin{table}[htbp]
\centering
\caption{Documented impacts and delayed satellite observations for the two detectability-limit case studies. Detected area represents standing water recorded within the corresponding HANZE regions during the first relevant post-peak acquisition and should not be interpreted as the maximum event extent.}
\label{si:tab:limits}
\small
\begin{tabular}{@{}lP{2.4cm}P{3.1cm}rr@{}}
\toprule
Event & Process & Documented impact (HANZE) & Detected area (\unit{\km\squared}) & Lag (days) \\
\midrule
Valencia DANA, Oct. 2024 & Flash / pluvial & 236 fatalities; 36,803 affected; \qty{226}{\km\squared} & 35.2 & 6 \\
\addlinespace[3pt]
Baltic surge, Oct. 2023 & Coastal surge & 2,140 affected across Germany and Denmark & 54.5 & 9 \\
\bottomrule
\end{tabular}
\end{table}

\paragraph{Evaluation.}
The case study extends the Valencia finding to a second rapid-recession regime and shows that coastal storm-surge flooding may also be represented only by delayed residual water.
However, the evidence is less conclusive than for Valencia because the HANZE records do not provide a documented affected area against which the \qty{54.5}{\km\squared} satellite footprint can be compared.
The small and fragmented detected footprint is consistent with substantial recession before the next relevant Sentinel-1 acquisition, but the available data do not establish how much of the maximum surge extent was missed.
The result should therefore be interpreted as corroborating evidence for the archive's temporal detectability limit rather than as an independent estimate of event completeness.

The case also illustrates the importance of distinguishing a documented hydrological event from an archive-defined cluster.
The matched source record includes Baltic surge flooding together with other coastal and inland observations outside the documented regions, while the broader autumn composite combines several events over nearly two months.
Analyses requiring event-specific attribution must therefore restrict the spatial domain, inspect acquisition and cluster dates, and avoid interpreting a multi-raster maximum composite as the state of one flood at a single time.

% ==========================================================================
\subsection{Cross-dataset corroboration audit against HANZE}
\label{si:sec:hanze}

\paragraph{Motivation.}
Section~6 of the main text reports the principal corroboration results, including overall corroboration and variation by flood type, year, and country.
This section documents the full audit underlying those results and examines how the apparent correspondence between the satellite archive and HANZE changes across these three dimensions.
It first documents the coarser same-country criterion in full and then the region-level criterion on which the results in the main text are based.
The analysis is designed as a fitness-for-use assessment of the source archive rather than as a validation of individual satellite footprints.

\subsubsection{Country-Level}

\paragraph{Data and method.}
The reference dataset is HANZE v3.0.1b, which contains 397 documented European flood records with start dates between 2015 and 2025.
Under the same-country criterion, a HANZE record is classified as corroborated when the EuroFlood index contains at least one archived event in the same country whose date interval falls within a seven-day extension of the documented event period.
This criterion measures same-country temporal co-occurrence between the two records and does not establish that the matched satellite footprint corresponds spatially to the documented flood.
The binary event-level results are aggregated by flood type, calendar year, and country.
Two additional analyses assess the robustness and interpretation of the raw corroboration rates.
First, sensitivity to the temporal matching criterion is evaluated using padding windows of three, seven, and fourteen days.
Second, a permutation null estimates the corroboration expected from temporal coincidence alone.
For each of 1,000 permutations, the start date of every HANZE record is redrawn uniformly within 2015--2025 while preserving its country and duration, after which the same matching procedure is repeated.
This null is necessary because the satellite archive contains several hundred events per year, making same-country date matches relatively common even when the records are unrelated.

\paragraph{Results.}
Using the seven-day criterion, 84.6\% of the 397 HANZE records are corroborated by at least one archived event.
Corroboration varies substantially by flood type (Table~\ref{si:tab:hanze}a).
River floods have the highest rate among the well-sampled categories, with 91.8\% of 195 records corroborated, whereas flash floods have a rate of 77.1\% across 192 records.
The river/coastal and coastal categories contain only two and eight records, respectively, and their percentages are therefore too uncertain for substantive interpretation.
Corroboration also varies across the archive period (Table~\ref{si:tab:hanze}b).
The annual rate increases from 76.1\% in 2015 to 90.0\% in 2025, although the progression is not monotonic: it peaks at 97.6\% in 2024 and includes a pronounced minimum of 68.0\% in 2022.
These values demonstrate that correspondence between the two records is temporally non-stationary, but they do not by themselves identify whether the variation originates from satellite acquisition, archive processing, flood characteristics, HANZE reporting, or sampling uncertainty.
Country-level rates are similarly heterogeneous (Table~\ref{si:tab:hanze}c).
Ireland, Romania, the United Kingdom, and Bosnia and Herzegovina have 100\% corroboration among countries with at least six documented records.
The lowest rates occur in Switzerland at 37.5\% and Austria at 44.4\%, although both estimates are based on fewer than ten HANZE records.
The geographical pattern is consistent with reduced detectability of rapid and topographically confined floods, but the permissive country-level matching criterion and small samples preclude a definitive terrain-based attribution.
The overall corroboration rate is moderately sensitive to the matching window, increasing from 80.9\% with a three-day extension to 84.6\% with seven days and 89.2\% with fourteen days.
The corresponding excess above the permutation null remains positive across all three windows, showing that the result is not produced solely by the selected seven-day criterion.
For the headline seven-day analysis, the mean null corroboration rate is 78.8\%, with a central 95\% interval of 75.3--82.4\%.
The observed value of 84.6\% exceeds all but one of the 1,000 permuted values, corresponding to an empirical $p=0.002$.
The excess above the null is concentrated in river floods.
River-flood corroboration is 91.8\% compared with a null expectation of 79.8\%, whereas flash-flood corroboration is 77.1\% compared with a null expectation of 77.3\% and is not statistically distinguishable from the permuted record ($p=0.58$).
The raw flash-flood rate therefore provides little evidence of event-specific archive sensitivity beyond the level expected from same-country temporal coincidence.
Figure~5 of the main text presents the flood-type rates of the region-level criterion relative to its own null distributions, with the same-country rates shown as open circles for comparison, and summarises the annual and country-level results.
One dependence affects the uncertainty intervals shown in the main-text figure.
HANZE records are country-specific episodes, so one synoptic flood can contribute several correlated rows when it affects multiple countries, as occurred during Storm Boris in September 2024.
The Wilson intervals treat rows as independent and may consequently be slightly too narrow.

\begin{table}[htbp]
\centering
\caption{Corroboration of 397 HANZE flood records from 2015 to 2025 by the satellite archive. A documented record is classified as corroborated when at least one archived event occurs in the same country within a seven-day extension of its documented dates. Panel~(a) reports results by flood type, panel~(b) by year, and panel~(c) by country for countries with at least six documented records.}
\label{si:tab:hanze}
\small
\begin{tabular}{@{}l@{\hskip 1.2em}l@{\hskip 1.2em}l@{}}
\begin{tabular}[t]{@{}lrr@{}}
\multicolumn{3}{@{}l}{(a) By flood type} \\
\toprule
Type & Rate (\%) & $n$ \\
\midrule
River & 91.8 & 195 \\
River/coastal & 100.0 & 2 \\
Coastal & 87.5 & 8 \\
Flash & 77.1 & 192 \\
\bottomrule
\end{tabular}
&
\begin{tabular}[t]{@{}lrr@{}}
\multicolumn{3}{@{}l}{(b) By year} \\
\toprule
Year & Rate (\%) & $n$ \\
\midrule
2015 & 76.1 & 46 \\
2016 & 78.6 & 42 \\
2017 & 81.5 & 27 \\
2018 & 78.8 & 33 \\
2019 & 86.5 & 37 \\
2020 & 93.5 & 31 \\
2021 & 94.9 & 39 \\
2022 & 68.0 & 25 \\
2023 & 82.6 & 46 \\
2024 & 97.6 & 41 \\
2025 & 90.0 & 30 \\
\bottomrule
\end{tabular}
&
\begin{tabular}[t]{@{}lrr@{}}
\multicolumn{3}{@{}l}{(c) By country} \\
\toprule
Country & Rate (\%) & $n$ \\
\midrule
Ireland & 100.0 & 6 \\
Romania & 100.0 & 14 \\
United Kingdom & 100.0 & 23 \\
Bosnia and Herz. & 100.0 & 8 \\
France & 97.9 & 47 \\
Spain & 94.9 & 59 \\
Croatia & 90.0 & 10 \\
Norway & 85.7 & 14 \\
Italy & 84.9 & 73 \\
Serbia & 81.8 & 11 \\
Germany & 81.2 & 16 \\
Poland & 77.8 & 9 \\
Bulgaria & 75.0 & 8 \\
Greece & 66.7 & 30 \\
Albania & 61.5 & 13 \\
Austria & 44.4 & 9 \\
Switzerland & 37.5 & 8 \\
\bottomrule
\end{tabular}
\end{tabular}
\end{table}

\paragraph{Evaluation.}
The audit provides evidence that the archive contains meaningful information about documented river floods, while also showing that a high raw corroboration rate does not necessarily imply strong detection capability.
The permutation analysis is essential to this interpretation because same-country temporal matches occur frequently by chance in a dense continental event archive.
For river floods, the substantial excess above the null supports the archive's use for retrospective event discovery and screening.
For flash floods, the absence of an excess above the null is consistent with the detectability limitations demonstrated by the Valencia case and cautions against treating archive presence or absence as reliable evidence of event occurrence.
The country-level results identify potential geographical variation in fitness for use, but they should not be interpreted as precise national performance rankings.
Matching is performed at country scale rather than by spatial overlap, HANZE sample sizes differ considerably among countries, and several documented floods contribute correlated records across national borders.
The low rates in Austria and Switzerland are consistent with under-detection of rapid Alpine floods, and the region-level criterion below provides the spatially explicit test of this explanation.
The annual results show that the correspondence between the archive and the documentary record cannot be assumed constant over time.
However, the apparent increase should not be attributed solely to improvements in Sentinel-1 coverage or Global Flood Monitoring processing because changes in flood composition, documentary reporting, and small annual samples may also contribute.
The 2022 minimum is notable but should not receive a causal interpretation without a dedicated analysis of acquisition availability and event composition.
Corroboration is therefore best understood as an indicator of cross-dataset correspondence rather than a measure of completeness or accuracy.
HANZE contains damaging floods documented through impact records, while the satellite archive records water visible during available acquisitions, and neither dataset is a complete inventory of all European flooding.
The main practical implication is that archive-derived temporal, geographical, and process-based comparisons should account explicitly for the non-stationary and process-dependent corroboration quantified here.

\subsubsection{Region-Level}

\paragraph{Data and method.}

The same-country criterion requires only temporal coincidence within the same country.
The region-level criterion additionally requires spatial overlap: a HANZE record is corroborated when an archived event overlaps the documented period extended by seven days and contains index cells within at least one reported affected region.
Region geometries are taken from HANZE v3.0.1b, which resolves all 660 region codes represented in the 397 records.
Each region is queried directly from the index, so the audit requires no source-raster downloads.
The criterion measures spatio-temporal co-occurrence rather than event identity.
Administrative regions are much larger than individual flood footprints, archive files may combine spatially dispersed observations, and index footprints are enlarged by conservative resampling.
The criterion is therefore deliberately permissive.
Chance correspondence is estimated with three null models, each using 1,000 permutations.
The uniform null randomly reassigns flood dates within 2015--2025 while preserving region and duration.
The seasonal null moves each record to the same time of year in another archive year, with a random shift of up to 15~days.
The archive-shift null applies a common circular shift to all archived-event dates, preserving their temporal clustering.

\paragraph{Results.}

Using the seven-day window, 50.9\% of the 397 HANZE records are corroborated, compared with 35.4--35.9\% under the three null models (Table~\ref{si:tab:hanze-regions-type}).
The observed rate exceeds all 1,000 permutations under each null.
River floods show the strongest signal: 68.7\% are corroborated compared with null expectations of 42.7--46.6\%.
Flash floods are corroborated at 30.7\%, compared with 23.3--26.7\% under the null models.
Their excess is significant only under the seasonal null, whereas the river-flood excess is robust across all three null models.
Of the 336 records matched under the same-country criterion, 202 (60.1\%) remain matched at region level, including 74.9\% of river-flood matches but only 39.9\% of flash-flood matches.
No record is matched at region level without also being matched at country level.
The excess above the uniform null remains positive across temporal windows of 3--14 days and minimum detected-area thresholds up to \qty{5}{\km\squared} (Table~\ref{si:tab:hanze-regions-grid}).
For the seven-day window, the excess increases from 15.4 percentage points for any detected cell to 20.0 percentage points when at least \qty{5}{\km\squared} of inundation is required.
Annual corroboration ranges from 37.0\% in 2015 to 68.3\% in 2024 and is 53.3\% in 2025.

\begin{table}[htbp]
\centering
\caption{Region-level corroboration of 397 HANZE records by flood type using a seven-day matching window. Corroborated shares include Wilson 95\% intervals. Null columns report the mean of 1,000 permutations with empirical $p$-values in parentheses. The final column gives the share of same-country matches retained at region level.}
\label{si:tab:hanze-regions-type}
\footnotesize
\setlength{\tabcolsep}{4pt}
\begin{tabular}{@{}lrlllll@{}}
\toprule
Flood type & $n$ & Corroborated (\%) & Uniform null & Seasonal null & Archive-shift null & Persisting matches (\%) \\
\midrule
River & 195 & 68.7 (61.9--74.8) & 42.8 (0.001) & 46.6 (0.001) & 42.7 (0.001) & 74.9 (134/179) \\
Flash & 192 & 30.7 (24.6--37.6) & 26.6 (0.07) & 23.3 (0.003) & 26.7 (0.13) & 39.9 (59/148) \\
Coastal & 8 & 87.5 (52.9--97.8) & 61.0 (0.10) & 67.5 (0.18) & 60.4 (0.10) & 100.0 (7/7) \\
River/Coastal & 2 & 100.0 (34.2--100.0) & 68.8 (0.47) & 71.5 (0.50) & 68.9 (0.45) & 100.0 (2/2) \\
\midrule
All types & 397 & 50.9 (46.0--55.8) & 35.5 (0.001) & 35.9 (0.001) & 35.4 (0.001) & 60.1 (202/336) \\
\bottomrule
\end{tabular}
\end{table}

\begin{table}[htbp]
\centering
\caption{Sensitivity of region-level corroboration to the temporal window and minimum detected area. Values give the observed share, mean uniform-null share, and their difference in percentage points. All observed shares exceed the 1,000 permuted values ($p=0.001$).}
\label{si:tab:hanze-regions-grid}
\footnotesize
\setlength{\tabcolsep}{4pt}
\begin{tabular}{@{}lrrrrrrrrr@{}}
\toprule
 & \multicolumn{3}{c}{Three days} & \multicolumn{3}{c}{Seven days} & \multicolumn{3}{c}{Fourteen days} \\
\cmidrule(lr){2-4}\cmidrule(lr){5-7}\cmidrule(l){8-10}
Detected area in regions & obs. & null & excess & obs. & null & excess & obs. & null & excess \\
\midrule
Any index cell & 49.6 & 31.8 & 17.8 & 50.9 & 35.5 & 15.4 & 52.4 & 40.8 & 11.6 \\
At least \qty{1}{\km\squared} & 43.6 & 23.6 & 20.0 & 44.8 & 26.2 & 18.6 & 46.3 & 30.2 & 16.2 \\
At least \qty{5}{\km\squared} & 36.5 & 16.1 & 20.4 & 37.8 & 17.8 & 20.0 & 38.8 & 20.5 & 18.3 \\
\bottomrule
\end{tabular}
\end{table}

\paragraph{Evaluation.}

Adding the spatial criterion substantially reduces chance matching: the uniform-null expectation falls from 78.8\% for the same-country test to 35.5\% at region level.
Most corroborated records are also relatively unambiguous, with a median of one candidate archived event and a 90th percentile of two.
For 81\% of corroborated records, the documented onset falls within an archived observation window.
For the remainder, the first matched window begins a median of three days after onset, consistent with delayed satellite observations such as the Valencia and Baltic cases.
The spatial criterion nevertheless remains coarse.
Detected inundation covers a median of only 0.16\% of the reported affected regions.
For the 26 corroborated records with a HANZE affected-area estimate, the largest matched archive footprint has a median area ratio of 0.52 relative to the documented area (interquartile range 0.21--1.20).
This ratio should not be interpreted as completeness because index resampling enlarges the detected footprint and the two datasets use different spatial definitions.
Across the 17 countries with at least six records, corroboration decreases as the share of flash floods increases ($r=-0.51$).
Rates are at least 75\% in Ireland, the United Kingdom, Norway, Romania, and Bosnia and Herzegovina, but substantially lower in Austria, Poland, Bulgaria, Greece, Italy, and Switzerland.
Several country estimates are based on small samples, and records from the same transboundary flood are not statistically independent.

% ==========================================================================
\subsection{Synthesis: a fitness-for-use profile of the observed archive}
\label{si:sec:synthesis}

Taken together, the case studies and corroboration audit define a consistent fitness-for-use profile for the observed archive, complementing the discussion in Section~8 of the main text.
The archive is best suited to large, persistent, lowland river floods for which inundation remains visible during one or more Sentinel-1 acquisitions.
This regime is represented by the continental recurrence analysis in Section~7.1 of the main text, the repeated flooding of the Shannon callows (Section~\ref{si:sec:shannon}), the transboundary cluster associated with Storm Boris (Section~\ref{si:sec:boris}), and the strong excess above the permutation null for documented river floods (Section~\ref{si:sec:hanze}).
The archive is less reliable for short-lived or spatially confined processes, including flash and pluvial floods (Section~\ref{si:sec:dana}), coastal storm surges (Section~\ref{si:sec:baltic}), and floods in steep terrain.
The geographical pattern is consistent with this interpretation, although small samples and differences in flood composition prevent a definitive attribution to topography.

Several independent findings support this profile.
Under the region-level criterion used in the main text, river-flood corroboration exceeds chance across all three null models, whereas the smaller flash-flood excess is significant only under the seasonal null.
The Valencia and Baltic case studies show directly how rapid inundation may peak and recede before the next relevant satellite acquisition, leaving only delayed residual water in the archive.
The exposure studies further demonstrate that a spatially detailed detected footprint can omit important parts of a documented event, particularly in steep valleys and urban areas.
These results consistently indicate that archive presence provides useful evidence of detected inundation, while archive absence does not establish that flooding did not occur.

Two additional properties qualify all quantitative uses of the archive.
First, index-derived extents are affected by opposing sources of spatial bias: synthetic-aperture radar under-detection can omit inundation, while conservative any-wet resampling to the approximately \qty{90}{\m} index grid enlarges the detected footprint by a median factor of 1.50 relative to the native \qty{20}{\m} wet area.
The index is therefore appropriate for discovery, recurrence screening, and spatial selection, whereas inundated area and depth statistics should be calculated from the native-resolution rasters and interpreted with the uncertainties of the source product.
Second, the corroboration of documented floods is temporally non-stationary.
The region-level results vary across years and do not reproduce the pronounced 2022 minimum of the same-country criterion.
Interannual changes in event counts, recurrence, or mapped area consequently combine variation in flood occurrence with variation in acquisition opportunities, detectability, archive processing, and documentary coverage.

The archive should therefore be used primarily for retrospective event discovery, observation-based screening, and comparison with complementary hazard and impact datasets, rather than as an uncorrected record of flood frequency or complete event extent.
Its strongest applications are those that acknowledge the archive's event definition, inspect acquisition timing, distinguish detected from total inundation, and combine satellite observations with documentary or modelled information where completeness is important.

\section{Verification details}
\label{si:sec:si-verification}

\subsection{Index verification}
\label{si:sec:si-verify}

The verification sample contains 110 events, selected as ten random events from each archive year using seed 42.
The same sample is used for the observed--modelled comparison in Section~7.2 of the main text.
Each event footprint reconstructed from the published index was compared with two derivations from the native \qty{20}{\m} source raster.
The first re-runs the EuroFlood ingest routine.
The second implements the ingest rule independently, without importing EuroFlood code: positive-depth source pixels are transformed to geographic coordinates and assigned to the corresponding index cell.
All 110 events are reproduced exactly by both methods ($\mathrm{IoU}=1.000$), with no omitted or additional cells across 5.74 million compared memberships.
Eight additional events representing archive extremes, including the largest source file and largest footprint, are also reproduced exactly.
A complete count over all 3,611 events provides a further release-wide check: for every event, the number of indexed cells matches the count recorded during source ingest.
Across the verification sample, conservative resampling enlarges the native wet area by a median factor of 1.50 (10th--90th percentile: 1.28--1.94).
The ingest imposes no upper depth threshold.
In the verification sample, 0.34\% of wet source pixels exceed \qty{20}{\m} reconstructed depth.
Depth-based case studies exclude these values as implausible, so their native-resolution extents may be slightly smaller than the corresponding index footprints.

\subsection{Systematic query-cost sweep}
\label{si:sec:si-sweep}

\begin{figure}[htbp]
\centering
\includegraphics[width=\linewidth]{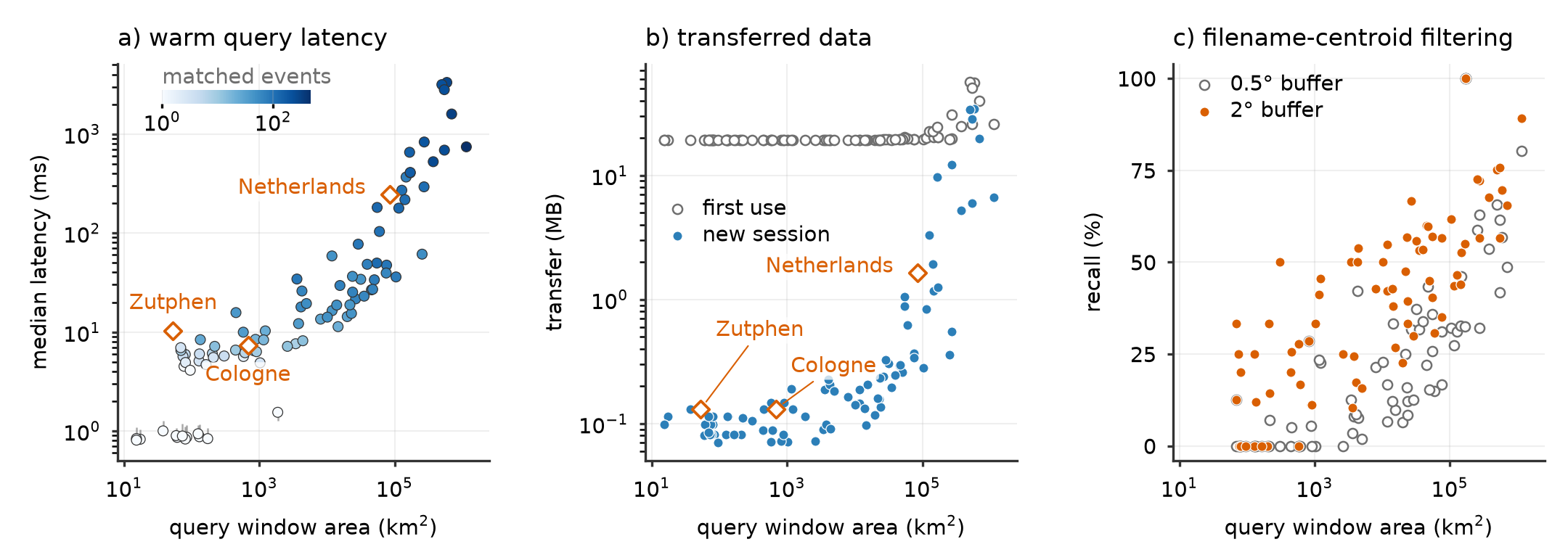}
\caption{Query performance across 90 spatial windows. (a)~Warm-query latency, with interquartile ranges and points coloured by the number of matched events. (b)~Data transferred by the first query and by a new process with a warmed cache. (c)~Recall of filename-centroid filtering with \ang{0.5} and \ang{2} buffers. Diamonds mark the three illustrative queries reported in the main text.}
\label{si:fig:sweep}
\end{figure}

Query performance was evaluated using 90 prespecified windows sampled from the bounding boxes of 1,765 GISCO NUTS regions.
The regions were divided into six logarithmic area classes, with up to 15 windows sampled from each class using seed 42.
Window areas range from \qty{15}{\km\squared} to \qty{1.1e6}{\km\squared}; 13 contain no archived events.
For each window, a local HTTP range server recorded transferred bytes for a first query with an empty cache and for a new process with a warmed disk cache.
Each condition was repeated five times, followed by ten repeated queries to measure warm latency.
Nine windows were also tested against the public endpoint.
Across the 90 windows, the first query transferred a median of \qty{19.3}{\mega\byte}, a new process with a warmed cache transferred \qty{0.14}{\mega\byte}, and repeated queries completed in \qty{14}{\ms}.
Warm transfer and latency increased sublinearly with query area, with fitted exponents of 0.38 and 0.59, respectively (Fig.~\ref{si:fig:sweep}).
Filename-centroid filtering recovered a median of 15\% of intersecting events with a \ang{0.5} buffer and 42\% with a \ang{2} buffer.
Complete recall was achieved for only 2.6\% of windows containing events.
A footprint catalogue recovered all events by construction but, unlike the inverted index, does not retain cell-level event membership.

% ==========================================================================
\section{Related software tools}
\label{si:sec:si-tools}

EuroFlood sits within a broader software ecosystem for accessing, processing, and analysing flood and Earth-observation data.
The tools summarised in Table~\ref{si:tab:tools} address related but distinct tasks, including hydrological data retrieval, modelled-hazard access, satellite-based flood mapping, depth reconstruction, and generic catalogue discovery.
The comparison is intended to clarify these functional differences rather than to rank the tools or present them as direct substitutes for one another.
The relationship between these tools and the specific contribution of EuroFlood is discussed in Sections~3 and~5 of the main text.

\begin{table}[htbp]
\centering
\caption{Selected software tools related to flood and Earth-observation data access. The table identifies the principal data or product handled by each tool and its primary role within the wider software ecosystem.}
\label{si:tab:tools}
\footnotesize
\setlength{\tabcolsep}{4pt}
\begin{tabular}{@{}P{0.24\linewidth}P{0.33\linewidth}P{0.35\linewidth}@{}}
\toprule
Tool or interface & Data or product accessed & Primary role \\
\midrule
FLEXTH & Flood depth reconstructed from observed extent and terrain & Depth-reconstruction method used to generate the CEMS-EFAS flood-depth maps \\
\addlinespace[2pt]
\texttt{CLIMADA-petals} \texttt{rf\_glofas} & GLOFAS discharge and modelled flood depth through CDS & Flood-hazard retrieval within a climate-risk modelling framework \\
\addlinespace[2pt]
GEE \texttt{FloodHazard} with \texttt{geemap} & Modelled GLOFAS flood-hazard layers & Cloud-based raster access, visualisation, and analysis \\
\addlinespace[2pt]
Planetary Computer (Deltares) & Modelled coastal flood-depth layers distributed through STAC and COG & Cloud-native catalogue and raster delivery \\
\addlinespace[2pt]
\texttt{climetlab-cems-flood} & EFAS and GLOFAS river-discharge data through CDS & Programmatic retrieval of hydrological time series \\
\addlinespace[2pt]
GFM through openEO & Sentinel-1-derived flood extent from Global Flood Monitoring & Standardised Earth-observation discovery and processing workflows \\
\addlinespace[2pt]
\texttt{HyRiver} and \texttt{pygeohydro} & Hydrological and geospatial datasets for the United States & Domain-specific access to multiple hydrological data providers \\
\addlinespace[2pt]
\texttt{hydrafloods} and \texttt{FLOODPY} & Flood extent derived from raw synthetic-aperture radar imagery & Satellite flood-mapping and image-processing workflows \\
\addlinespace[2pt]
\texttt{cdsapi}, \texttt{earthaccess}, and \texttt{pystac-client} & Provider catalogues and archives, including CDS, NASA, and STAC services & Generic discovery and retrieval of Earth-observation and environmental data \\
\bottomrule
\end{tabular}
\end{table}

\end{document}